\documentclass{aa}
\usepackage{graphicx}
\usepackage{amssymb}
\usepackage{amsmath}
\usepackage{natbib}
\usepackage{mathrsfs}
\usepackage{ulem}
\usepackage{txfonts}
\usepackage{lmodern}
\usepackage{booktabs}

\usepackage[switch]{lineno} 

\newcommand{\corot}{{\textsc{CoRoT}}}

\newcommand{\kepler}{\textit{Kepler}}
\newcommand{\otto}{KIC7341231}
\newcommand{\duree}{650}

\newcommand{\ind}[1]{_{\rm #1}}
\newcommand{\ex}[1]{^{\rm #1}}

\def\m2s2{\,m$^{2}$\,s$^{-2}$} 

\newcommand{\vect}[1]{\boldsymbol{\rm #1}}

\newcommand{\vaisala}{Brunt-V\"ais\"al\"a}
\newcommand{\cesam}{\textsc{Cesam2k}}
\newcommand{\astec}{\textsc{ASTEC}}

\newenvironment{itemize*}%
  {\begin{itemize}%
    \setlength{\itemsep}{1pt}%
    \setlength{\parskip}{1pt}}%
  {\end{itemize}}

\newcommand\T{\rule{0pt}{2.6ex}}
\newcommand\B{\rule[-1.2ex]{0pt}{0pt}}

\begin{document}
\title{Seismic constraints on the radial dependence of the internal rotation profiles of six \textit{Kepler} subgiants and young red giants}
\titlerunning{Constraints on the internal rotation profiles of six \textit{Kepler} red giants}
\author{
S. Deheuvels\inst{1,2}
\and G. Do\u{g}an\inst{3,4}
\and M.~J. Goupil\inst{5}
\and T. Appourchaux\inst{6}
\and O. Benomar\inst{7}
\and H. Bruntt\inst{4,8}
\and T.~L. Campante\inst{9,4}
\and L. Casagrande\inst{10}
\and T. Ceillier\inst{11}
\and G.~ R. Davies\inst{9,11,4}
\and P. De Cat\inst{12}
\and J.~N. Fu\inst{13}
\and R.~A. Garc{\'{\i}}a\inst{11}
\and A. Lobel\inst{12}
\and B. Mosser\inst{5}
\and D.~R. Reese\inst{14}
\and C. Regulo\inst{15,16}
\and J. Schou\inst{20}
\and T. Stahn\inst{17}
\and A.~O. Thygesen\inst{18}
\and X.~H. Yang\inst{13}
\and W.~J. Chaplin\inst{9,4}
\and J. Christensen-Dalsgaard\inst{4}
\and P. Eggenberger\inst{19}
\and L. Gizon\inst{17,20}
\and S. Mathis\inst{11}
\and J. Molenda-\.{Z}akowicz\inst{21}
\and M. Pinsonneault\inst{22}
}

\institute{Universit\'e de Toulouse; UPS-OMP; IRAP; Toulouse, France
\and CNRS; IRAP; 14, avenue Edouard Belin, F-31400 Toulouse, France
\and High Altitude Observatory, National Center for Atmospheric Research, P.O. Box 3000, Boulder, CO 80307, USA 
\and Stellar Astrophysics Centre, Department of Physics and Astronomy, Aarhus University, Ny Munkegade 120, DK-8000 Aarhus C, Denmark 
\and LESIA, UMR8109, Observatoire de Paris, Universit\'e Pierre et Marie Curie, Universit\'e Denis Diderot, CNRS, 5 Place Jules Janssen 92195 Meudon Cedex, France 
\and Institut d'Astrophysique Spatiale, UMR8617, Universit\'e Paris XI, B\^atiment 121, 91405 Orsay Cedex, France 
\and Sydney Institute for Astronomy (SIfA), School of Physics, University of Sydney, NSW 2006, Australia 
\and Aarhus Katedralskole, Skolegyde 1, DK-8000 Aarhus C, Denmark
\and School of Physics and Astronomy, University of Birmingham, Edgbaston, Birmingham, B15 2TT, UK 
\and Research School of Astronomy and Astrophysics, Mount Stromlo Observatory, The Australian National University, ACT 2611, Australia 
\and Laboratoire AIM Paris-Saclay, CEA/DSM-CNRS-Universit\'e Paris Diderot, IRFU/SAp, Centre de Saclay, 91191 Gif-sur-Yvette cedex, France 
\and Royal Observatory of Belgium, Ringlaan 3, B-1180 Ukkel, Belgium 
\and Department of Astronomy, Beijing Normal University, Beijing, China 
\and Institut d'Astrophysique et de G\'eophysique de l'Universit\'e de Li\`ege, All\'ee du 6 Ao\^ut 17, 4000 Li\`ege, Belgium 
\and Instituto de Astrof\`isica de Canarias, 38205, La Laguna, Tenerife, Spain 
\and Universidad de La Laguna, Dpto de Astrof\`isica, 38206, La Laguna, Tenerife, Spain 
\and Institut f\"ur Astrophysik, Georg-August-Universit\"at G\"ottingen, 37077 G\"ottingen, Germany 
\and Zentrum f\"ur Astronomie der Universit\"at Heidelberg, Landessternwarte, K\"onigstuhl 12, 69117 Heidelberg, Germany 
\and Observatoire de Gen\`eve, Universit\'e de Gen\`eve, 51 Ch. des Maillettes, CH-1290 Sauverny, Suisse 
\and Max-Planck-Institut f\"ur Sonnensystemforschung, 37191 Katlenburg-Lindau, Germany 
\and Instytut Astronomiczny Uniwersytetu Wroc{\l}awskiego, ul. Kopernika 11, 51-622 Wroc{\l}aw, Poland 
\and Department of Astronomy, the Ohio State University, Columbus, OH, 43210 USA 
}

\offprints{S. Deheuvels\\ \email{sebastien.deheuvels@irap.omp.eu}
}


\abstract{We still do not understand which physical mechanisms are responsible for the transport of angular momentum inside stars. The recent detection of mixed modes that contain the clear signature of rotation in the spectra of \kepler\ subgiants and red giants gives us the opportunity to make progress on this issue.}
{Our aim is to probe the radial dependance of the rotation profiles for a sample of \kepler\ targets. For this purpose, subgiants and early red giants are particularly interesting targets because their rotational splittings are more sensitive to the rotation outside the deeper core than is the case for their more evolved counterparts.}
{We first extract the rotational splittings and frequencies of the modes for six young \kepler\ red giants. We then perform a seismic modeling of these stars using the evolutionary codes \cesam\ and \astec. By using the observed splittings and the rotational kernels of the optimal models, we perform inversions of the internal rotation profiles of the six stars.}
{We obtain estimates of the core rotation rates for these stars, and upper limits to the rotation in their convective envelope. We show that the rotation contrast between the core and the envelope increases during the subgiant branch. Our results also suggest that the core of subgiants spins up with time, while their envelope spins down. For two of the stars, we show that a discontinuous rotation profile with a deep discontinuity reproduces the observed splittings significantly better than a smooth rotation profile. Interestingly, the depths that are found most probable for the discontinuities roughly coincide with the location of the H-burning shell, which separates the layers that contract from those that expand.}
{We characterized the differential rotation pattern of six young giants with a range of metallicities, and with both radiative and convective cores on the main sequence. This will bring observational constraints to the scenarios of angular momentum transport in stars. Also, if the existence of sharp gradients in the rotation profiles of young red giants is confirmed, it should help discriminate between the physical processes that could transport angular momentum in the subgiant and red giant branches.}

\keywords{Stars: oscillations -- Stars: rotation -- Stars: evolution}

\maketitle

\section{Introduction}

Rotation is a key element to understanding stellar structure and evolution. However, the way angular momentum (AM) is transported inside stars remains uncertain. Hydrodynamic mechanisms and meridional circulation as currently implemented in 1D stellar evolution codes are not efficient enough to account for the solid-body rotation of the solar radiative interior (\citealt{zahn92}, \citealt{mathis04}), which has been found through helioseismology (\citealt{schou98}, \citealt{chaplin99}, \citealt{garcia04}, \citealt{effdarwich13}). Other processes are probably at work, such as transport through internal gravity waves (e.g. \citealt{charbonnel05}), through a fossil magnetic field (e.g. \citealt{gough98}), or through magnetic instabilities (e.g. \citealt{spruit99}). However, the importance of the contributions of these processes, as well as the timescales over which they operate, are still unclear.

Asteroseismology can bring a significant contribution to this debate by providing observational constraints on the internal rotation profiles of stars. Indeed, rotation is known to lift the degeneracy between non-radial modes of same radial orders and degrees, but different azimuthal orders, and the frequency splitting between these modes (known as the \textit{rotational splitting}) is directly linked to the internal rotation. The space missions \corot\ (\citealt{baglin06b}) and \kepler\ (\citealt{borucki10}) are providing us with unprecedentedly long time series, which have already made it possible to measure rotational splittings for certain stars. For instance, an average of the internal rotation of the main sequence \corot\ target HD52265 was successfully estimated by interpreting the observed rotational splittings (\citealt{ballot11}, \citealt{gizon13}).
The \kepler\ satellite, by observing stars over several years, gave the opportunity to measure the rotational splittings of the modes for hundreds of red giants (\citealt{mosser12b}). This result is all the more interesting, since in these stars the non-radial modes have a mixed character: they behave as pressure modes (p modes) in the envelope and as gravity modes (g modes) in the core (\citealt{osaki75}, \citealt{aizenman77}). Mixed modes were first detected from the ground (\citealt{kjeldsen95b}) and then from space with \corot\ (\citealt{rome}) and \kepler\ (e.g. \citealt{metcalfe10}, \citealt{mathur11}, \citealt{appourchaux12}). They have already made it possible to probe the core structure of subgiants (\citealt{deheuvels11}, \citealt{benomar13}) and red giants (\citealt{beck11}, \citealt{mosser12a}), thus allowing us to distinguish RGB stars from clump stars (\citealt{bedding11}, \citealt{mosser11}). The interpretation of the rotational splittings of mixed modes in several red giants showed that there exists a strong radial differential rotation in these stars, with the core rotating at least five times faster than the envelope (\citealt{beck12}, \citealt{deheuvels12}). \cite{mosser12b} found that this is a general feature of red giants, and showed that the core of these stars spins down as they ascend the red giant branch (RGB), in spite of the contraction of the central layers which should spin it up if AM were conserved. This implies effective AM transport during the ascent of the RGB.

Different classes of theoretical models predict radically different core rotation rates for giants depending on the efficiency of AM transport. The limiting case of an instantaneous exchange of AM (strong core-envelope coupling) for first ascent giant stars would imply that they rotate rigidly, which is clearly inconsistent with the detected level of differential rotation. On the other hand, for higher-mass secondary clump stars, \cite{tayar13} found that the measured core rotation rates are consistent with strongly coupled models. This is probably linked to the fact that such stars are expected to leave the main sequence with much more rapid rotation than their low-mass counterparts. Core rotation for first ascent giants obtained from asteroseismology is both much faster than predicted from strongly coupled models and slower by several orders of magnitude than expected from models with hydrodynamic AM transport (\citealt{eggenberger12}, \citealt{marques13}, \citealt{ceillier13}), showing the need for a more efficient source of AM transport in these stars. Attempts were recently made to estimate the timescale of AM exchange on the RGB. \cite{eggenberger12} found that the ad hoc diffusion coefficients that are required to explain the timescale for core-envelope decoupling in young MS stars (e.g. \citealt{denissenkov10}) can also reproduce the core rotation rates of some giants, suggesting that similar mechanisms might be at play. Assuming a solar-like rotation profile on the MS, \cite{tayar13} showed that the detected core rotation is consistent with post-MS decoupling during the first dredge-up phase. However, the nature of the physical process responsible for core-envelope coupling remains unknown.

Until now, the interpretation of rotational splittings of red giants brought information exclusively on the rotation of the innermost layers of the star because the rotational kernels of red giants are most sensitive to these regions (\citealt{goupil13}). However, constraining the radial dependence of the rotation profile would undoubtedly provide useful information on the processes of AM transport that are at work. For this purpose, subgiants and young red giants are particularly interesting targets because their rotational splittings are more sensitive to the rotation of the envelope than is the case in more evolved stars. We selected six \kepler\ subgiants and young red giants that seemed most favorable to probing the internal rotation profile. The selection process is described in Sect. \ref{sect_selection}. We first extracted the frequencies and rotational splittings of the oscillation modes by analyzing the power spectra of the six targets (Sect. \ref{sect_analysis}). In Sect. \ref{sect_spectro_photo}, we present the atmospheric parameters that were available for these stars prior to this study. Since two of the stars had not been observed spectroscopically before, we observed them from the ground and the results are presented in Sect. \ref{sect_spectro_photo}. In order to interpret the observed splittings, we searched for stellar models that reproduce both the surface observables and the observed mode frequencies in Sect. \ref{sect_model}. By using those, we performed inversions of the internal rotation profiles of these stars, which are presented in Sect. \ref{sect_inversion}. We obtained precise estimates of the core and the envelope rotation rates of these stars and we showed that for two of them a discontinuous rotation profile with a discontinuity located near the H-burning shell reproduces the observed splittings significantly better than a smooth rotation profile.

\section{Selection of targets \label{sect_selection}}

\begin{table*}
  \begin{center}
  \caption{Global seismic parameters of the selected targets and estimates of the stars' masses, radii and surface gravity inferred from scaling relations. \label{tab_global_param}}
\begin{tabular}{l c c c c c c}
\hline \hline
\T \B Star & Ref. letter & $\Delta\nu$ ($\mu$Hz) & $\nu\ind{max}$ ($\mu$Hz) & $M$ & $R$ & $\log g$ \\
\hline
\T KIC12508433 & A & $45.3\pm0.2$ &$ 793\pm21$ & $1.20\pm0.16$ &  $2.20\pm0.10$ &  $3.83\pm0.04$ \\
 KIC8702606 & B & $39.9\pm0.4$ &$ 664\pm14$ & $1.27\pm0.15$ &  $2.44\pm0.11$ &  $3.77\pm0.02$ \\
 KIC5689820 & C & $41.0\pm0.3$ &$ 695\pm15$ & $1.11\pm0.16$ &  $2.29\pm0.12$ &  $3.76\pm0.04$ \\
 KIC8751420 & D & $34.7\pm0.4$ &$ 598\pm14$ & $1.50\pm0.20$ &  $2.83\pm0.15$ &  $3.71\pm0.03$ \\
 KIC7799349 & E & $33.7\pm0.4$ &$ 561\pm8$ & $1.33\pm0.14$ &  $2.77\pm0.12$ &  $3.68\pm0.02$ \\ 
\B KIC9574283 & F & $30.0\pm0.5$ &$ 455\pm8$ & $1.24\pm0.17$ &  $2.92\pm0.17$ &  $3.60\pm0.02$ \\
\hline 
\end{tabular}
\end{center}
\end{table*}

Among the \kepler\ targets, we searched for stars for which the internal rotation profile can be probed by following a similar procedure as the one led by \cite{deheuvels12} (hereafter noted D12) for \otto. For this purpose, the stars must satisfy the following criteria:
\begin{itemize}
\item The stars need to have been observed over a long enough period so that the frequency resolution is much smaller than the rotational splittings. We selected stars that were observed over at least 5 quarters ($\sim470$ days), which corresponds to a frequency resolution below $0.02\,\mu$Hz.
\item Their modes should have a linewidth significantly smaller than the rotational splitting, to ensure that the $m$-components of the rotational multiplets are well separated. It has been shown by \cite{appourchaux12} that the mode linewidths increase very rapidly with increasing temperature ($\Gamma\propto T\ind{eff}^s$, with $s\sim16$). As a result, only the cooler targets have narrow enough modes so that the rotational splitting of the modes is clearly visible.
\item We restricted ourselves to stars that are not too evolved. Indeed, the core of subgiants and young red giants is less dense than that of more evolved star, which makes their rotational splittings more sensitive to the rotation in other regions than the innermost layers. Besides, young red giants can be modeled by using existing fitting procedures. For these stars, the combined knowledge of the large separation of acoustic modes $\Delta\nu$ and the period spacing $\Delta\Pi_1$ of $l=1$ gravity modes can yield precise estimates of the stellar mass and age for a given set of input physical parameters. This can be used to model these stars (\citealt{deheuvels11}). For more evolved red giants, the relation between $\Delta\nu$, $\Delta\Pi_1$, and the stellar mass becomes degenerate (for a given large separation $\Delta\nu$, a large change in mass induces almost no change in $\Delta\Pi_1$). This degeneracy occurs for stars whose mean large separation is below a threshold limit which varies between 30 and 40 $\mu$Hz depending on the stellar mass (\citealt{mosser12a}). We therefore retained only stars with $\langle\Delta\nu\rangle>30\,\mu$Hz.
\end{itemize}

\begin{figure}
\begin{center}
\includegraphics[width=9cm]{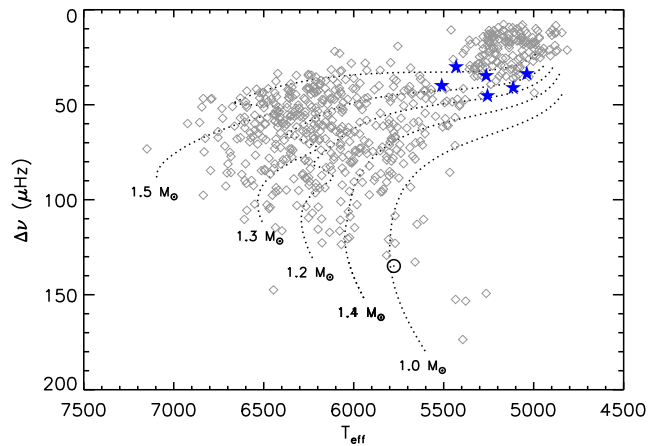}
\end{center}
\caption{Location of the selected targets in a seismic HR diagram (mean large separation $\Delta\nu$ against effective temperature). 
The blue filled stars indicate the six targets selected in our sample and the gray diamonds correspond to the set of \kepler\ targets studied by \cite{chaplin13}. The dashed lines indicate evolutionary tracks of models of different masses and solar metallicity.
\label{fig_HR}}
\end{figure}

We found six \kepler\ targets that satisfy these criteria simultaneously. They are listed in Table \ref{tab_global_param}. For clarity, these stars will be further referred to with letters A through F as specified in Table \ref{tab_global_param}. The seismic properties of these targets will be discussed in detail in Sect. \ref{sect_analysis}. However, preliminary information on the stars can already be obtained from their mean large separation $\langle\Delta\nu\rangle$ and the frequency of maximum power of their oscillations $\nu\ind{max}$. Indeed, scaling relations were proposed between these global seismic parameters and stellar properties such as the mass, radius and surface gravity (\citealt{brown91}). These scaling relations rely on the hypothesis that there exists a relation between $\nu\ind{max}$ and the acoustic cut-off frequency. From observations, these scaling relations have been empirically verified to work at the level of a few percent at least (\citealt{huber11}, \citealt{silva12}). \cite{belkacem11} recently proposed a theoretical explanation for this relation. We applied these scaling relations to the stars of our sample using the values of $\langle\Delta\nu\rangle$ and $\nu\ind{max}$ that were obtained by \cite{chaplin13} for these stars (see Table \ref{tab_global_param}). We  thus obtained first rough estimates of the masses, radii, and $\log g$ of the six stars, which are given in Table \ref{tab_global_param}\footnote{We note that to apply these scaling relations, estimates of the effective temperatures of the stars are also required. We here used the spectroscopic estimates that are obtained in Sect. \ref{sect_spectro_photo} of this paper.}. Fig. \ref{fig_HR} shows the location of the selected targets in an asteroseismic HR diagram (large separation plotted as a function of the effective temperature). The stars of our sample are roughly in the same evolutionary state ($3.59 \leqslant\log g\leqslant 3.83$) and they lie either at the end of the subgiant branch or at the base of the RGB. The absence of younger subgiants is caused by the fact that they are hotter. As a result, their modes have larger linewidths and it is much harder to extract their rotational splittings.

\section{Seismic properties \label{sect_analysis}}

The frequencies of the oscillation modes bear information about the internal structure of a star, and in particular about its internal rotation. Indeed, rotation is known to lift the degeneracy between the non-radial modes of same radial order $n$ and degree $l$ but different azimuthal order $m$, thus forming rotational multiplets. For slow rotators, the effects of the centrifugal force can be neglected, and if we further assume that the rotation profile is spherically symmetric, the frequency of the $(n,l,m)$ mode can be written as $\nu_{n,l,m} = \nu_{n,l,0}+m\delta\nu_{n,l}$, where $\delta\nu_{n,l}$ is known as the \textit{rotational splitting} and can be expressed as a weighted average of the rotation profile $\Omega(r)$
\begin{linenomath*}
\begin{equation}
\delta\nu_{n,l}\equiv\int \frac{K_{n,l}(r)\Omega(r)}{2\pi}\,dr.
\label{eq_split}
\end{equation}
\end{linenomath*}
The functions $K_{n,l}(r)$, known as the rotational kernels, essentially depend on the mode eigenfunctions. 

Our goal in the analysis of the oscillation spectra of the stars was twofold: 
\begin{enumerate}
\item estimating the mode frequencies in order to use them as observables for the modeling of the stars (see Sect. \ref{sect_model}). This requires first to identify the modes in the oscillation spectra.
\item extracting the rotational splittings to perform inversions of the internal rotation profiles (see Sect. \ref{sect_inversion}). 
\end{enumerate}

\subsection{\kepler\ observations}

The targets that were selected for this study have been observed with \kepler\ over periods ranging from 470 to \duree\ days\footnote{Stars A, B, E, and F were observed during \duree\ days from quarters Q5 to Q11. Star D was observed during 560 days (no observations during Q6), and star C during 470 days (no observations during Q5 and Q9).} with the short cadence mode (integration time of 58.84876 s).  Corrections have been applied to the raw \kepler\ time series: the light curves were processed using the \kepler\ pipeline developed by \cite{jenkins10}, and they were further corrected from outliers, occasional jumps, and drifts following \cite{garcia11}. Long-period instrumental drifts were also corrected by subtracting a smoothed version of the light curve over a width of 1 day.

The power density spectra of the selected stars were obtained by using the Lomb-Scargle periodogram (\citealt{lomb76}, \citealt{scargle82}). They all clearly show the signature of solar-like oscillations in frequency intervals that range from about $[300;550]\,\mu$Hz for the most evolved target (star F) to approximately $[550;950]\,\mu$Hz for the least evolved one (star A).

\subsection{Identification of the modes}

\begin{figure*}
\begin{center}
\includegraphics[width=9cm]{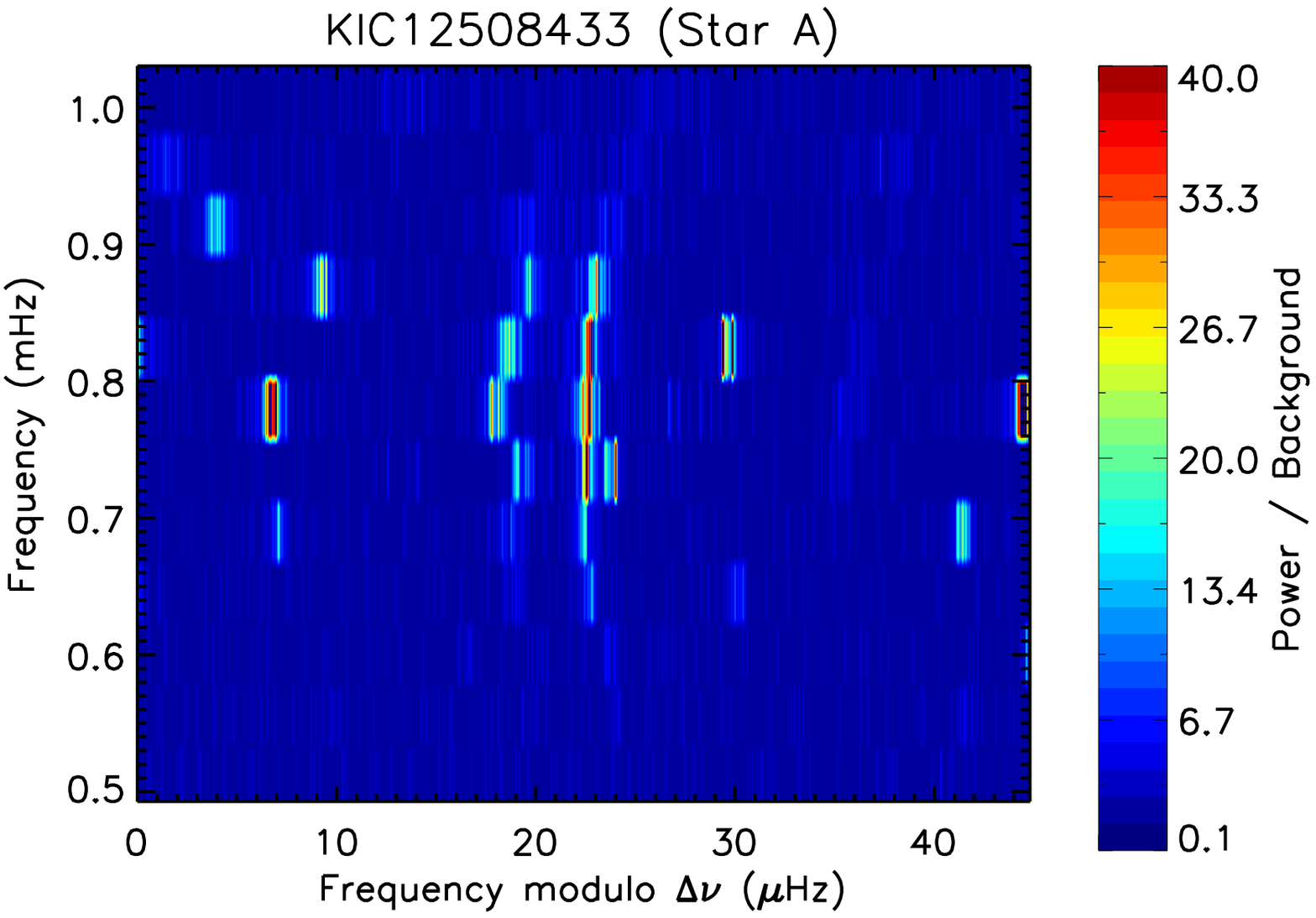}
\includegraphics[width=9cm]{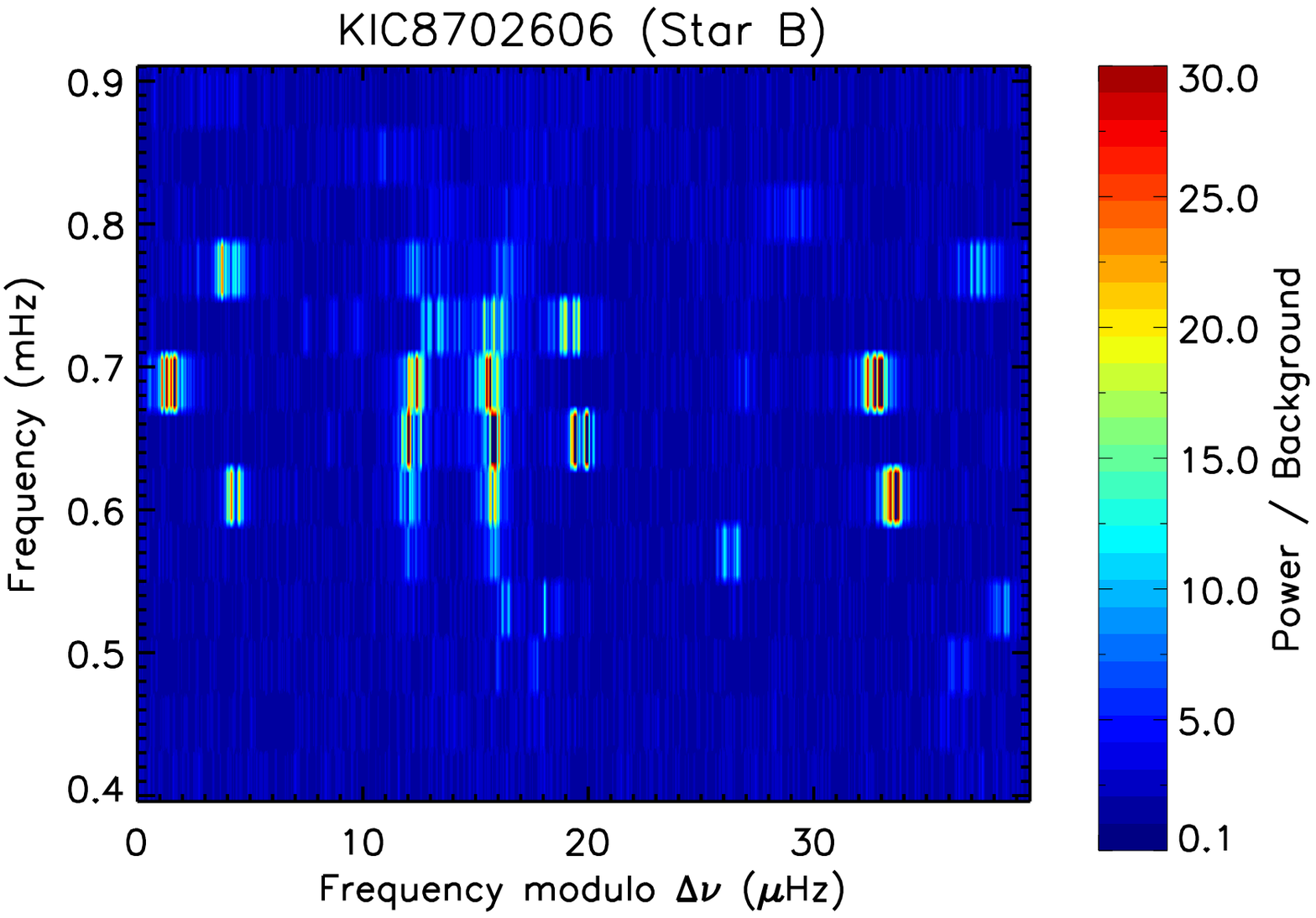}
\includegraphics[width=9cm]{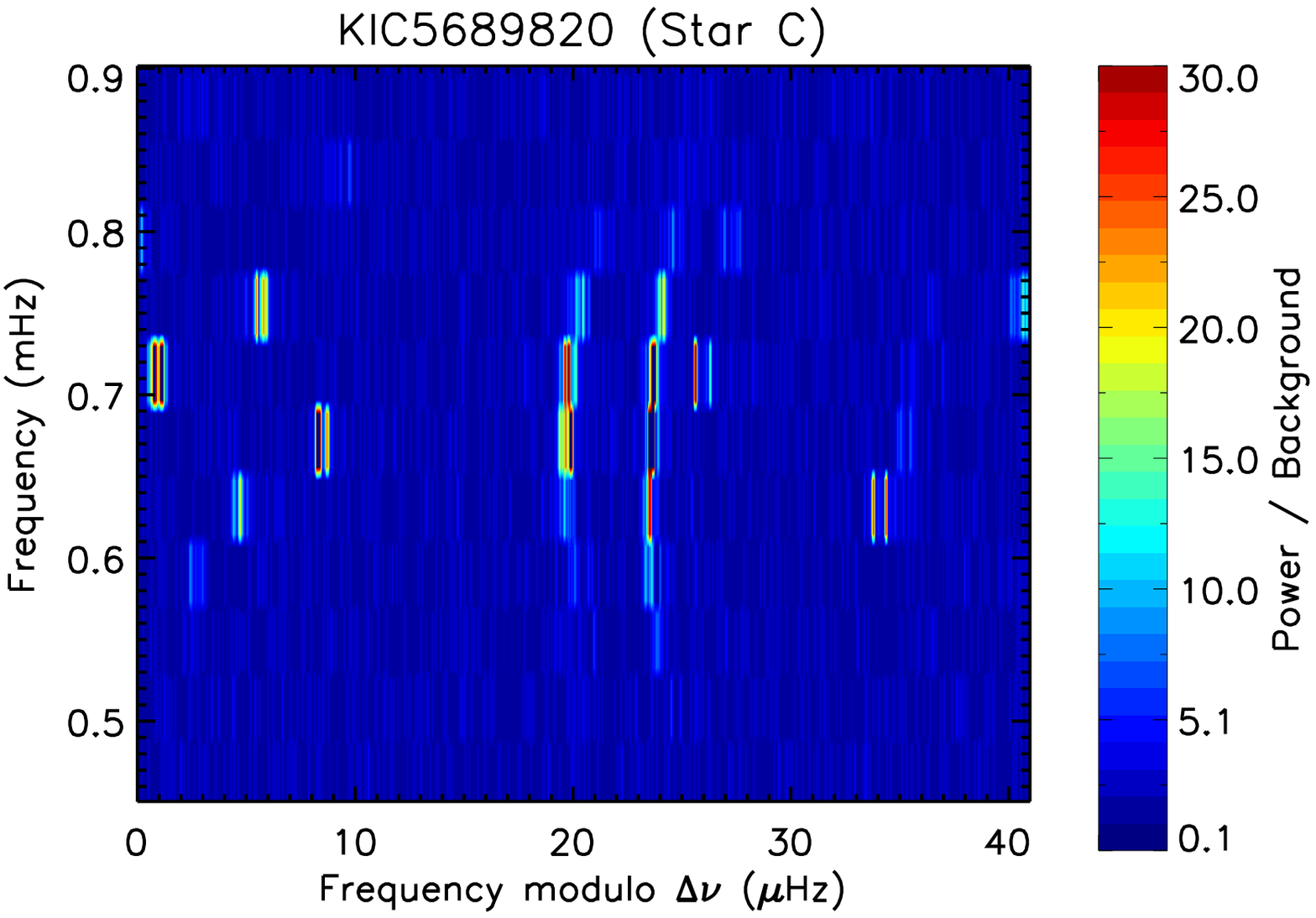}
\includegraphics[width=9cm]{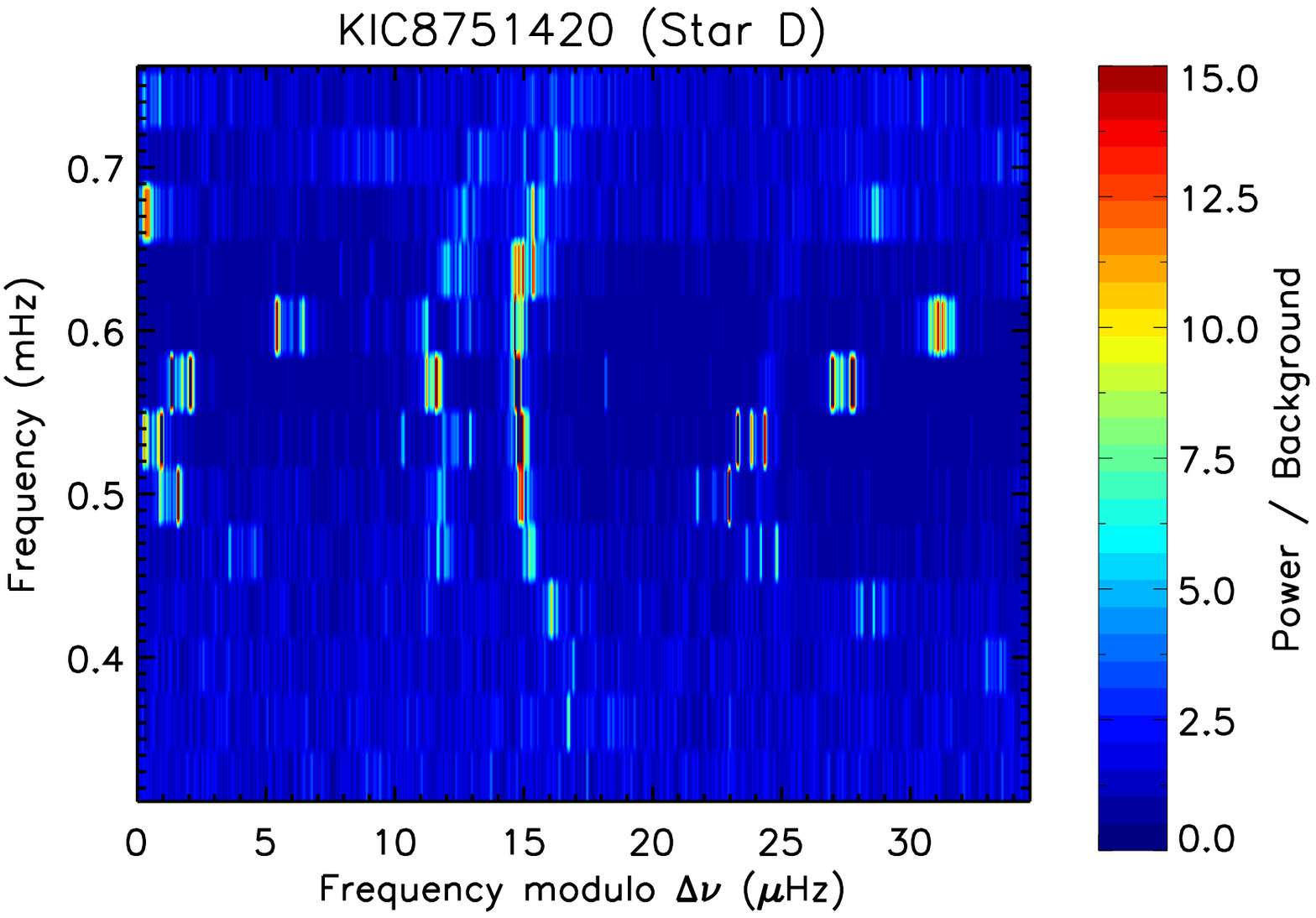}
\includegraphics[width=9cm]{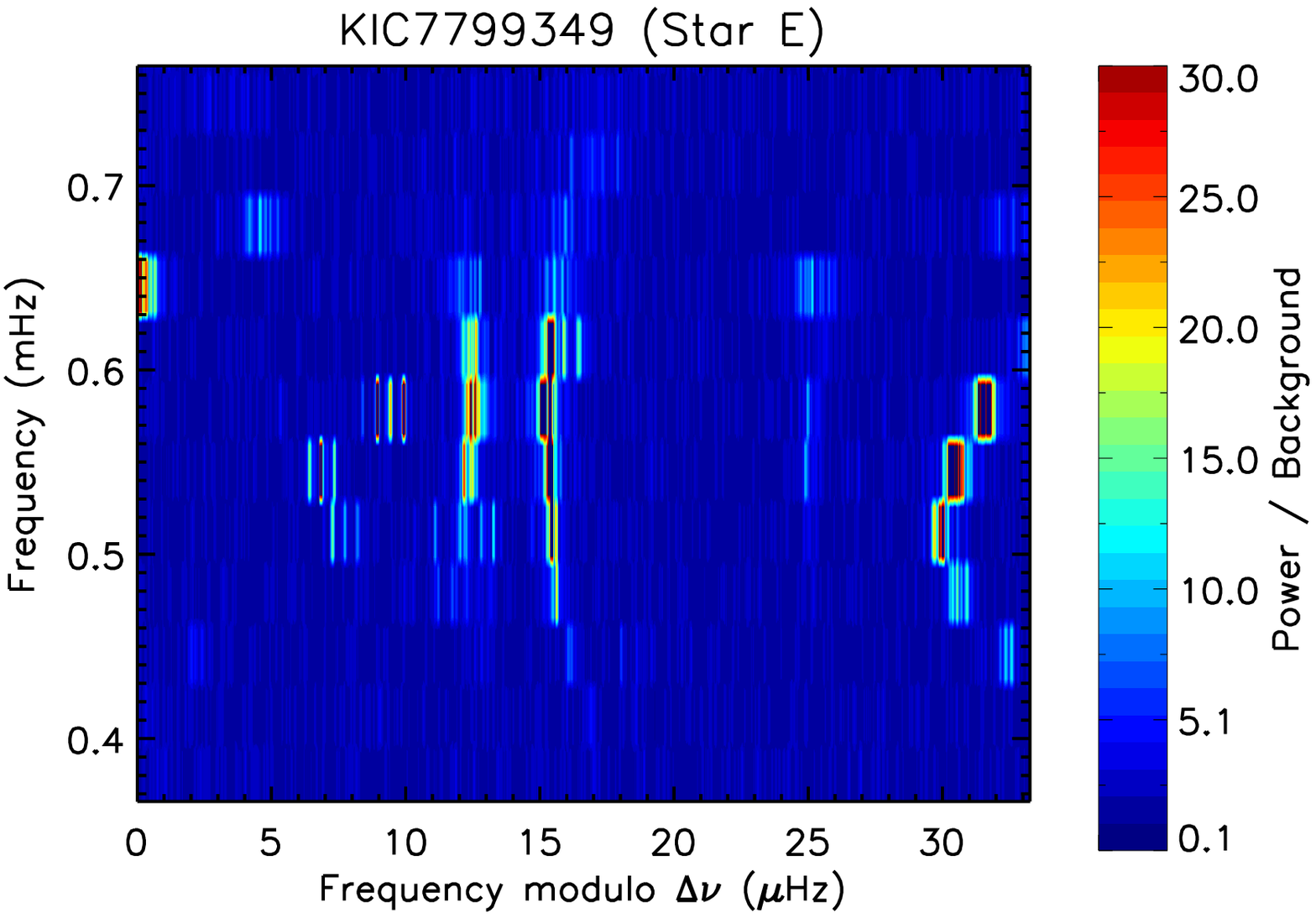}
\includegraphics[width=9cm]{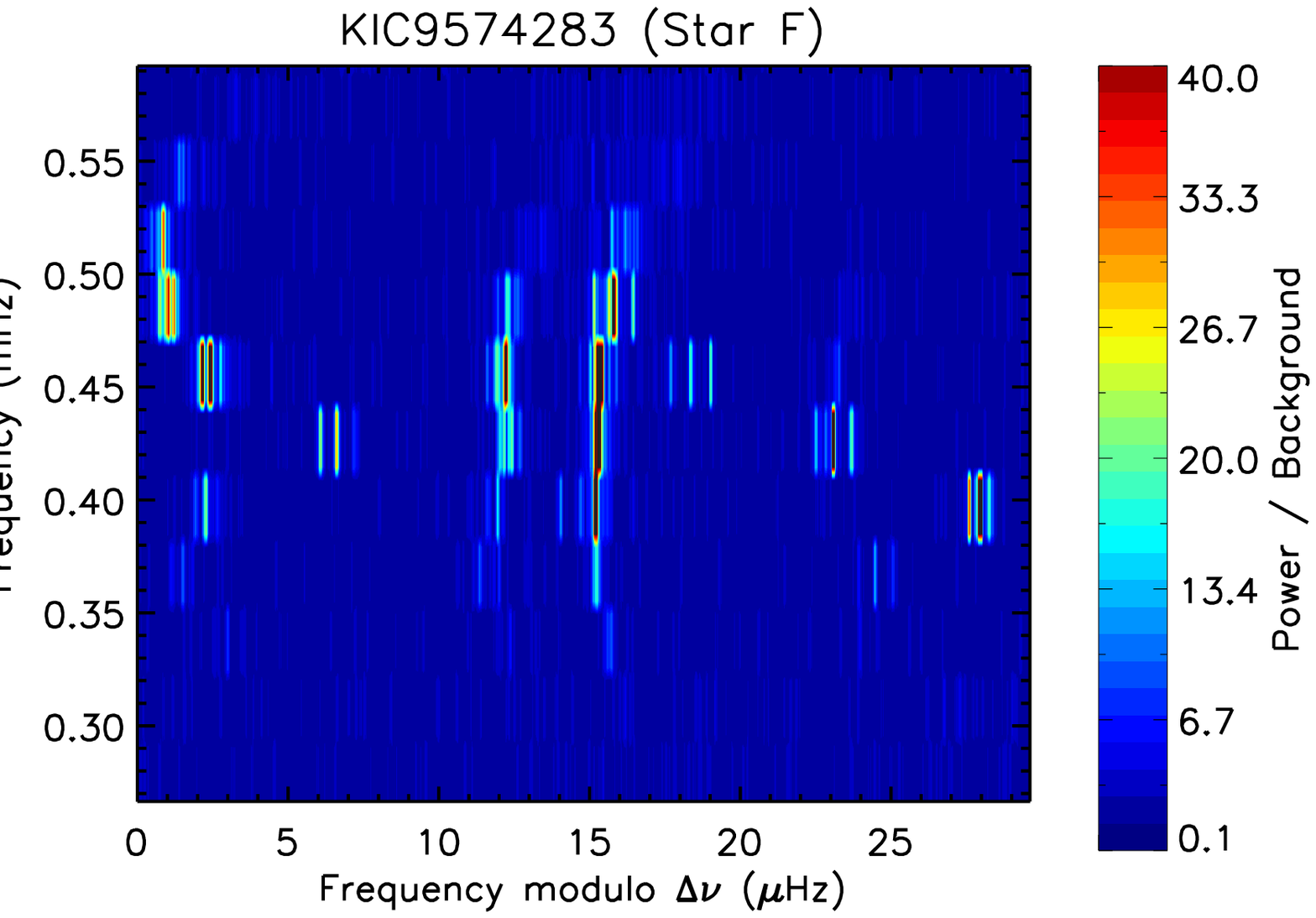}
\end{center}
\caption{\'Echelle diagrams of the stars from our sample. The mean large separations that were used to build the diagrams are specified in Table \ref{tab_global_param}. For more clarity, the power spectra were binned over a $0.25\mu$Hz boxcar and clipped at a maximum of 20 to 40 times the noise level.
\label{fig_echelle_obs}}
\end{figure*}

First estimates of the mean large separation of acoustic modes $\langle\Delta\nu\rangle$ were obtained by computing an autocorrelation of the power spectra. We built \'echelle diagrams for the six stars using these estimates of the large separation. In all of them, the neighboring ridges that correspond to $l=0$ and $l=2$ modes can easily be identified. We then fine-tuned our estimates of $\langle\Delta\nu\rangle$ so that the $l=0$ ridge is as vertical as possible in the \'echelle diagram (see Fig. \ref{fig_echelle_obs}). The corresponding values of $\langle\Delta\nu\rangle$ are given in Table \ref{tab_global_param}. We note that these estimates could be refined by taking into account the curvature of the $l=0$ ridge (\citealt{mosser13}), but in this study the models will be constrained using the individual frequencies of the oscillation modes, which contain more precise information on the structure than the mean large separation (see Sect. \ref{sect_MLE_results} and \ref{sect_model}). The $l=1$ modes all have a mixed behavior, which makes their identification harder. The stars that were selected in this study have an evolutionary status that is intermediate between the subgiants, for which the g modes that lie in the frequency range of observations have low radial orders ($n\sim1$), and typical RGB stars, for which the radial orders of g modes are huge ($n\sim100$, \citealt{mosser12a}). 
\cite{mosser12a} proposed a method to identify the degree of the detected mixed modes based on asymptotic relations, which they successfully applied to hundreds of red giants. We used this method to obtain first estimates of the frequencies of $l=1$ modes for the six stars.



\subsection{Extraction of the mode frequencies and rotational splittings}



To obtain estimates of the frequencies and rotational splittings of the observed modes, we fitted a model of the power spectral density (PSD) to the power spectra of the six stars. We followed a procedure that is very similar to the one adopted by D12. We here only briefly summarize it, with an emphasis on the few differences.

\subsubsection{Model of the PSD}

\begin{figure}
\begin{center}
\includegraphics[width=9cm]{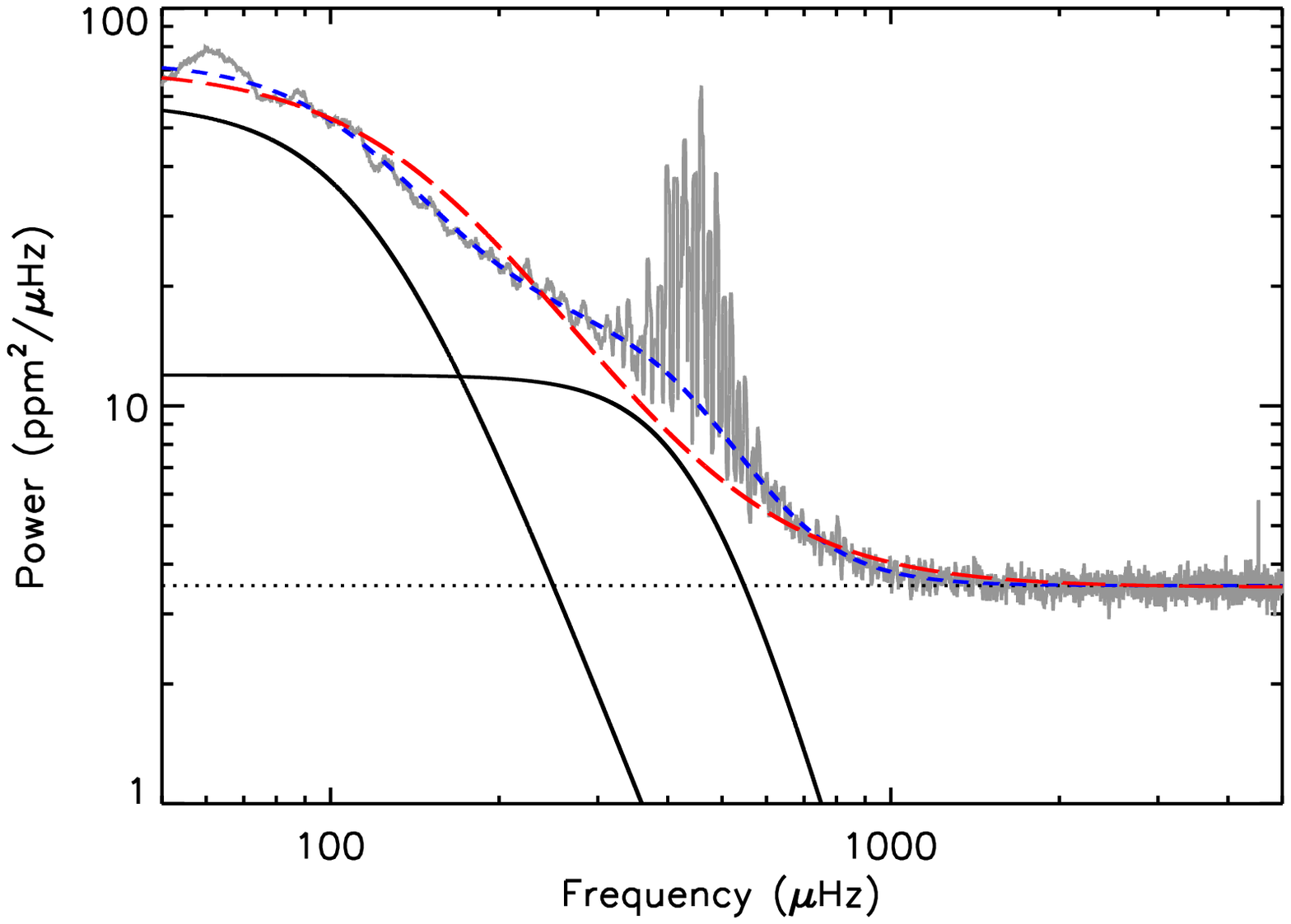}
\end{center}
\caption{Power spectrum of KIC9574283 (smoothed over a 10 $\mu$Hz boxcar) computed with \duree\ days of \kepler\ data (gray curve). Two Harvey-type laws (black solid lines) had to be considered to reproduce the shape of the background. The fitted background is represented by the blue bashed line, while the red long-dashed line shows the background that is obtained with only one Harvey profile. The dotted line shows the contribution from the photon noise.
\label{fig_background}}
\end{figure}

The background was fitted prior to the extraction of the mode parameters by using a maximum likelihood estimation (MLE) method in the same way as described in D12. The contribution from granulation to the background was modeled as a Harvey profile (\citealt{harvey85}), and a white noise was added, corresponding to photon noise. \cite{karoff13} have recently shown that an additional component with a timescale intermediate between that of granulation and the periods of the acoustic modes is needed. This component could be attributed to bright points (\citealt{aigrain04}), a second granulation population (\citealt{vazquez05}), or more likely faculae based on its timescale (\citealt{karoff12}, \citealt{karoff13}). In this study, we also found that a background including only the contribution from granulation poorly reproduces the observations (see the example of KIC9574283 in Fig. \ref{fig_background}). We thus included an additional Harvey profile, which greatly improved the agreement with the observations (Fig. \ref{fig_background}). 
In the following, the background parameters were held fixed to their fitted values when extracting the mode parameters. We note that in order to fit the background, the component of the PSD that corresponds to solar-like oscillations was modeled as a Gaussian function. Its central frequency provides an estimate of the frequency of maximum power of the oscillations $\nu\ind{max}$. The values that were obtained are listed in Table \ref{tab_global_param}.

The stochastically excited oscillation modes were modeled as Lorentzian functions. We assumed that the modes are split by rotation following Eq. \ref{eq_split}. We note that in fact, theoretical models predict a fast rotation for the cores of red giants, which could in certain cases invalidate the linear dependence of the splittings on the rotation rate expressed by Eq. \ref{eq_split} (e.g. \citealt{marques13}, \citealt{ceillier13}). In this case, a non-perturbative approach is needed (\citealt{ouazzani13}). However, \kepler\ observations have shown that the core rotation of red giants is in fact much slower, making the use of Eq. \ref{eq_split} relevant (D12, \citealt{mosser12b}). For the stars of our sample, the $l=1$ rotational multiplets show clear symmetry, which justifies the use of a linear splitting. The case of $l=2$ modes is more complex and will be discussed in Sect. \ref {sect_MLE_results}.

\begin{table}
\caption{Characteristics of the fit performed by each fitting group to extract the mode frequencies and rotational splittings from the power spectra.}
\begin{center}
\begin{tabular}{l c c c}
\hline \hline
\T\B Fitter & Method & Stars fitted & Final fit\\
\hline
\T Appourchaux & MLE Global$^a$ & A$\rightarrow$ F & - \\
 Benomar & MCMC$^b$ & D, F & D, F \\
 Campante & MLE Global$^a$ & E & - \\
 Davies & MLE Global$^a$ & A$\rightarrow$ F & - \\
 Deheuvels & MLE Global$^a$ & A$\rightarrow$ F & A, B, C, E \\
 & MLE Local$^c$ & A$\rightarrow$ F & - \\
 Regulo & MLE Global$^a$ & A$\rightarrow$ F & - \\
\B Stahn & MLE Local$^c$ & A$\rightarrow$ F & - \\
\hline
\end{tabular}
\end{center}
{\small \textbf{Notes: } The first column provides the name of the fitter. The second column gives the method followed by the fitter. The third column lists the stars that were fitted. The fourth column lists the stars for which the fitter performed the final fit.} \\
{\small $^a$\cite{appourchaux08}, $^b$\cite{benomar09a}, $^c$\cite{anderson90}}
\label{tab_analysis}
\end{table}%

\begin{table}
\begin{center}
\caption{Estimates of the frequencies and rotational splittings of the detected modes for KIC12508433 (star A), obtained by fitting Lorentzian functions to the mode profiles. The rotational splittings are given only for the modes for which at least six of the seven teams agreed to within 1-$\sigma$.
\label{tab_mle_1250}}
\begin{tabular}{c c c}
\hline \hline
\T \B $l$ & $\nu_{n,l}$ ($\mu$Hz) & $\delta\nu_{n,l}$ ($\mu$Hz) \\
\hline
\T 0 & $606.085\pm0.078$ & n.a. \\
0 & $649.934\pm0.034$ & n.a. \\
0 & $694.407\pm0.036$ & n.a. \\
0 & $739.336\pm0.031$ & n.a. \\
0 & $784.165\pm0.031$ & n.a. \\
0 & $829.072\pm0.036$ & n.a. \\
0 & $874.216\pm0.045$ & n.a. \\
\B 0 & $919.672\pm0.092$ & n.a. \\
\hline
\T 1 & $579.158\pm0.027$ & $0.256\pm0.036$ \\
1 & $627.052\pm0.023$ & $0.189\pm0.024$ \\
1 & $657.311\pm0.024$ & $0.282\pm0.024$ \\
1 & $679.081\pm0.031$ & $0.223\pm0.027$ \\
1 & $713.436\pm0.039$ & $0.236\pm0.025$ \\
1 & $740.554\pm0.013$ & $0.240\pm0.011$ \\
1 & $768.335\pm0.026$ & $0.210\pm0.031$ \\
1 & $806.093\pm0.031$ & $0.193\pm0.042$ \\
1 & $836.040\pm0.015$ & $0.246\pm0.011$ \\
1 & $860.483\pm0.048$ & $0.178\pm0.102$ \\
1 & $900.001\pm0.056$ & $0.336\pm0.050$ \\
\B 1 & $942.258\pm0.066$ & - \\
\hline
\T 2 & $690.651\pm0.073$ & $0.206\pm0.067$ \\
2 & $736.118\pm0.044$ & $0.223\pm0.014$ \\
2 & $779.601\pm0.035$ & $0.194\pm0.024$ \\
2 & $825.116\pm0.049$ & $0.217\pm0.025$ \\
\B 2 & $915.505\pm0.143$ & - \\
\hline
\hline 
\end{tabular}
\\
{\small $^\sharp$n.a.: not applicable.}
\end{center}
\end{table}

\begin{table}
\begin{center}
\caption{Same as Table \ref{tab_mle_1250} for KIC8702606 (star B).
\label{tab_mle_870}}
\begin{tabular}{c c c}
\hline \hline
\T \B $l$ & $\nu_{n,l}$ ($\mu$Hz) & $\delta\nu_{n,l}$ ($\mu$Hz) \\
\hline
\T 0 & $491.070\pm0.024$ & n.a. \\
0 & $531.002\pm0.050$ & n.a. \\
0 & $570.074\pm0.050$ & n.a. \\
0 & $609.620\pm0.037$ & n.a. \\
0 & $649.219\pm0.034$ & n.a. \\
0 & $688.624\pm0.039$ & n.a. \\
0 & $768.568\pm0.084$ & n.a. \\
\B 0 & $808.595\pm0.192$ & n.a. \\
\hline
\T 1 & $511.519\pm0.026$ & $0.327\pm0.026$ \\
1 & $533.045\pm0.014$ & $0.311\pm0.016$ \\
1 & $552.948\pm0.040$ & $0.262\pm0.035$ \\
1 & $580.559\pm0.024$ & $0.289\pm0.022$ \\
1 & $598.164\pm0.025$ & $0.231\pm0.023$ \\
1 & $627.328\pm0.032$ & $0.202\pm0.032$ \\
1 & $653.089\pm0.014$ & $0.283\pm0.014$ \\
1 & $674.472\pm0.036$ & $0.228\pm0.042$ \\
1 & $705.729\pm0.035$ & $0.272\pm0.041$ \\
1 & $731.742\pm0.042$ & $0.303\pm0.040$ \\
1 & $756.272\pm0.038$ & $0.369\pm0.038$ \\
1 & $789.662\pm0.080$ & - \\
\B 1 & $820.804\pm0.139$ & - \\
\hline
\T 2 & $526.581\pm0.018$ & - \\
2 & $566.494\pm0.054$ & $0.154\pm0.029$ \\
2 & $605.910\pm0.035$ & $0.192\pm0.025$ \\
2 & $645.614\pm0.038$ & $0.159\pm0.027$ \\
2 & $685.296\pm0.040$ & $0.124\pm0.030$ \\
\B 2 & $764.843\pm0.245$ & $0.281\pm0.165$ \\
\hline
\hline 
\end{tabular}
\\
{\small $^\sharp$n.a.: not applicable.}
\end{center}
\end{table}

\begin{table}
\begin{center}
\caption{Same as Table \ref{tab_mle_1250} for KIC5689820 (star C).
\label{tab_mle_568}}
\begin{tabular}{c c c}
\hline \hline
\T \B $l$ & $\nu_{n,l}$ ($\mu$Hz) & $\delta\nu_{n,l}$ ($\mu$Hz) \\
\hline
\T 0 & $516.270\pm0.011$ & n.a. \\
0 & $556.610\pm0.040$ & n.a. \\
0 & $597.242\pm0.047$ & n.a. \\
0 & $638.197\pm0.027$ & n.a. \\
0 & $679.266\pm0.033$ & n.a. \\
0 & $720.358\pm0.048$ & n.a. \\
0 & $761.716\pm0.045$ & n.a. \\
\B 0 & $803.158\pm0.100$ & n.a. \\
\hline
\T 1 & $576.439\pm0.053$ & $0.256\pm0.024$ \\
1 & $619.386\pm0.045$ & $0.092\pm0.090$ \\
1 & $648.779\pm0.012$ & $0.298\pm0.010$ \\
1 & $664.182\pm0.012$ & $0.213\pm0.008$ \\
1 & $697.615\pm0.014$ & $0.172\pm0.016$ \\
1 & $722.626\pm0.012$ & $0.339\pm0.010$ \\
1 & $743.321\pm0.049$ & $0.190\pm0.019$ \\
1 & $778.322\pm0.060$ & - \\ 
1 & $805.884\pm0.033$ & $0.336\pm0.027$ \\
\B 1 & $828.908\pm0.027$ & $0.435\pm0.024$ \\
\hline
\T 2 & $511.857\pm0.011$ & - \\
2 & $675.341\pm0.062$ & $0.129\pm0.006$ \\
2 & $716.503\pm0.038$ & $0.108\pm0.017$ \\
\B 2 & $757.937\pm0.095$ & $0.188\pm0.019$ \\
\hline
\hline 
\end{tabular}
\\
{\small $^\sharp$n.a.: not applicable.}
\end{center}
\end{table}


\begin{table}
\begin{center}
\caption{Same as Table \ref{tab_mle_1250} for KIC8751420 (star D).
\label{tab_mle_875}}
\begin{tabular}{c c c}
\hline \hline
\T \B $l$ & $\nu_{n,l}$ ($\mu$Hz) & $\delta\nu_{n,l}$ ($\mu$Hz) \\
\hline
\T 0 & $397.838\pm0.125$ & n.a. \\
0 & $431.742\pm0.033$ & n.a. \\
0 & $465.562\pm0.028$ & n.a. \\
0 & $499.896\pm0.024$ & n.a. \\
0 & $534.472\pm0.021$ & n.a. \\
0 & $568.986\pm0.016$ & n.a. \\
0 & $603.621\pm0.027$ & n.a. \\
0 & $638.313\pm0.054$ & n.a. \\
0 & $673.558\pm0.058$ & n.a. \\
0 & $708.983\pm0.136$ & n.a. \\
\B 0 & $744.280\pm0.159$ & n.a. \\
\hline
\T \B 1 & $414.335\pm0.065$ & - \\
\T \B 1 & $444.246\pm0.025$ & $0.477^{+0.030}_{-0.028}$ \\
\T \B 1 & $454.377\pm0.032$ & $0.482^{+0.034}_{-0.034}$ \\
\T \B 1 & $474.526\pm0.010$ & $0.599^{+0.014}_{-0.013}$ \\
\T \B 1 & $486.236\pm0.019$ & $0.343^{+0.019}_{-0.018}$ \\
\T \B 1 & $507.300\pm0.010$ & $0.627^{+0.010}_{-0.010}$ \\
\T \B 1 & $520.228\pm0.016$ & $0.296^{+0.017}_{-0.016}$ \\
\T \B 1 & $543.422\pm0.008$ & $0.520^{+0.010}_{-0.010}$ \\
\T \B 1 & $555.963\pm0.013$ & $0.368^{+0.014}_{-0.012}$ \\
\T \B 1 & $581.601\pm0.012$ & $0.404^{+0.014}_{-0.014}$ \\
\T \B 1 & $594.836\pm0.015$ & $0.474^{+0.016}_{-0.016}$ \\
\T \B 1 & $619.974\pm0.028$ & $0.230^{+0.034}_{-0.032}$ \\
\T \B 1 & $658.595\pm0.041$ & - \\
\T \B 1 & $686.796\pm0.088$ & - \\
\T \B 1 & $702.832\pm0.180$ & - \\
\hline
\T \B 2 & $462.124\pm0.043$ & $0.144^{+0.034}_{-0.035}$ \\
\T \B 2 & $496.619\pm0.067$ & $0.120^{+0.045}_{-0.065}$ \\
\T \B 2 & $565.727\pm0.023$ & $0.108^{+0.015}_{-0.017}$ \\
\T \B 2 & $635.865\pm0.038$ & -  \\
\hline 
\end{tabular}
\\
{\small $^\sharp$n.a.: not applicable.}
\end{center}
\end{table}

\begin{table}
\begin{center}
\caption{Same as Table \ref{tab_mle_1250} for KIC7799349 (star E).
\label{tab_mle_779}}
\begin{tabular}{c c c}
\hline \hline
\T \B $l$ & $\nu_{n,l}$ ($\mu$Hz) & $\delta\nu_{n,l}$ ($\mu$Hz) \\
\hline
\T 0 & $416.037\pm0.034$ & n.a. \\
0 & $448.499\pm0.032$ & n.a. \\
0 & $481.218\pm0.027$ & n.a. \\
0 & $514.341\pm0.021$ & n.a. \\
0 & $547.540\pm0.022$ & n.a. \\
0 & $580.633\pm0.025$ & n.a. \\
0 & $614.069\pm0.071$ & n.a. \\
0 & $647.589\pm0.072$ & n.a. \\
0 & $681.043\pm0.099$ & n.a. \\
\B 0 & $715.405\pm0.139$ & n.a. \\
\hline
\T 1 & $434.586\pm0.021$ & $0.202\pm0.020$ \\
1 & $450.967\pm0.003$ & $0.566\pm0.003$ \\
1 & $464.741\pm0.061$ & $0.214\pm0.025$ \\
1 & $496.250\pm0.034$ & $0.286\pm0.020$ \\
1 & $506.636\pm0.009$ & $0.466\pm0.009$ \\
1 & $528.865\pm0.018$ & $0.243\pm0.022$ \\
1 & $539.024\pm0.008$ & $0.458\pm0.011$ \\
1 & $562.658\pm0.030$ & $0.212\pm0.029$ \\
1 & $574.859\pm0.005$ & $0.491\pm0.005$ \\
1 & $596.986\pm0.025$ & $0.191\pm0.023$ \\
1 & $614.593\pm0.023$ & $0.553\pm0.022$ \\
1 & $632.243\pm0.035$ & $0.255\pm0.040$ \\
1 & $657.096\pm0.069$ & $0.485\pm0.061$ \\
1 & $669.829\pm0.042$ & $0.470\pm0.036$ \\
\B 1 & $697.785\pm0.033$ & $0.537\pm0.037$ \\
\hline
\T 2 & $511.478\pm0.032$ & - \\
2 & $544.490\pm0.026$ & $0.141\pm0.030$ \\
2 & $578.009\pm0.032$ & $0.143\pm0.020$ \\
2 & $611.144\pm0.052$ & $0.134\pm0.029$ \\
\B 2 & $644.174\pm0.164$ & - \\
\hline
\hline 
\end{tabular}
\\
{\small $^\sharp$n.a.: not applicable.}
\end{center}
\end{table}


\begin{table}
\begin{center}
\caption{Same as Table \ref{tab_mle_1250} for KIC9574283 (star F).
\label{tab_mle_957}}
\begin{tabular}{c c c}
\hline \hline
\T \B $l$ & $\nu_{n,l}$ ($\mu$Hz) & $\delta\nu_{n,l}$ ($\mu$Hz) \\
\hline
\T 0 & $312.285\pm0.100$ & n.a. \\
0 & $341.349\pm0.024$ & n.a. \\
0 & $370.549\pm0.018$ & n.a. \\
0 & $400.137\pm0.017$ & n.a. \\
0 & $429.818\pm0.018$ & n.a. \\
0 & $459.493\pm0.015$ & n.a. \\
0 & $489.424\pm0.042$ & n.a. \\
0 & $519.536\pm0.047$ & n.a. \\
\B 0 & $581.219\pm0.233$ & n.a. \\
\hline

\T \B 1 & $328.715\pm0.025$ & $0.487^{+0.032}_{-0.037}$ \\
\T \B 1 & $356.821\pm0.013$ & $0.334^{+0.020}_{-0.020}$ \\
\T \B 1 & $379.806\pm0.011$ & $0.545^{+0.022}_{-0.010}$ \\
\T \B 1 & $387.209\pm0.013$ & $0.347^{+0.018}_{-0.020}$ \\
\T \B 1 & $398.982\pm0.021$ & $0.702^{+0.012}_{-0.020}$ \\
\T \B 1 & $412.872\pm0.008$ & $0.345^{+0.009}_{-0.010}$ \\
\T \B 1 & $421.160\pm0.007$ & $0.557^{+0.009}_{-0.009}$ \\
\T \B 1 & $437.630\pm0.009$ & $0.589^{+0.012}_{-0.013}$ \\
\T \B 1 & $446.599\pm0.016$ & $0.317^{+0.022}_{-0.023}$ \\
\T \B 1 & $462.516\pm0.009$ & $0.661^{+0.016}_{-0.009}$ \\
\T \B 1 & $474.775\pm0.017$ & $0.219^{+0.022}_{-0.025}$ \\
\T \B 1 & $489.561\pm0.010$ & $0.647^{+0.010}_{-0.011}$ \\
\T \B 1 & $504.201\pm0.030$ & - \\
\T \B 1 & $534.500\pm0.046$ & - \\
\hline
\T \B 2 & $366.917\pm0.051$ & - \\
\T \B 2 & $396.731\pm0.022$ & $0.161^{0.014}_{-0.023}$ \\
\hline 
\end{tabular}
\\
{\small $^\sharp$n.a.: not applicable.}
\end{center}
\end{table}

\begin{table}
\caption{Inclination angles obtained for the stars of the sample from the final fit (see text).}
\begin{center}
\begin{tabular}{l c}
\hline \hline
\T\B Star & Inclination angle (degrees)\\
\hline
\T KIC12508433 (A) & $53\pm2^\circ$ \\
KIC8702606 (B) & $70\pm2^\circ$ \\
KIC5689820 (C) & $70\pm2^\circ$ \\
KIC8751420 (D) & $72\pm2^\circ$ \\
KIC7799349 (E) & $53\pm1^\circ$ \\
\B KIC9574283 (F) & $44\pm2^\circ$ \\
\hline
\end{tabular}
\end{center}
\label{tab_incl}
\end{table}%

Within the rotational multiplets, the modes were assumed to have a common width. The ratios between their heights $h_{l,m}$ are given by a visibility factor that depend only on the inclination angle of the star (\citealt{gizon03}, \citealt{ballot06}). For global fits, each rotational multiplet thus contributes four free parameters (frequency, height, width, and rotational splitting). One additional free parameter is needed: the inclination angle, which is common to all the modes. For local fits, the inclination angle is left free for all the multiplets, bringing to five the number of free parameters per multiplet.

Usually, for main-sequence stars, the ratio between the height $h_l$ of a multiplet of degree $l$ (defined as the sum of the heights of its components, i.e. $h_l\equiv\sum_m h_{l,m}$) and the height of the closest radial mode $h_0$ is given by a geometric factor obtained by integration over the stellar disk, taking into account the limb darkening profile. For \kepler\ stars, typical ratios are $h_1/h_0=1.5$ and $h_2/h_0=0.5$ (\citealt{ballot11}). These ratios do not necessarily hold for stars with mixed modes such as the targets of our sample. Indeed, some non-radial modes are mainly trapped in the core and their longer lifetimes prevent us from resolving them, even with \duree\ days of data. In this case, the mode height depends on the mode inertia (\citealt{dupret09}) and the theoretical visibility ratios are inappropriate. Besides, recent observations have shown that these visibility ratios vary from one star to another for red giants (\citealt{mosser12c}, \citealt{benomar13}). The mode heights were thus left free in our fits.

\subsubsection{Results \label{sect_MLE_results}}

Two types of fit of the PSD were performed to estimate the mode frequencies and splittings. Six independent teams followed a frequentist approach and used the MLE method as is commonly done for the analysis of stochastically excited modes. The main difference between these analyses lies in the initial guesses taken for the mode parameters and the type of fitting that was chosen: either a \textit{global} fit (all the modes are fitted simultaneously) as prescribed by \cite{appourchaux08} or a \textit{local} one (modes are fitted individually) as was done for the Sun (\citealt{anderson90}). Apart from the computational time (which is much shorter for local fits) the only difference between the two approaches is that local fits consider the inclination angle as a free parameter for each mode, whereas the angle is common to all modes in global fits. This enabled us to check the robustness of the optimal angle that is obtained from global fits. One other team fitted the PSD by using a Bayesian approach coupled with a Markov chain Monte Carlo (MCMC) algorithm, following the method described by \cite{benomar09a}. One of the advantages of the latter approach is that it gives access to the probability density function of the fitted parameters. Its computational time is however much longer than MLE methods. Therefore, only two stars of the sample could be fitted this way (stars D and F). These two stars were chosen because they are the ones for which the radial dependence of the rotation profile can be constrained at best (see Sect. \ref{sect_inversion})\footnote{The Bayesian fits were performed a posteriori, which allowed us to apply this method to the most interesting targets.}. The characteristics of all the fits that were performed are summed up in Table \ref{tab_analysis}.

To derive a reliable set of mode frequencies and rotational splittings for the six stars from the results of the different teams, we adopted the following procedure. For each fitted mode, we rejected outliers by applying the Peirce criterion to both the mode frequencies and the rotational splittings in the same way as described by \cite{mathur11} (see also \citealt{campante11}, \citealt{appourchaux12}). We then selected only the modes for which at least two fitters agreed within 1-$\sigma$ error bars (\textit{maximal} mode set, as introduced by \citealt{metcalfe10}). We note that prior to applying the Peirce criterion, a first selection had to be made. Indeed, for a few modes, the splitting found by one team corresponds to half or twice the value that is found by the other teams. This can arise when the initial guess for the splitting is too far off, which can lead to a wrong identification of the $m$ components of the multiplet (for instance an $m=-1/m=+1$ pair is mistaken for an $m=0/m=+1$ pair). If these wrong splitting estimates were kept, the mean value of the data set would be significantly altered and the first iteration of the Peirce criterion would reject all the data points. We thus found it necessary to discard them first.

For the modes that were rejected by the Peirce criterion, we identified several sources of disagreement:
\begin{itemize}
\item At higher frequency, the mode linewidths increase and for several stars, they become larger than the rotational splittings, which prevents us from measuring them. For these modes, the agreement between the teams on the estimated frequencies remains good in most cases, but there are large disagreements on their splittings.
\item Problems were found to arise when the $m$-components of an $l=1$ mode overlap another mode. If this other mode is a radial mode, then the $l=1$ mode could usually be recovered correctly. However, if it is an $l=2$ mode, the components of the two modes are very hard to disentangle and the results obtained from the different teams vary a lot, and they are rejected by the Peirce criterion.
\item Large disagreements between the teams were also found for the splittings of $l=2$ modes when two neighboring $l=2$ mixed modes are observed instead of one p-dominated $l=2$ mode. This was already pointed out by D12 for KIC7341231, and \cite{deheuvels13} showed that for early red giants, the $l=2$ rotational multiplets that are undergoing an avoided crossing are not symmetric with respect to their central component. This means that Eq. \ref{eq_split} does not apply to these modes, which are thus not considered in the following.
\item Even though a few $l=3$ modes were detected in several stars of the sample, no reliable frequency or rotational splitting could be estimated for these modes. This is mostly caused by their very low SNR. 
\end{itemize}

Starting from the maximal set obtained with the Peirce criterion, a final fit was performed by one fitter, using a Bayesian approach for stars D and F, and the MLE method for the other stars. The obtained parameters for the modes are given in Tables \ref{tab_mle_1250} to \ref{tab_mle_957}. The inclination angles derived from the fits are given in Table \ref{tab_incl}.

\begin{figure}
\begin{center}
\includegraphics[width=9cm]{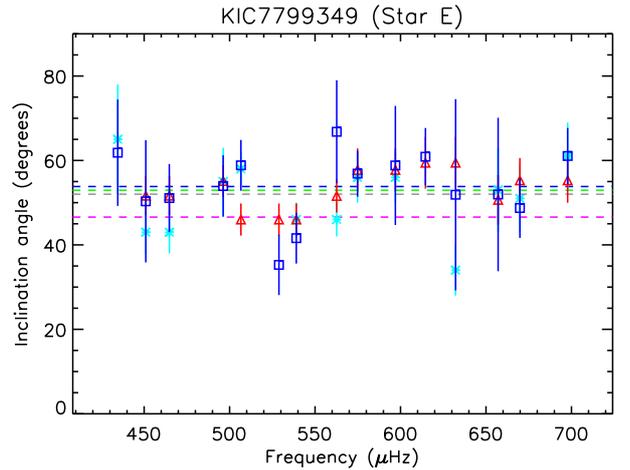}
\end{center}
\caption{Inclination angles obtained for star E (KIC7799349). The dashed lines indicate the results of global fits (gray: Campante, green: Davies, blue: Deheuvels, purple: Regulo), and the open symbols correspond to the results of local fits which allow for different angles for each mode (red triangles: Appourchaux, blue squares: Deheuvels, cyan stars: Stahn).
\label{fig_angle}}
\end{figure}

The question of the inclination angle is interesting. So far, all the studies that extracted mode parameters from solar-like pulsators assumed that the inclination angle is common for all modes. This is true if the whole star rotates about the same axis. While this assumption seems reasonable, it has been questioned before (e.g. \citealt{bai93} for the Sun). If the core rotates about a different angle than the envelope, we would expect the height ratios within g-dominated multiplets to correspond to an inclination angle that differs from the p-dominated multiplets (\citealt{gough93}). In this study, three of the six teams performed local fits, leaving the inclination angle free for each mode. This gave us the opportunity to check the common assumption of a single rotation axis in the stars of our sample. We found that the obtained inclination angles vary very little from one mode to another and are in good agreement with the angle that is found from global fits. An example is given for star E (KIC7799349) in Fig. \ref{fig_angle}. This shows that the rotation axis of the core is not significantly inclined from the one of the envelope in these stars.

\section{Surface parameters \label{sect_spectro_photo}}

To characterize the stars of the sample, we derived estimates of their effective temperature and surface metallicity using both spectroscopy and photometry. These estimates will serve as surface observables for our modeling of the stars which is presented in Sect. \ref{sect_model}.

\subsection{Spectroscopic observations \label{sect_spectro}}

\begin{table*}
  \begin{center}
  \caption{Surface parameters. \label{tab_surface_param}}
\begin{tabular}{l c c c c c c }
\hline \hline
\T Star & \multicolumn{3}{c}{Spectroscopy} & & \multicolumn{2}{c}{Photometry} \\
\cline{2-4}
\cline{6-7}
\T \B  & $T\ind{eff}$ (K) & $[$Fe/H$]$ (dex) & Reference & & $T\ind{eff, SDSS}$ (K) & $T\ind{eff, IRFM}$ (K) \\ 
\hline
\T A (KIC12508433) & $5248\pm130$ & $0.25\pm0.23$ & this study (FIES) & & $5139\pm55$ & $5302\pm124$ \\ 
 B (KIC8702606) & $5540\pm60$ & $-0.09\pm0.06$ & \cite{bruntt12} & & $5640\pm71$ & $5576\pm125$ \\ 
 C (KIC5689820) & $4978\pm167^a$ & $0.24\pm0.16^a$ & \cite{yang13} & & $5092\pm66$ & $5047\pm97$ \\ 
 D (KIC8751420) & $5264\pm60$ & $-0.15\pm0.06$ & \cite{bruntt12} & & n.a.$^b$ & $5205\pm98$ \\ 
 E (KIC7799349) & $5115\pm60$ & $0.41\pm0.06$ & \cite{bruntt12} & & $5108\pm50$ & $5020\pm101$ \\ 
\B F (KIC9574283) & $5120\pm55$ &$-0.40\pm0.08$ & this study (HERMES) & & $5354\pm61$ & $5174\pm114$ \\ 
\hline 
\end{tabular}
\end{center}
{\small $^a$mean measurement errors for the LAMOST spectra.} \\
{\small $^b$not applicable: no KIC photometry is available for this star.} \\
\end{table*}

Before this study was started, four stars of the sample (stars B, C, D, and E) had already been observed spectroscopically. Since then, star A was also observed by \cite{molenda13}, whose results are mentioned below. Star C was observed with the Guo Shou Jing Telescope, also known as the Large sky Area Multi-Object fibre Spectroscopic Telescope, (LAMOST) (\citealt{yang13}). Stars B, D, and E were among the \kepler\ targets that were observed by \cite{bruntt12} with the ESPaDOnS spectrometer at the 3.6-m Canada-France-Hawaii Telescope (CFHT) in the USA and with the NARVAL spectrometer mounted on the 2-m Bernard Lyot Telescope at the Pic du Midi Observatory in France. The atmospheric parameters obtained by the authors for these stars are given in Table \ref{tab_surface_param}.


Specific campaigns of observations were led for the two other stars. Star F was observed with the HERMES instrument of the 1.2-m Mercator Telescope at the Roque de los Muchachos Observatory on La Palma Island (Canary Islands, Spain), which is a high-efficiency bench-mounted {\'e}chelle spectrograph that observes the complete wavelength range from 420 nm to 900 nm in a single exposure (\citealt{raskin11}). We used the High Resolution Fiber mode ($R\simeq80,000$) for two concatenated observations of 1800 s. The spectra have a maximum S/N $\sim60$ around 600 nm ($\sim35$ around 500 nm). The science exposures were co-added to minimize the amount of CCD cosmic ray hits. The spectra were calibrated using the latest version of the HERMES pipeline (release 4.0) developed at the Royal Observatory of Belgium in collaboration with the HERMES Consortium. Typical calibration steps were performed, including spectral order tracing and extraction, average flat-fielding, Th-Ar lamp wavelength calibration, and hot pixel removal using cross-order profiling. The wavelength scale was corrected to the barycentric rest frame, which includes a wavelength scale rebinning to 0.0015625 nm. The analysis of the spectrum was performed with the semi-automatic software package VWA (\citealt{bruntt09}) and the atmospheric parameters that were obtained are listed in Table \ref{tab_surface_param}.

Star A was observed with the FIber-fed Echelle Spectrograph (FIES) mounted on the 2.56-m Nordic Optical Telescope (NOT) in October 2011. The obtained spectrum has a maximum S/N $\sim80$ with a resolving power of 25,000. The reduction of the data and the analysis of the spectrum were performed using the semi-automatic software package VWA by following the same procedure as in \cite{thygesen12}. The atmospheric parameters that were derived for this star are given in Table \ref{tab_surface_param}. We note that since the present study was started, star A was also observed with the HERMES instrument by \cite{molenda13}. The authors analyzed its spectrum using two different codes: ROTFIT (\citealt{frasca03}) and a combination of the codes ARES (\citealt{sousa07}) and MOOG (\citealt{sneden73}). They obtained a temperature of $T\ind{eff}=5134\pm121$ K and a metallicity of $[$Fe/H$]=0.08\pm0.22$ dex with ROTFIT and $T\ind{eff}=5281\pm76$ K and $[$Fe/H$]=0.21\pm0.06$ dex with ARES+MOOG. These values are in good agreement with the results obtained with the FIES data quoted in Table \ref{tab_surface_param}.

We note that our data include spectra with very different native resolutions and analysis techniques, and it is difficult to estimate the resulting systematic errors, which are not included here.  We have, however, adopted star-by-star uncertainties that reflect the random errors in our measurements.

\subsection{Photometric estimates of $T\ind{eff}$}

The \kepler\ Input Catalogue (KIC) provides \textit{griz} photometry for all the selected targets, except for star D. We followed the recipe prescribed by \cite{pinsonneault12} to obtain photometric estimates of the effective temperatures of the stars. The KIC \textit{griz} photometry was corrected to be consistent with SDSS photometry. The colors were dereddened using the extinctions obtained from the reddening map of \cite{drimmel03} and the extinction coefficients given by \cite{An09}. We then used the polynomials given by \cite{pinsonneault12} to obtain temperature estimates. Finally, the spectroscopic estimates of metallicity were used to correct our $T\ind{eff}$ estimates from metallicity effects by interpolating between the tabulated values given by \cite{pinsonneault12}. The errors on the $T\ind{eff}$ estimates were computed by varying the metallicity within the spectroscopic error bars and assuming an uncertainty of 0.02 mag for the extinction. The obtained temperatures, referred to as $T\ind{eff, SDSS}$ are given in Table \ref{tab_surface_param}. We note that no KIC photometry is available for star D (KIC8751420), so the effective temperature of the star could not be estimated by this method.

We also combined the optical \textit{griz} photometry with the infrared $JHK\ind{S}$ photometry available from the Two Micron All Sky Survey catalog (2MASS, \citealt{skrutskie06}) and applied the InfraRed Flux Method (IRFM) as prescribed by \cite{casagrande10} to obtain an additionnal estimate of the effective temperature. Error bars on these measurements are dominated by the uncertainty on the reddening, which can at present not be estimated to better than 0.01 mag. The values of $T\ind{eff}$ coming from the IRFM are given in Table \ref{tab_surface_param}. For star D, we used $BV$ photometry instead of \textit{griz} photometry, which is missing, and applied the polynomials of \cite{casagrande10} to obtain an estimate of $T\ind{eff}$. We note from Table \ref{tab_surface_param} that the IRFM values of $T\ind{eff}$ all agree with the \textit{griz}-color $T\ind{eff}$ within less than 1.4 $\sigma$. The largest difference occurs for star F, where the SDSS temperature is higher. However, the IRFM $T\ind{eff}$ is in good agreement with the spectroscopic estimate for this star.

For all the stars except one (star F), three distinct measurements of $T\ind{eff}$ (spectroscopy, SDSS photometry, and IRFM photometry) yielded values that agree within less than 1 $\sigma$, which shows that these estimates are very reliable. For star F, the IRFM temperature and the spectroscopic one also agree within less than 1 $\sigma$, but the SDSS temperature is higher than the two other estimates (1.4-$\sigma$ difference with IRFM $T\ind{eff}$ and 2.8-$\sigma$ difference with spectroscopic $T\ind{eff}$). In the following, we adopt the spectroscopic temperatures and examine the consequences of an alternative (larger) $T\ind{eff}$ for Star F (see Sect. \ref{sect_model}).

\section{Seismic modeling \label{sect_model}}

To derive information on the internal rotation profile of the stars from the rotational splittings that were obtained in Sect. \ref{sect_analysis}, we needed to have access to the rotational kernels of the modes. We thus performed a modeling of the six stars of the sample. To establish that our conclusions on the rotation profiles of the stars do not critically depend on the choice of an optimal model, we modeled the stars using two different evolutionary codes: \cesam\ (\citealt{cesam}) and \astec\ (\citealt{astec}). In these two codes, the effects of rotation on the structure and evolution was neglected.

\subsection{\cesam\ models \label{sect_model_cesam}}

\subsubsection{Properties of the models}

The models that were computed with \cesam\ use the OPAL 2005 equation of state and opacity tables as described in \cite{lebreton08}. The nuclear reaction rates were computed using the NACRE compilation (\citealt{angulo99}). The atmosphere was described by Eddington's gray law. We assumed the classical solar mixture of heavy elements of \cite{grevesse93}. Convection was treated using the Canuto-Goldman-Mazzitelli (CGM) formalism (\citealt{canuto96}). This description involves a free parameter, the mixing length, which is taken as a fraction $\alpha\ind{CGM}$ of the pressure scale height $H_p$. 
The effects of microscopic diffusion were neglected in this study. We did not include any overshooting at the boundary of convective cores during the main sequence.

The mode frequencies of the models were computed using the oscillation code LOSC (\citealt{losc}). It is well known that the lack of a satisfactory way of modeling surface convection in stellar models induces shifts in the absolute mode frequencies, which is known as \textit{near-surface effects}. To correct for these effects, we used the recipe advocated by \cite{kjeldsen08}, which consists of adding to the mode frequencies a power law whose exponent is calibrated on the Sun. With \cesam\ and the CGM formalism for convection, we found an exponent of 4.25, which was used for the models computed with \cesam\ in this work. Mixed modes are less sensitive to near-surface effects than pure acoustic modes because the contribution to the kinetic energy from the core is larger. To take this into account, the surface correction of non-radial modes are multiplied by a factor $Q_{n,l}^{-1}$, where $Q_{n,l}$ corresponds to the ratio of the mode inertia to the inertia of the closest radial mode (e.g. \citealt{asteroseismology}, Chap. 7).

\subsubsection{Fitting procedure}

\begin{figure*}
\begin{center}
\includegraphics[width=6cm]{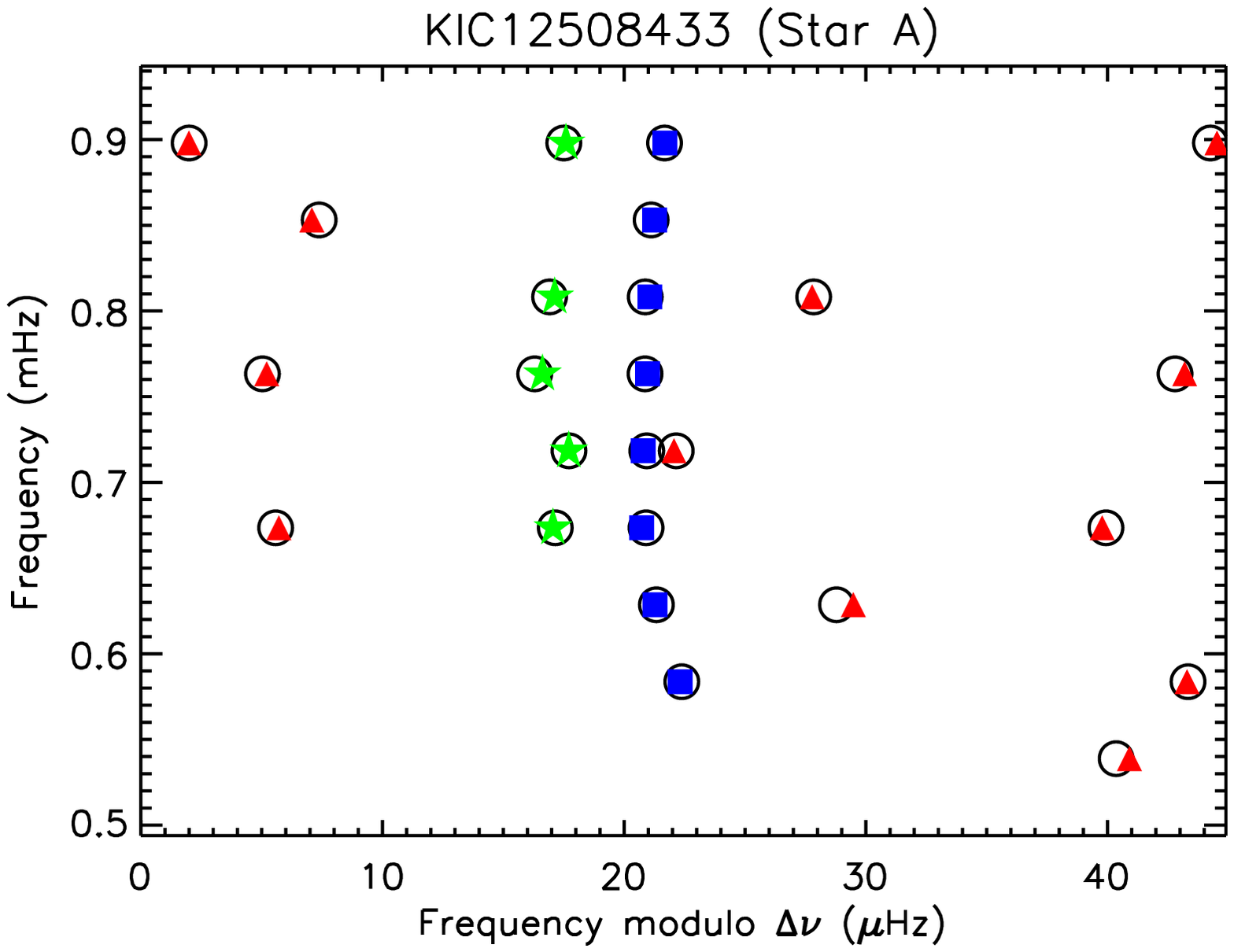}
\includegraphics[width=6cm]{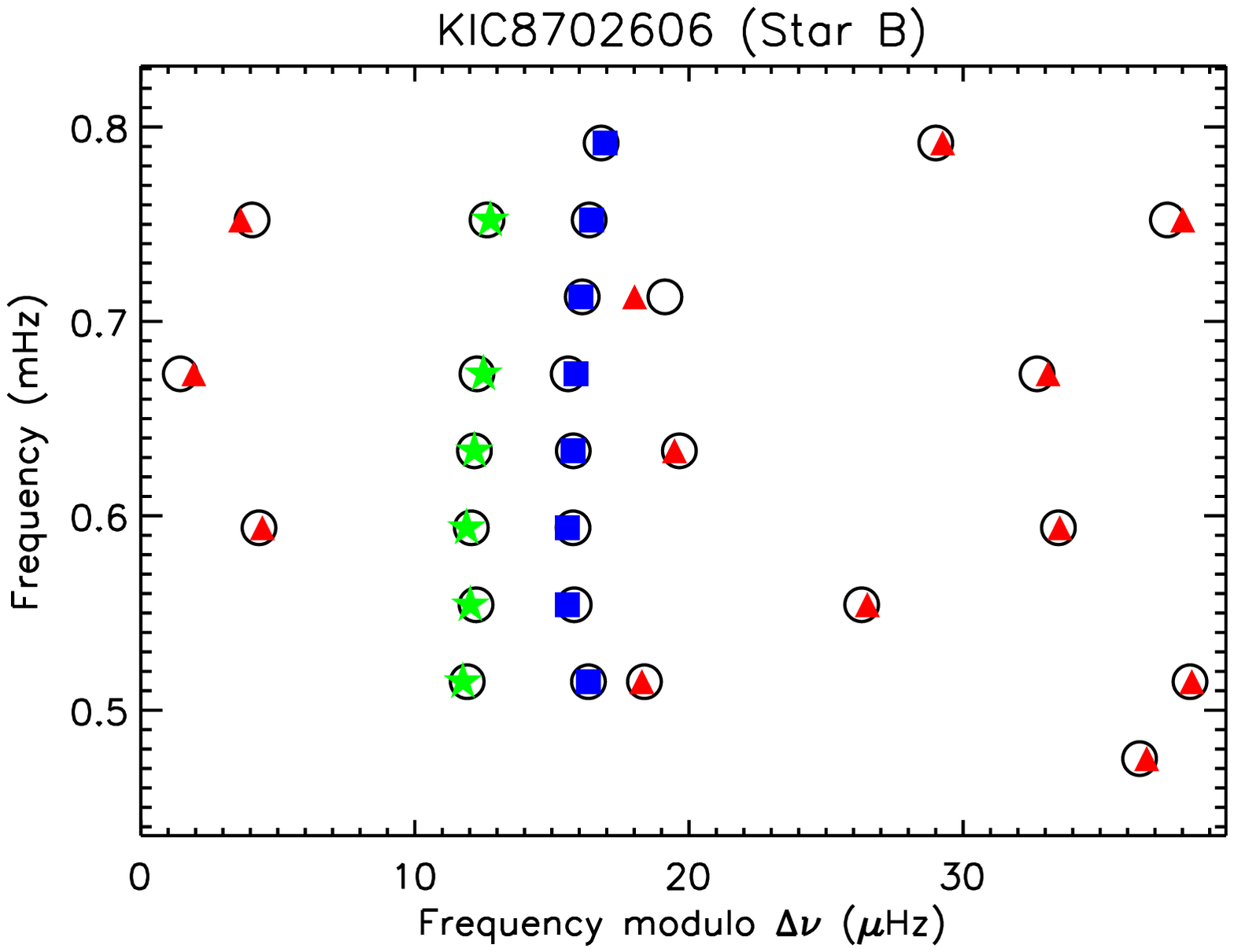}
\includegraphics[width=6cm]{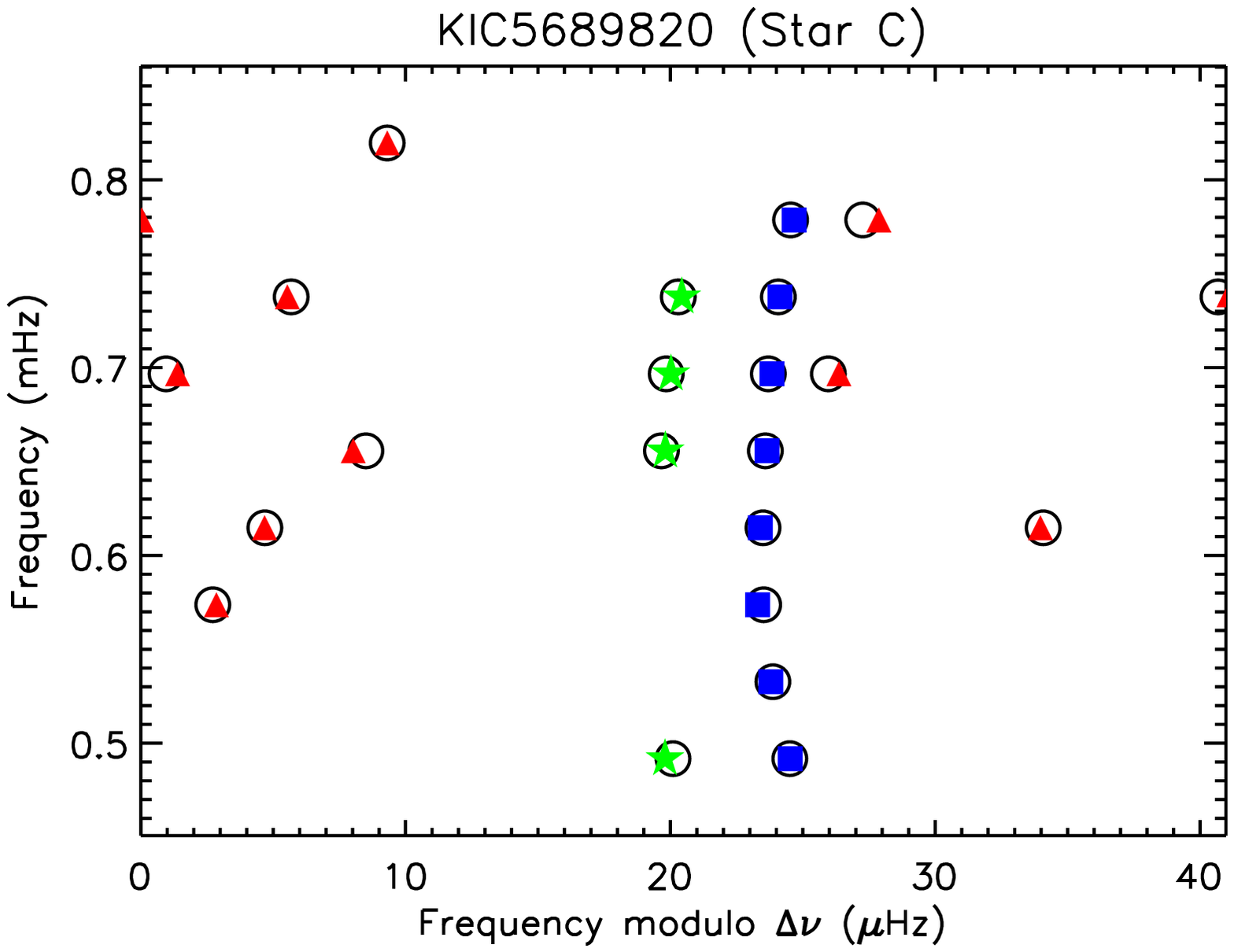}
\includegraphics[width=6cm]{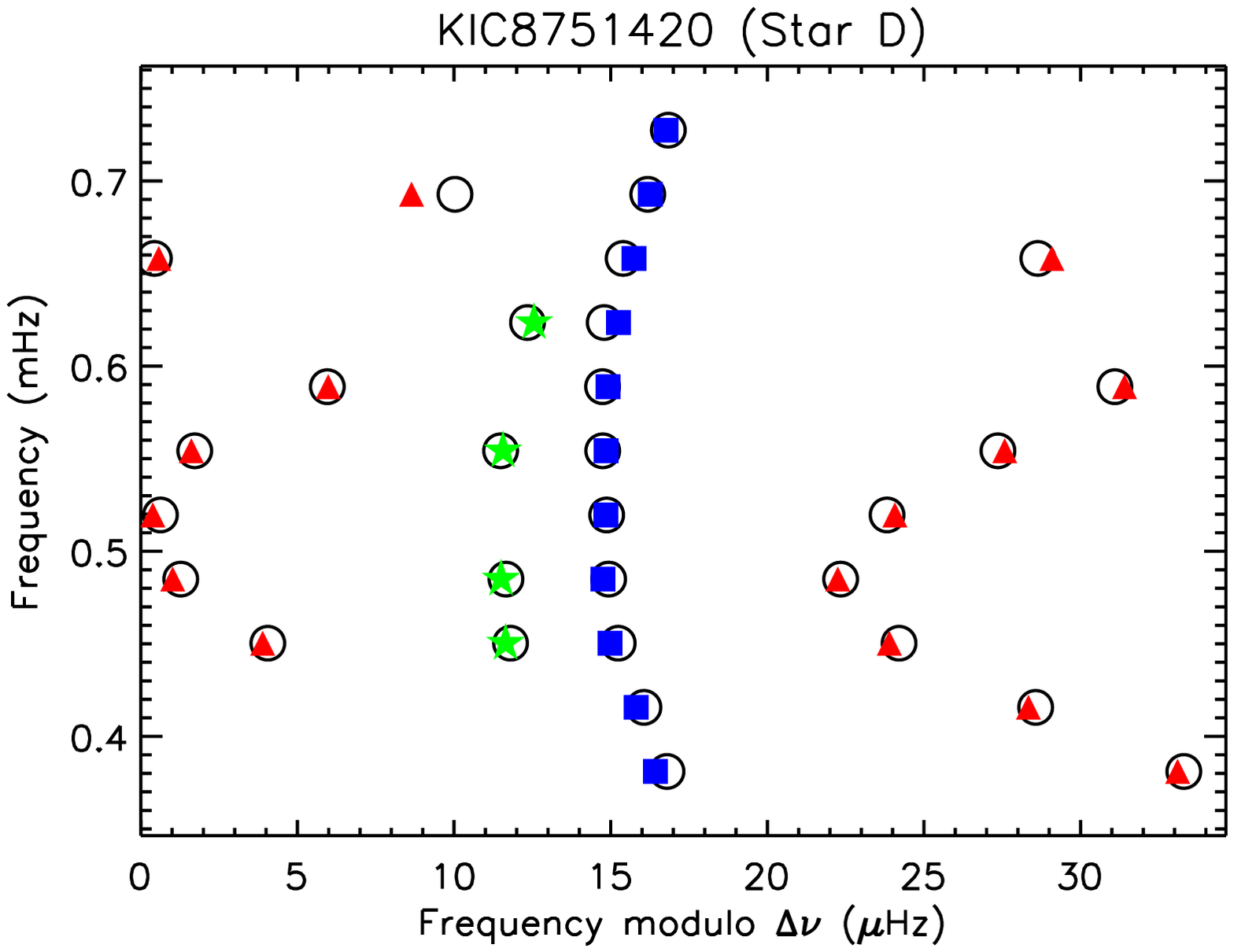}
\includegraphics[width=6cm]{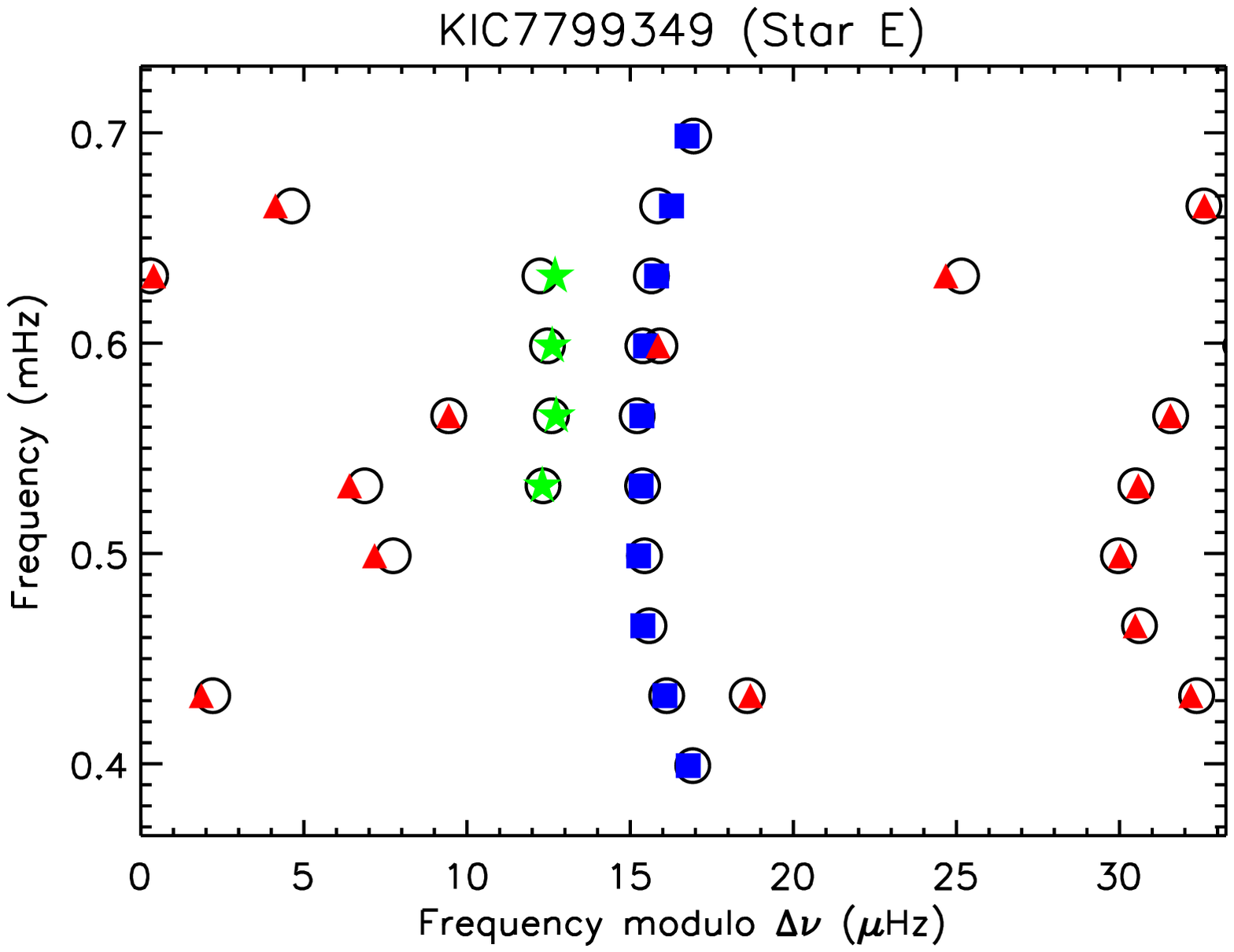}
\includegraphics[width=6cm]{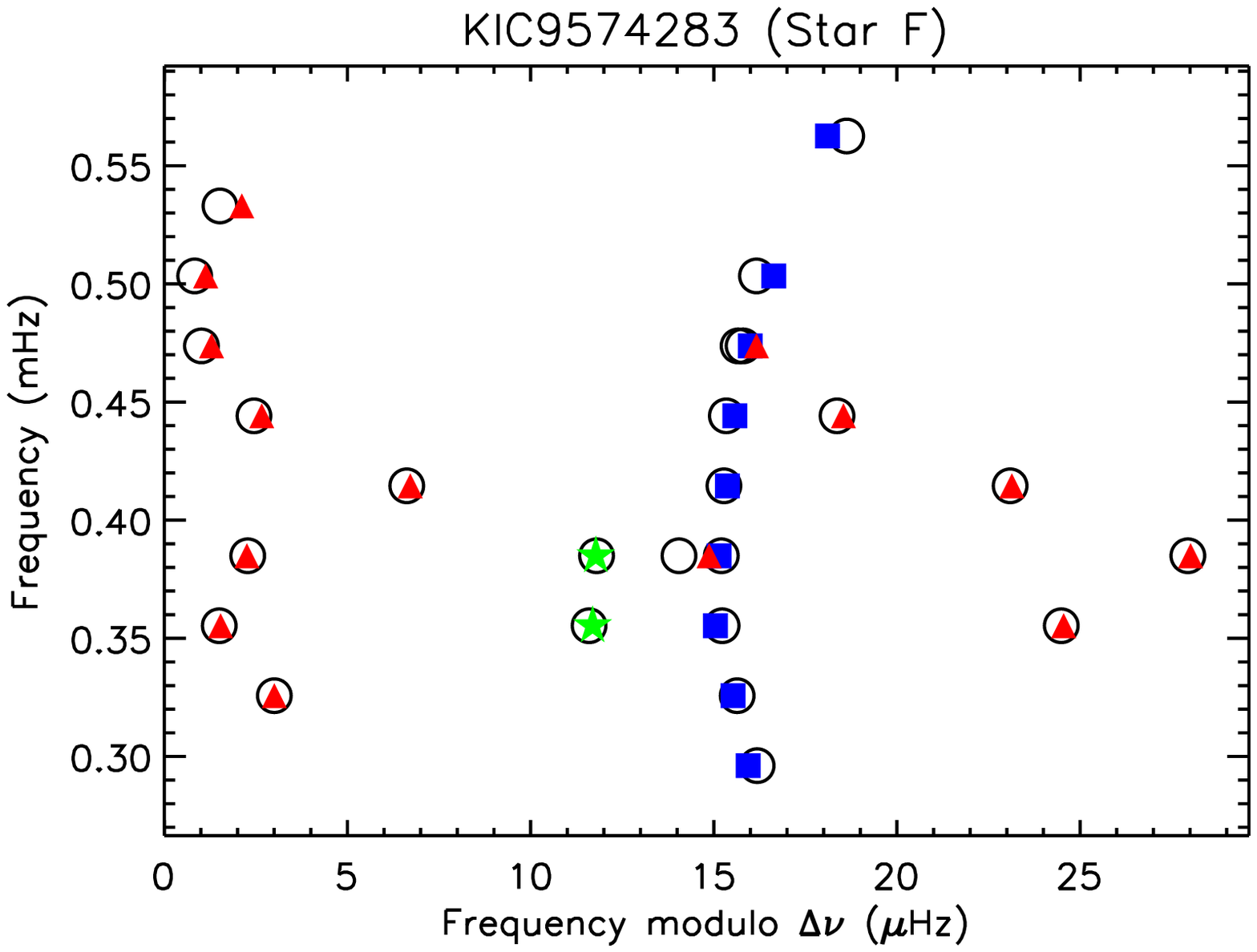}
\end{center}
\caption{\'Echelle diagrams of the stars from our sample. The open circles represent the observed frequencies and the colored filled symbols stand for the best model (blue squares: $l=0$, red triangles: $l=1$, green stars: $l=2$).
\label{fig_echelle_mod}}
\end{figure*}

To model the stars, we used the method that was first proposed by \cite{deheuvels11} to model subgiants with mixed modes, and which was later adapted to the case of early red giants by D12. This method uses the fact that for these stars, the combined knowledge of the mean large separation $\langle\Delta\nu\rangle$ and the mean period spacing of $l=1$ modes $\langle\Delta\Pi_1\rangle$ can yield very precise estimates of the stellar mass and age when other input physical parameters are fixed (see \citealt{deheuvels11} and D12 for more details). 

To apply this method, we first needed to determine which values of $\langle\Delta\nu\rangle$ and $\langle\Delta\Pi_1\rangle$ should be used. An estimate of $\langle\Delta\nu\rangle$ can be obtained observationally from the mean large separation of p modes. However, this value is not directly comparable to the one computed from stellar models because it is affected by near-surface effects. One way of circumventing this problem is to instead require the models to match the observed frequencies of the lowest-order radial modes (which are the least affected by near-surface effects). Two methods have been proposed to estimate $\langle\Delta\Pi_1\rangle$ from the observations (\citealt{benomar12}, \citealt{mosser12a}). The estimates obtained with the method of \cite{mosser12a} for the six stars (denoted $\Delta\Pi_1^{\rm obs}$) are given in Table \ref{tab_deltap1}. In principle, these methods can also be applied to the oscillation spectra of stellar models. However, automating them is not straightforward. Besides, it would require to compute the frequencies of $l=1$ mixed modes a large number of times for each model during the optimization process, which is time-consuming. We therefore chose to use the period spacing obtained from an asymptotic expansion, which is further denoted $\Delta\Pi_1^{\rm mod}$. We note that for red giants, $\Delta\Pi_1^{\rm mod}$ is usually calculated by assuming that the \vaisala\ frequency $N_{\rm BV}$ is much larger than the mode frequency in the whole g-mode cavity, thus yielding the approximate expression $\Delta\Pi_1^{\rm mod}\approx\pi^2\sqrt{2}(\int_{r_a}^{r_b}N_{\rm BV}/r \,dr)^{-1}$. For subgiants and young red giants this approximation is not justified because the frequencies of the observed mixed modes are larger than those of their more evolved counterparts and the \vaisala\ frequency is smaller owing to the less dense core. Making this approximation for the studied targets yields estimates of $\Delta\Pi_1^{\rm mod}$ that are overestimated by 3 to 11\%. Appendix \ref{app_deltap} explains how the $\Delta\Pi_1^{\rm mod}$ are computed from stellar models.

We then computed a grid of models by varying the initial helium content ($Y_0=0.24$ to 0.30), the mixing length parameter ($\alpha\ind{CGM}=0.55$ to $0.65$), and the metallicity in the range $(Z/X)\ind{obs}\pm\sigma\ind{obs}$. For each considered set of parameters, an automatic search was performed to determine the stellar mass and age that simultaneously reproduce the frequency of the lowest-order radial mode and the observed value of $\Delta\Pi_1$. Finally, the mass and age were fine-tuned to reproduce the observations as closely as possible. This last step causes the period spacing of the models to slightly differ from $\Delta\Pi_1^{\rm obs}$.
We computed the value of $\Delta\Pi_1^{\rm mod}$ for the optimal models by following the procedure described in Appendix \ref{app_deltap}. The values that were obtained for the selected stars are given in Table \ref{tab_deltap1}. The values of $\Delta\Pi_1^{\rm mod}$ agree with the ones estimated directly from the observations following the method of \cite{mosser12a} ($\Delta\Pi_1^{\rm obs}$) within less than 6\%. This is similar to the level of agreement that was reported between the methods of \citealt{benomar12} and \citealt{mosser12a} to estimate $\Delta\Pi_1^{\rm obs}$ (\citealt{benomar13}).

\begin{table}[!hb]
\caption{Value of $\Delta\Pi_1$ for each star computed either from the optimal models ($\Delta\Pi_1^{\rm mod}$) or directly from the observations by following the method of \cite{mosser12a} ($\Delta\Pi_1^{\rm obs}$).}
\begin{center}
\begin{tabular}{l c c}
\hline \hline
\T\B Star & $\Delta\Pi_1^{\rm mod}$ (s) & $\Delta\Pi_1^{\rm obs}$ (s)\\
\hline
\T A (KIC12508433) & 175.1 & 179.0 \\
B (KIC8702606) & 176.1 & 178.8 \\
C (KIC5689820) & 141.5 & 145.8 \\
D (KIC8751420) & 134.2 & 126.2 \\
E (KIC7799349) & 116.8 & 120.2 \\
\B F (KIC9574283) & 117.1 & 111.0  \\
\hline
\end{tabular}
\end{center}
\label{tab_deltap1}
\end{table}%

\begin{table*}
\caption{Parameters of the optimal models obtained for the stars of the sample with the codes \cesam\ and \astec.}
\begin{center}
\begin{tabular}{l c c c c c c c c | c c}
\hline \hline
\T\B Star & Code & $M/M_\odot$ & Age (Gyr) & $(Z/X)$ & $Y_0$ & $\alpha_{\rm conv}$ & $T_{\rm eff}$ (K) & $\log g$ & $\chi^2_{\rm atm}$ & $\chi^2_{\rm seis}$ \\
\hline
\T A (KIC12508433) & 	\cesam\ &	1.22 & 	5.9 & 	0.0500 & 	0.30 & 	$0.60^a$ & 	5026 & 	3.826 & 	1.06 & 	55 \\	
\B & 				\astec\ &	1.35 &	5.1 & 	0.0600 & 	0.28 & 	$1.80^b$ & 	5042 & 	3.841 & 	0.96 &	2007 \\
\T B (KIC8702606) & 	\cesam\ &	1.27 & 	3.8 & 	0.0173 & 	0.27 & 	$0.65^a$ & 	5575 & 	3.758 & 	0.43 & 	69 \\	
\B & 				\astec\ &	1.24 & 	4.1 & 	0.0200 & 	0.28 & 	$1.80^b$ & 	5380 & 	3.756 & 	2.39 &	128 \\
\T C (KIC5689820) & 	\cesam\ &	1.14 & 	6.9 & 	0.0388 & 	0.30 & 	$0.65^a$ & 	4973 & 	3.772 & 	0.03 &	215 \\	
\B & 				\astec\ &	1.25 & 	5.6 & 	0.0400 & 	0.30 & 	$1.80^b$ & 	4959 & 	3.785 & 	0.60 & 	355 \\
\T D (KIC8751420) & 	\cesam\ &	1.26 & 	3.8 & 	0.0151 & 	0.27 & 	$0.65^a$ & 	5281 & 	3.685 & 	0.31 & 	121 \\
\B & 				\astec\ &	1.32 &	3.8 & 	0.0150 & 	0.24 & 	$1.80^b$ & 	5164 & 	3.691 & 	1.31 &	271 \\
\T E (KIC7799349) & 	\cesam\ &	1.39 & 	3.8 & 	0.0548 & 	0.30 & 	$0.65^a$ & 	4898 & 	3.677 & 	4.6 & 	290 \\	
\B & 				\astec\ &	1.35 & 	3.6 & 	0.0400 & 	0.30 & 	$1.80^b$ & 	4944 & 	3.674 & 	6.3 &	 	350 \\
\T F (KIC9574283) & 	\cesam\ &	1.07 & 	6.0 &		0.0116 &	0.27 & 	$0.65^a$ & 	5194 &	3.574 &	0.95 &	195 \\	
\B & 				\astec\ &	1.10 &	6.5 & 	0.0100 & 	0.24 & 	$1.80^b$ & 	5034 & 	3.582 & 	1.69 &	597 \\
\hline
\end{tabular}
\end{center}
{\small $^a$computed with the CGM formalism for convection (\citealt{canuto96}).} \\
{\small $^b$computed with the classical MLT formalism for convection (\citealt{bohm58}).} \\
\label{tab_model}
\end{table*}%

\begin{figure}
\begin{center}
\includegraphics[width=9cm]{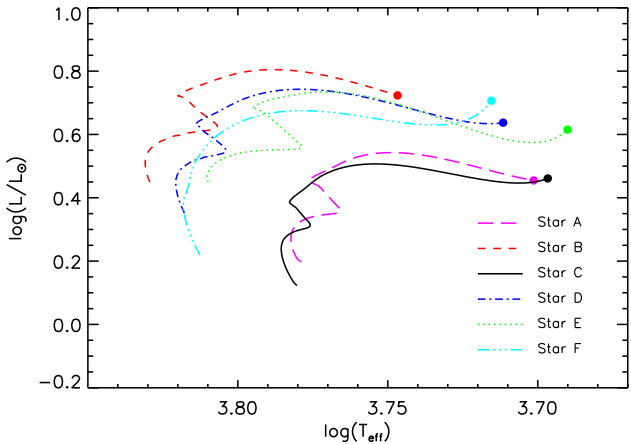}
\end{center}
\caption{Evolutionary tracks of the optimal models for stars A to F in the HR diagram. The current location of the model in the HR diagram is indicated by a filled circle.
\label{fig_HR_mod}}
\end{figure}

For each point of the grid, we estimated the agreement with the observations by computing a reduced $\chi^2$ function defined as
\begin{linenomath*}
\begin{equation}
\chi^2 = \frac{1}{N} \sum_{i=1}^N \left( \frac{\mathcal{O}_i\ex{mod}-\mathcal{O}_i\ex{obs}}{\sigma_i} \right)^2
\end{equation}
\end{linenomath*}
where $\mathcal{O}_i\ex{obs}$ $(i=1,N)$ are the observables, $\sigma_i$ their error bars, and $\mathcal{O}_i\ex{mod}$ the corresponding values in the computed models. As is usually done, we separated the contribution of the seismic constraints to the $\chi^2$ ($\chi^2\ind{seis}$) from the contribution of the atmospheric constraints ($\chi^2\ind{atm}$). Indeed, since the atmospheric constraints are much less numerous than the seismic ones, the total $\chi^2$ tends to drown their contribution. The values of $\chi^2\ind{seis}$ and $\chi^2\ind{atm}$ for the optimal models are listed in Table \ref{tab_model}. The evolutionary tracks of these models are shown in Fig. \ref{fig_HR_mod}. We thus confirm that our stars lie around the base of the RGB. Stars A and B, which have the highest value of $\log g$ are obviously still in the subgiant branch, while the 4 other stars just started their ascent of the RGB.

Except for star E, the atmospheric constraints are well reproduced, with values of $\chi^2\ind{atm}$ around 1. For star E, the higher value of $\chi^2\ind{atm}$ (4.6) is caused by the fact that the effective temperature of the model (4898 K) is lower than the spectroscopic one ($5115\pm60$ K) by about 3 $\sigma$. In fact for this star, no model in our grid has a $T\ind{eff}$ higher than 4900 K. We insist that each model in our grid is the result of an optimization of the stellar mass and age as described above, which means that this $T\ind{eff}$ upper limit cannot be increased by simply extending the grid. We note that the photometric estimate of $T\ind{eff}$ using the IRFM method is lower and would yield $\chi^2\ind{atm}=0.8$.

For star F, we found models that match the spectroscopic estimate of $T\ind{eff}$ ($5120\pm55$ K) and the IRFM temperature ($5174\pm114$ K), but no model in our grid could reproduce the higher value of $T\ind{eff}$ that was obtained from SDSS photometry ($5354\pm61$ K). We could obtain satisfactory fits with this higher $T\ind{eff}$ only by decreasing the metallicity to -1 dex, which is more than 7 $\sigma$ smaller than the spectroscopic value. 

For star D (KIC8751420), \citealt{huber12} obtained a precise estimate of the radius of the star from interferometric measurements with the CHARA array ($R/R_\odot=2.703\pm0.071$). The radius of our optimal model for this star ($R=2.668\;R_\odot$) is in perfect agreement with the interferometric value.

Even though the agreement between the observed mode frequencies and the frequencies of the models is visually good (see Fig. \ref{fig_echelle_mod}), the values of $\chi^2\ind{seis}$ are strikingly high (around 100 for the six stars). We found that the $l=1$ modes are by far the largest contributors to the value of $\chi^2\ind{seis}$ (they represent from 50\% to 97\% of $\chi^2\ind{seis}$). These disagreements are at least partly caused by imprecisions in the stellar models. With observation times as long as \duree\ days, the measurement errors on the mode frequencies reach values as low as $0.01\,\mu$Hz (see Tables \ref{tab_mle_568} to \ref{tab_mle_1250}), which certainly gives the possibility to test the physics that is used in current stellar models. Indeed, the frequencies of the $l=1$ mixed modes strongly depend on the coupling between the p-mode and the g-mode cavities, and thus on the evanescent zone that separates them (\citealt{rome}). However, these high values of $\chi^2\ind{seis}$ are most probably also caused by the crude way near-surface effects are dealt with in our models, by using the empirical correction that was prescribed by \cite{kjeldsen08}. It was already surprising that this correction, which was intended for main sequence stars, provides good fits to the observed frequencies of mixed modes for post main sequence stars --- provided it is weighted by the factor $Q_{n,l}^{-1}$ --- (e.g. D12, \citealt{dogan13}). We of course do not expect this correction to reproduce the observed absolute frequencies at a level of precision of $0.01\,\mu$Hz. To derive information on the core structure of these stars from the frequencies of mixed modes, this matter will need to be thoroughly studied. Such a study is however beyond the scope of the present paper. Indeed, variations in the mode frequencies of the order of the difference between models and observations induce very little change in the rotational kernels. This will be checked a posteriori in Sect. \ref{sect_inversion} by showing that our conclusions on the rotation profiles of the stars are not significantly modified when changing the reference model adopted for the inversions.

\subsection{\astec\ models \label{sect_model_astec}}

We also computed a grid of models using the Aarhus Stellar Evolution Code (ASTEC). The opacity tables, equation of state and nuclear reaction rates are the same as in the \cesam\ models. Convection was treated using the classical mixing length theory (MLT, \citealt{bohm58}) with a fixed value of the mixing length ($\alpha_{\rm MLT}=1.8$). Effects of diffusion on the stellar structure and core overshooting were neglected in the models. Models were computed with varying masses (from 1.0 to $1.6\,M_{\odot}$), initial helium content $Y_0$ (from 0.24 to 0.32), and metallicity with $(Z/X)$ from 0.01 to 0.07. The oscillation frequencies were calculated for the models whose atmospheric properties are within roughly 3 $\sigma$ of the observed values. For frequency calculations, we used the Aarhus adiabatic pulsation package (ADIPLS, \citealt{adipls}). The near-surface effects were corrected in the same way as described in Sect. \ref{sect_model_cesam}, with an exponent of 4.9 for the power law, as found by \cite{kjeldsen08} for the classical MLT. The best models were then selected as those that minimize the function $\chi^2\ind{seis}$, as defined in Sect. \ref{sect_model_cesam}. We note that here, we have not performed a pre-selection imposing the mean large separation of p modes and the period spacing of g modes as constraints to the models. After selecting the models that reproduced the observations at best, their parameters were fine-tuned by decreasing the mesh of the initial grid, and/or interpolating between the original time-steps to attempt to reduce the $\chi^2$ value. The resulting models are given in Table \ref{tab_model}. The fact that the selected \astec\ models have generally higher $\chi^2$ values than the \cesam\ models is probably caused by the fact the fitting method adopted here was initially designed for main-sequence stars and early subgiants, for which the density of the grid of models, particularly in terms of stellar age, is less crucial.

\section{Probing the internal rotation profile \label{sect_inversion}}

\subsection{Rotational splittings vs trapping of the modes \label{sect_splitvstrap}}

\begin{figure*}
\begin{center}
\includegraphics[width=8cm]{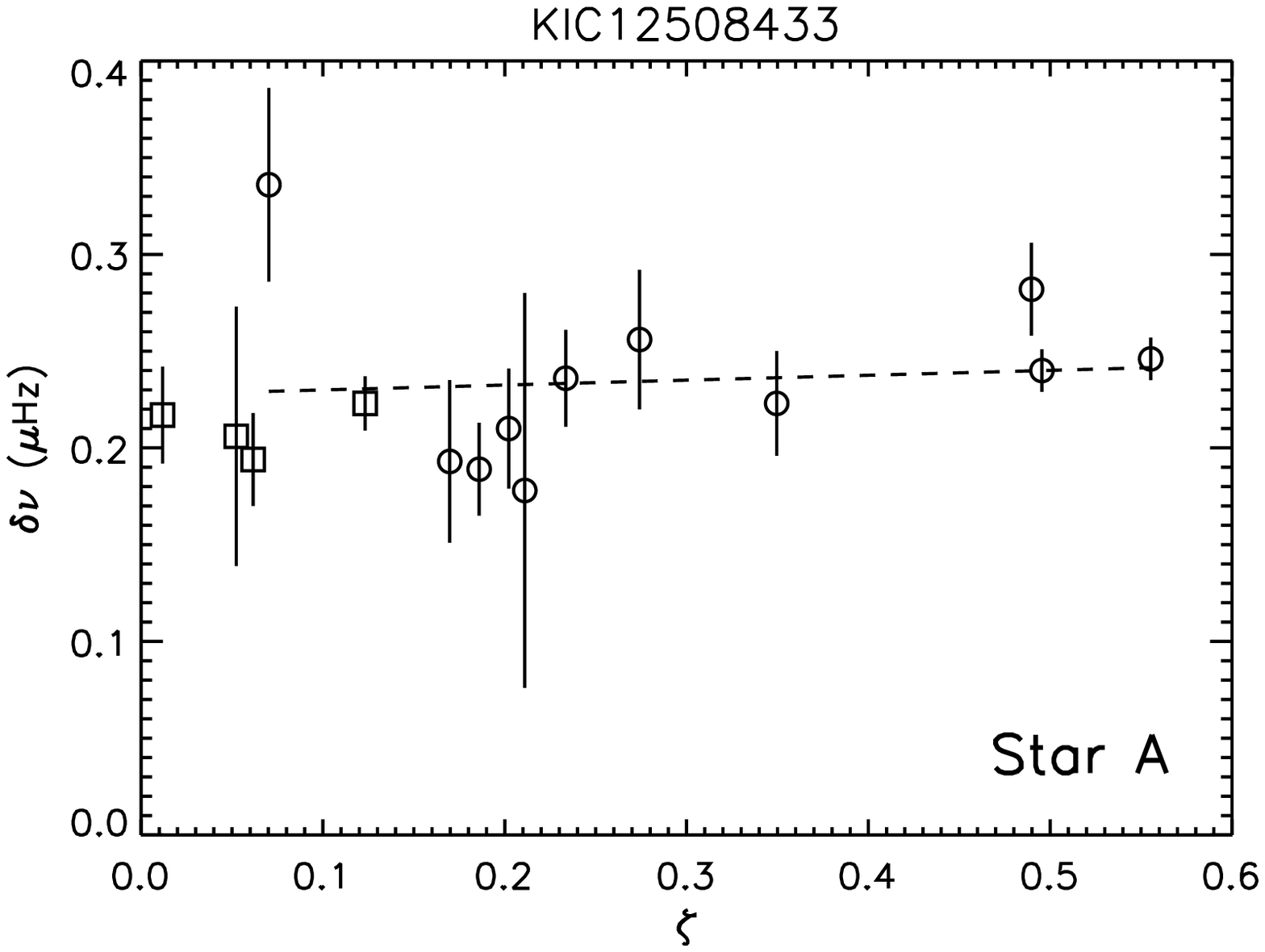}
\includegraphics[width=8cm]{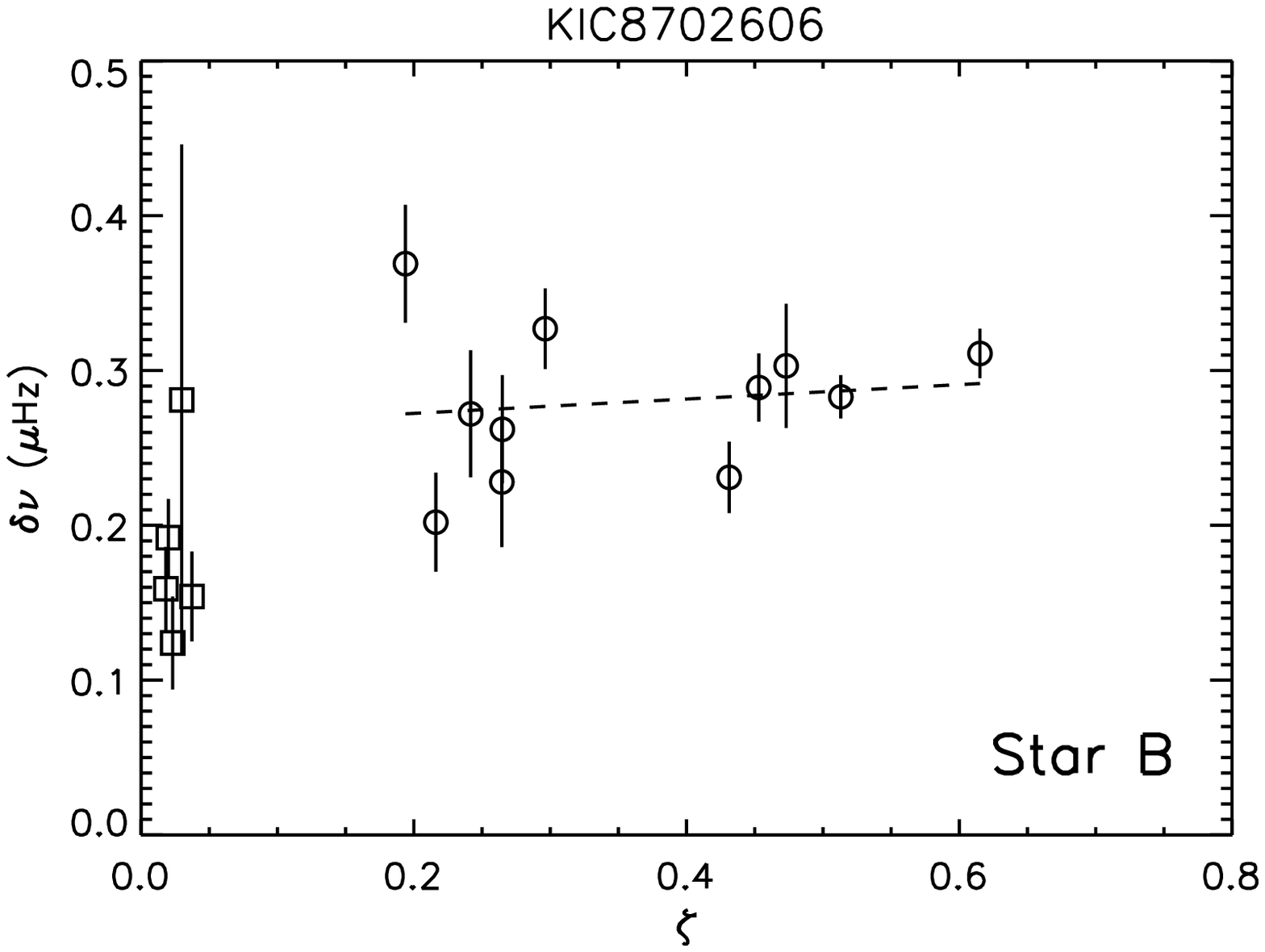}
\includegraphics[width=8cm]{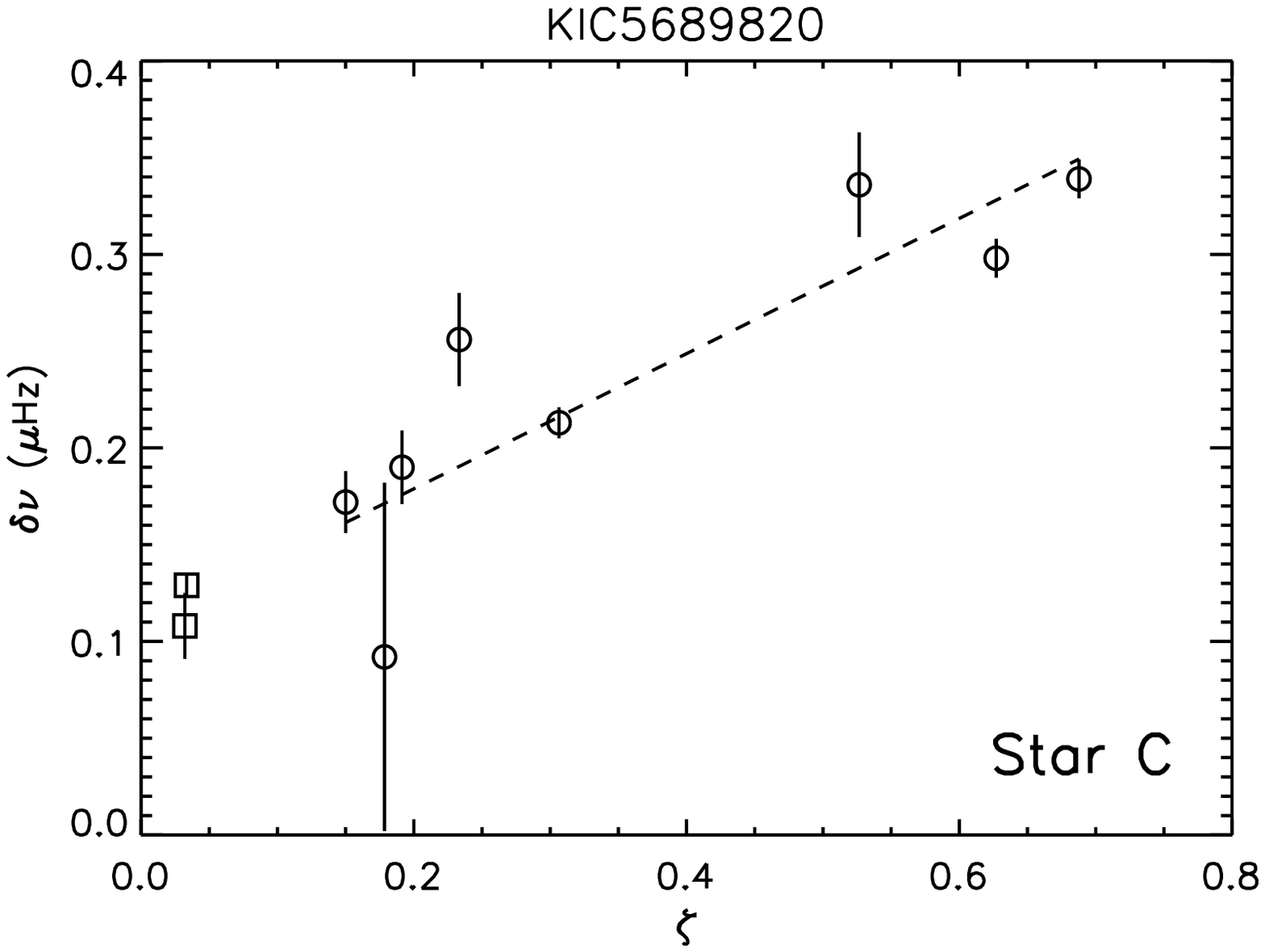}
\includegraphics[width=8cm]{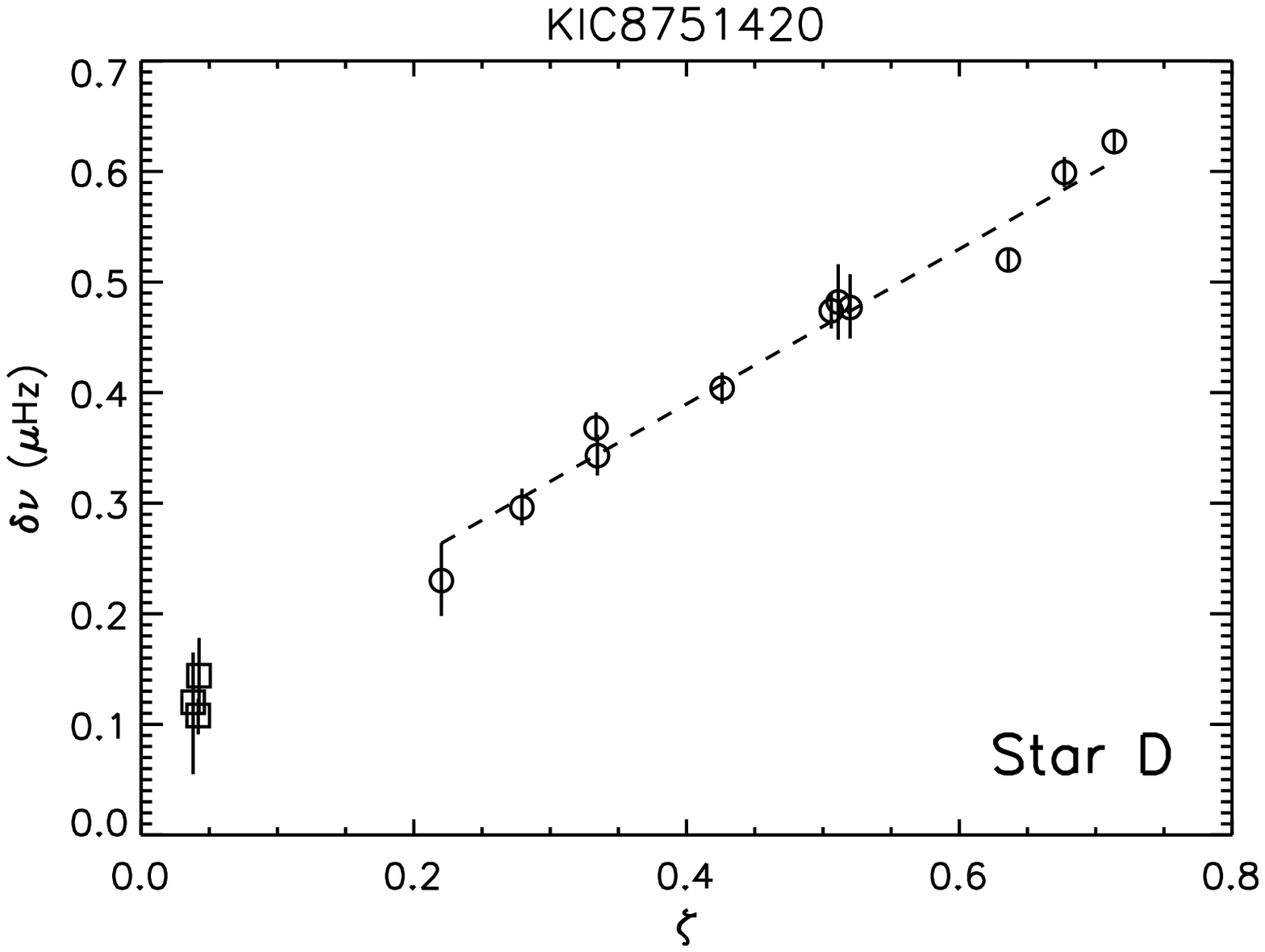}
\includegraphics[width=8cm]{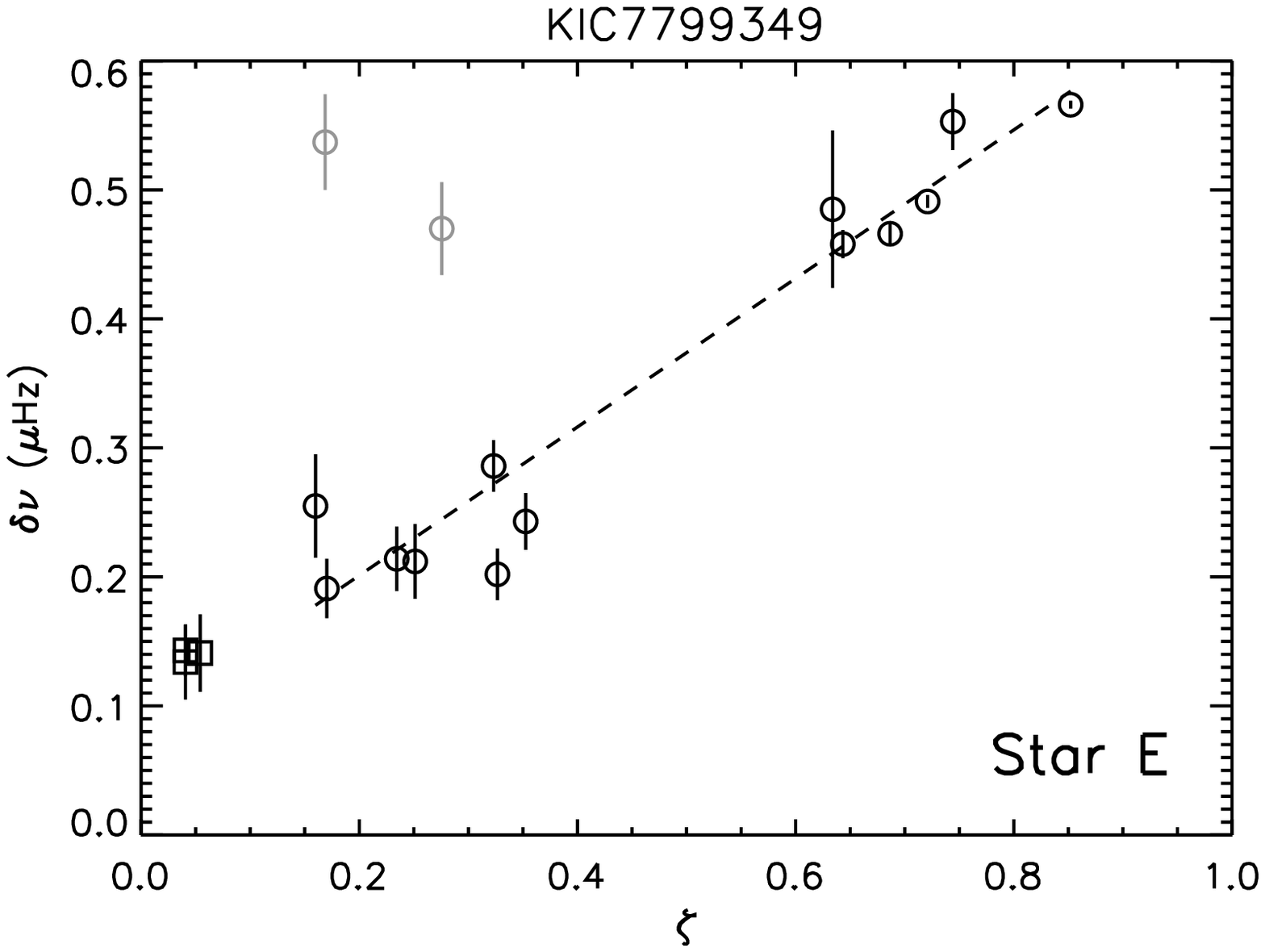}
\includegraphics[width=8cm]{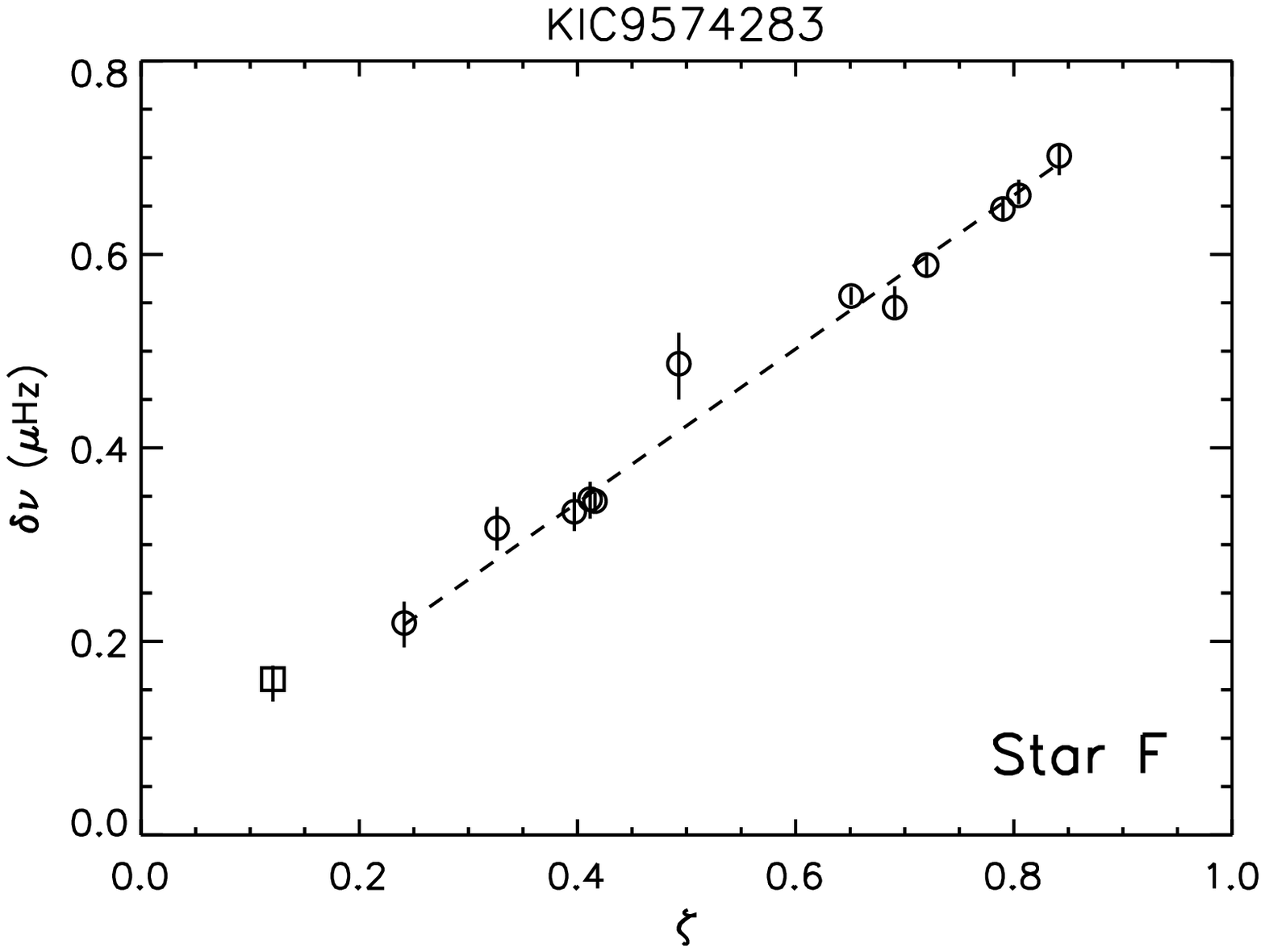}
\end{center}
\caption{Observed splittings (open symbols) for modes of degrees $l=1$ (circles) and $l=2$ (squares), plotted as a function of the parameter $\zeta$, which indicates the trapping of the modes (a value of $\zeta$ closer to 0 indicates a p-dominated mode, whereas a value of $\zeta$ closer to 1 indicates a g-dominated mode). The dashed lines indicate linear regressions of the relation between the splittings of $l=1$ modes and the parameter $\zeta$. Gray symbols indicate suspicious measurements that were not taken into account (see text).
\label{fig_zeta_split}}
\end{figure*}

Having access to stellar models for the stars of the sample, we could relate the observed splittings to the trapping of the corresponding modes. For this purpose, we plotted in Fig. \ref{fig_zeta_split} the observed splittings against the quantity $\zeta$, which was introduced by \cite{goupil13} and is defined as the ratio between the kinetic energy of the mode in the g-mode cavity and the total kinetic energy of the mode, i.e.
\begin{linenomath*}
\begin{equation}
\zeta\equiv\frac{\int_{r\ind{a}}^{r\ind{b}} \rho r^2 \left[\xi_r^2+l(l+1)\xi_h^2\right]\,dr}{\int_0^{R_\star} \rho r^2 \left[\xi_r^2+l(l+1)\xi_h^2\right]\,dr}.
\label{eq_zeta}
\end{equation}
\end{linenomath*}
where $l$ is the degree of the mode, $\xi_r$ and $\xi_h$ are the radial and horizontal displacements, and $r\ind{a}$ and $r\ind{b}$ are the inner and outer turning points of the g-mode cavity. A value of $\zeta$ close to 1 indicates that the mode is mainly trapped in the g-mode cavity (and thus in the core for our stars), and a value of $\zeta$ close to 0, that it is trapped in the p-mode cavity. It is clear from Fig. \ref{fig_zeta_split} that the splittings of g modes are larger than those of p modes, indicating that the core rotates faster than the envelope in these stars, as was found in previous studies of this type (\citealt{beck12}, D12). 

There is a roughly linear relation between the rotational splittings of $l=1$ modes and the ratio $\zeta$. This phenomenon was already theoretically explained by \cite{goupil13}. We note that for star E, the splittings that were obtained for the two highest-frequency $l=1$ modes (modes around 670 and 698 $\mu$Hz, colored in gray in Fig. \ref{fig_zeta_split}) lie well outside this linear relation. None of the rotation profiles that were tested in this study could account for the fitted splittings of these modes. In fact, these high-frequency modes have a large linewidth, and a possible explanation could be that they are too wide to reliably determine their rotational splittings. This should appear more clearly when longer time series are available from \kepler\ observations. In the following, the splittings of these two modes were excluded from the set of splittings that were used to perform the inversions.

\begin{table*}
\caption{Estimates of the mean rotation rate in the g-mode cavity $\langle\Omega\ind{g}\rangle$, the mean rotation rate in the p-mode cavity $\langle\Omega\ind{p}\rangle$, and the ratio between these quantities obtained from the coefficients of the $\delta\nu(\zeta)$ relation (see Sect. \ref{sect_splitvstrap}) or from OLA inversions (see Sect. \ref{sect_OLA}).}
\begin{center}
\begin{tabular}{l c c c c c c c c}
\hline \hline
\T\B Star & \multicolumn{2}{c}{$\langle\Omega\ind{g}\rangle/(2\pi)$ (nHz)} & & \multicolumn{2}{c}{$\langle\Omega\ind{p}\rangle/(2\pi)$ (nHz)} & & \multicolumn{2}{c}{$\langle\Omega\ind{p}\rangle/\langle\Omega\ind{g}\rangle$} \\
\cline{2-3}
\cline{5-6}
\cline{8-9}
\T\B &  $\delta\nu(\zeta)$ & OLA & & $\delta\nu(\zeta)$ & OLA & & $\delta\nu(\zeta)$ & OLA \\
\hline
\T A (KIC12508433) & 505 & $532\pm79$ & & 227 & $213\pm26$ & & 2.2 & $2.5\pm0.7$  \\
B (KIC8702606) & 619 & $629\pm109$ & & 263 & $164\pm17$ & & 2.4 & $3.8\pm1.1$ \\
C (KIC5689820) & 917 & $865\pm35$ & & 109 & $125\pm13$ & & 8.4 & $6.9\pm1.0$  \\
D (KIC8751420) & 1620 & $1540\pm50$ & & 109 & $102\pm24$ & & 14.8 & $15.1\pm4.0$   \\
E (KIC7799349) & 1323 & $1313\pm26$ & & 86 & $122\pm15$ & & 15.4 & $10.7\pm1.5$  \\
\B F (KIC9574283) & 1640 & $1556\pm22$ & & 26 & $74\pm28$ & & 64.1 & $21.0\pm8.2$ \\
\hline
\end{tabular}
\end{center}
{\small \textbf{Notes:} $\langle\Omega\ind{g}\rangle$ represents the average rotation rate in the g-mode cavity (innermost 1.5\% to 2.5\% of the star) and $\langle\Omega\ind{p}\rangle$ a weighted average of the rotation rate in the p-mode cavity, which can be regarded as an upper limit of the rotation rate in the convective envelope (see text).}
\label{tab_om_OLA}
\end{table*}%


\cite{goupil13} showed that the coefficients of the linear relation $\delta\nu(\zeta)$ \textbf{for $l=1$ modes} can be used to obtain estimates of the mean rotation rate in the g-mode cavity $\langle\Omega\ind{g}\rangle$ and the mean rotation rate in the p-mode cavity $\langle\Omega\ind{p}\rangle$. By combining their Eq. 21 and 22, we obtain
\begin{linenomath*}
\begin{equation}
\delta\nu = \zeta\left( \frac{1}{2} \left\langle \frac{\Omega\ind{g}}{2\pi} \right\rangle - \left< \frac{\Omega\ind{p}}{2\pi} \right> \right) + \left< \frac{\Omega\ind{p}}{2\pi} \right> 
\label{eq_goupil}
\end{equation}
\end{linenomath*}
We thus fitted a relation of the type $\delta\nu=A\zeta+B$ to the observed splittings of $l=1$ modes for the six stars (see Fig. \ref{fig_zeta_split}). From Eq. \ref{eq_goupil}, we derive $\langle\Omega\ind{g}/(2\pi)\rangle=2(A+B)$ and $\langle\Omega\ind{p}/(2\pi)\rangle=B$. We note that for more evolved red giants, the contribution from the envelope to the rotational splittings becomes negligible and reliable estimates of $\langle\Omega\ind{p}\rangle$ cannot be obtained from the $\delta\nu(\zeta)$ relation (\citealt{goupil13}). The obtained results are given in Table \ref{tab_om_OLA}. There are clear trends with the evolutionary status, suggesting that  $\langle\Omega\ind{g}\rangle$ increases and $\langle\Omega\ind{p}\rangle$ decreases as stars evolve at the base of the red giant branch, resulting in an increase of the ratio $\langle\Omega\ind{g}\rangle/\langle\Omega\ind{p}\rangle$. These trends are discussed in Sect. \ref{sect_discussion}.

By using the rotational splittings of the modes that were obtained in Sect. \ref{sect_analysis} and the rotational kernels of the best stellar models from Sect. \ref{sect_model}, we applied several inversion techniques to probe the rotational profiles of the six stars that were selected in our sample. The results that are presented below were obtained using the optimal models of \cesam. However, all the inversions were also performed using the best models of \astec\ and yielded results that are quantitatively very similar. 

\subsection{Core and envelope rotation \label{sect_OLA}}

We first tried to obtain localized constraints on the rotation profiles of the selected targets. For this purpose, the OLA (optimally localized averages) method is particularly well suited. The OLA method consists in forming combinations of the rotational kernels such that the resulting averaging kernels $\mathcal{K}(r;r_0)=\sum_k c_k(r_0)K_k(r)$ are as localized as possible around a target point $r_0$. Note that for clarity, we now use the subscript $k=1,M$ for the modes whose splitting could be determined, instead of their order $n$ and degree $l$. If the averaging kernel is sufficiently well localized around $r_0$, then it is straightforward to obtain an estimate of the rotation rate at depth $r_0$ through the relation
\begin{linenomath*}
\begin{align}
2\pi \sum_k  c_k(r_0) \delta\nu_{n,l} & = \sum_k c_k(r_0) \int_0^R K_k(r)\Omega(r)\,dr \nonumber \\
& = \int_0^R \mathcal{K}(r;r_0)\Omega(r)\,dr \nonumber \\
& \approx \Omega(r_0)
\label{eq_omega_OLA}
\end{align}
\end{linenomath*}
The coefficients $c_k(r_0)$ are searched so that the averaging kernel $\mathcal{K}(r;r_0)$ approaches the Dirac function $\delta(r-r_0)$ as closely as possible. For this purpose, \cite{backus68} advocated minimizing the function
\begin{linenomath*}
\begin{equation}
J=12 \int_0^R \mathcal{K}(r;r_0)^2 (r-r_0)^2\, dr + \gamma \sum_{k=1}^M \left[ c_k(r_0)\sigma_{\delta\nu_k} \right]^2
\label{eq_OLA}
\end{equation}
\end{linenomath*}
for each considered point $r_0$, by requiring that the integral of the averaging kernel be unity. Here, $\gamma$ is a trade-off parameter between resolution of the averaging kernel and error magnification, and the $\sigma_{\delta\nu_k}$ are the measurement errors of the rotational splitting estimates. As was pointed out by D12, it is very hard to obtain localized averaging kernels in our case, and we therefore took $\gamma=0$. 

Since the modes are mixed, their rotational kernels have a contribution both from the core, due to their g-mode character, and from the envelope where they behave as p modes. The ratio between these contributions depend on where the modes are predominantly trapped. Because of the shape of the mode kernels, it was impossible with the set of modes at our disposal to build averaging kernels that are well localized at intermediate depths inside the star and we could therefore not perform an inversion of the whole rotation profile throughout the star. We could however obtain estimates of the rotation rate in the core and the envelope.

\subsubsection{Core rotation \label{sect_OLA_core}}

\begin{figure}
\begin{center}
\includegraphics[width=9cm]{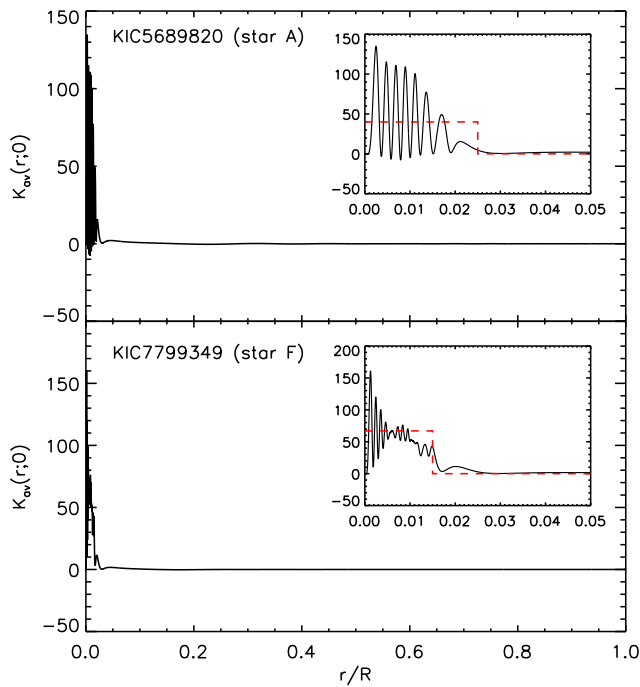}
\end{center}
\caption{Core averaging kernels obtained with the OLA method for the star A (top) and F (bottom). A zoom of the kernels in the core are also shown. Step functions between 0 and 0.025 $R$ (star A) or 0.015 (star F) are overplotted (red dashed lines).
\label{fig_core_kernels}}
\end{figure}

By minimizing the function $J$ defined by Eq. \ref{eq_OLA} for values of $r_0$ between 0 and 0.02 $R$, we obtained averaging kernels that efficiently cancel the contribution from the p-mode cavity. The more evolved the star is, the smaller the envelope contribution to the averaging kernel becomes. We show in Fig. \ref{fig_core_kernels} the core averaging kernels that were obtained for stars A and F, which are the less evolved and the most evolved star of the sample, respectively. It is clear that these averaging kernels are poor approximations to Dirac functions; however if we assume that the rotation profile varies smoothly in the core, they can be well approximated by step functions in the g-mode cavity (i.e. between 0 and 0.015 to 0.025 $R$), as can be seen in Fig. \ref{fig_core_kernels}. By applying Eq. \ref{eq_omega_OLA}, we thus obtained estimates of the average rotation rate in the innermost 1.5\% to 2.5\% for the stars of the sample. The results are summarized in Table \ref{tab_om_OLA}. The same inversions using the \astec\ models provide core rotation rates that agree with the values quoted in Table \ref{tab_om_OLA} within less than 1.3 $\sigma$. The core rotations obtained with the OLA inversions are in good agreement with the values obtained from the coefficients of the linear regression of the relation $\delta\nu(\zeta)$ (see Table \ref{tab_om_OLA}). This validates the prescription of \cite{goupil13} for stars at the base of the RGB.

\subsubsection{Envelope rotation \label{sect_OLA_env}} 

\begin{figure}
\begin{center}
\includegraphics[width=9cm]{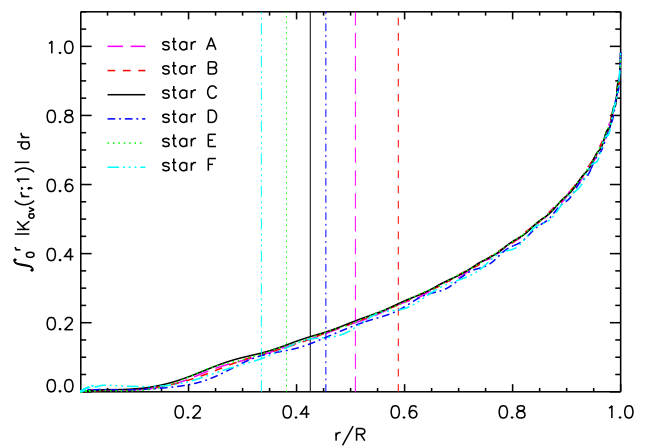}
\end{center}
\caption{Cumulative integral of the surface averaging kernel obtained with the OLA method for the stars of the sample. The vertical lines indicate the base of the convective envelope for each star.
\label{fig_surface_kernels}}
\end{figure}

\begin{figure*}
\begin{center}
\includegraphics[width=6cm]{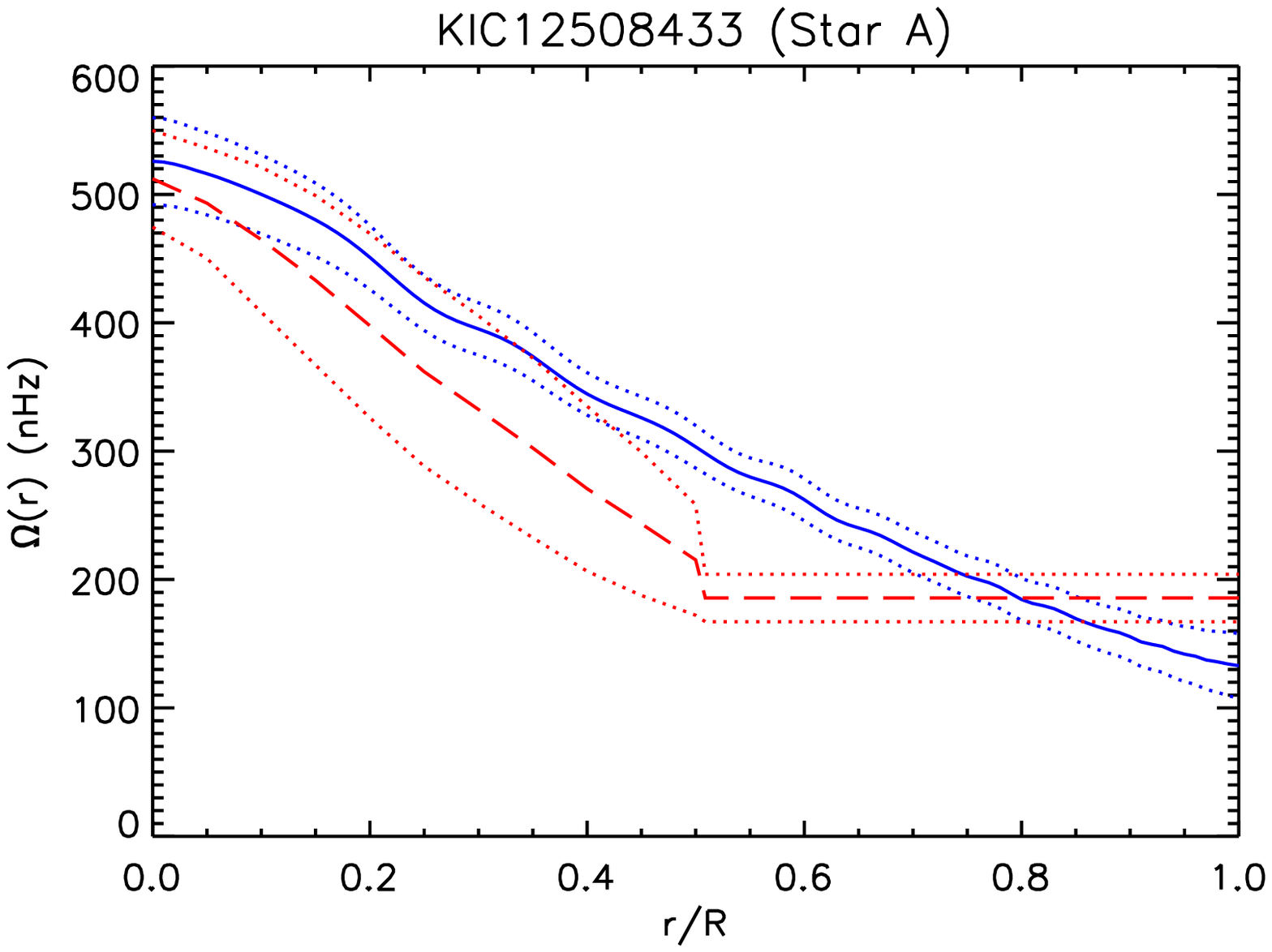}
\includegraphics[width=6cm]{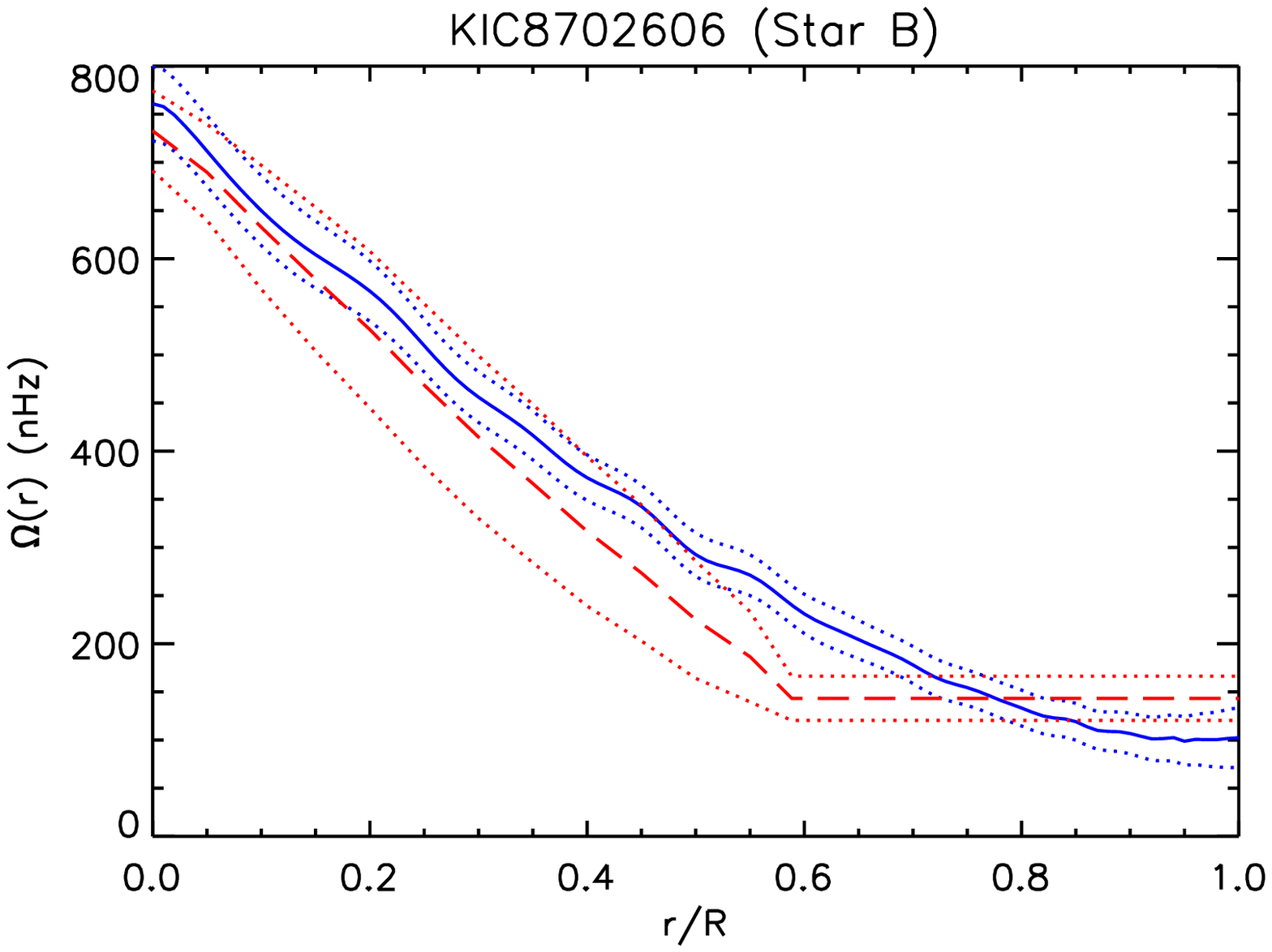}
\includegraphics[width=6cm]{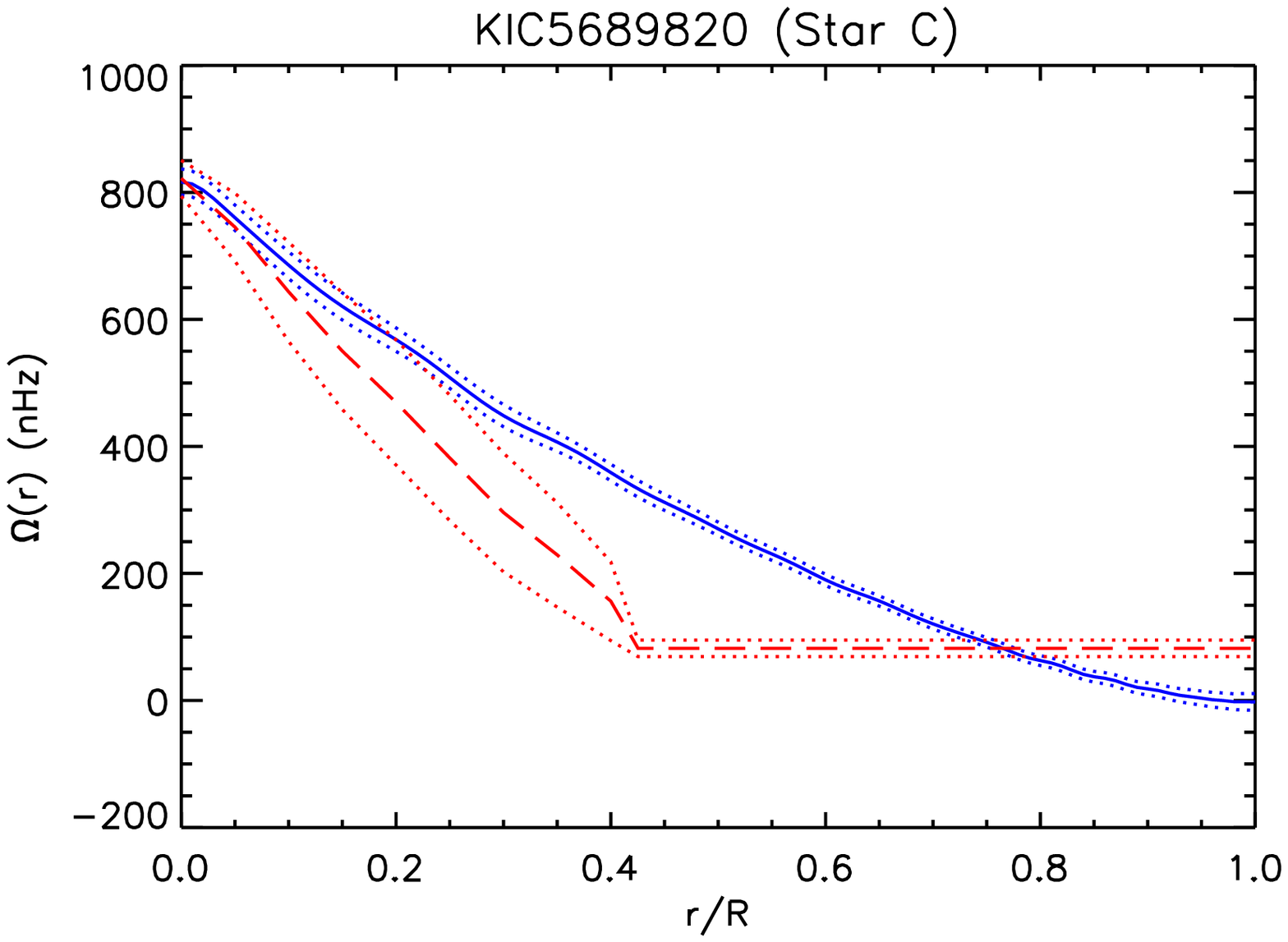}
\includegraphics[width=6cm]{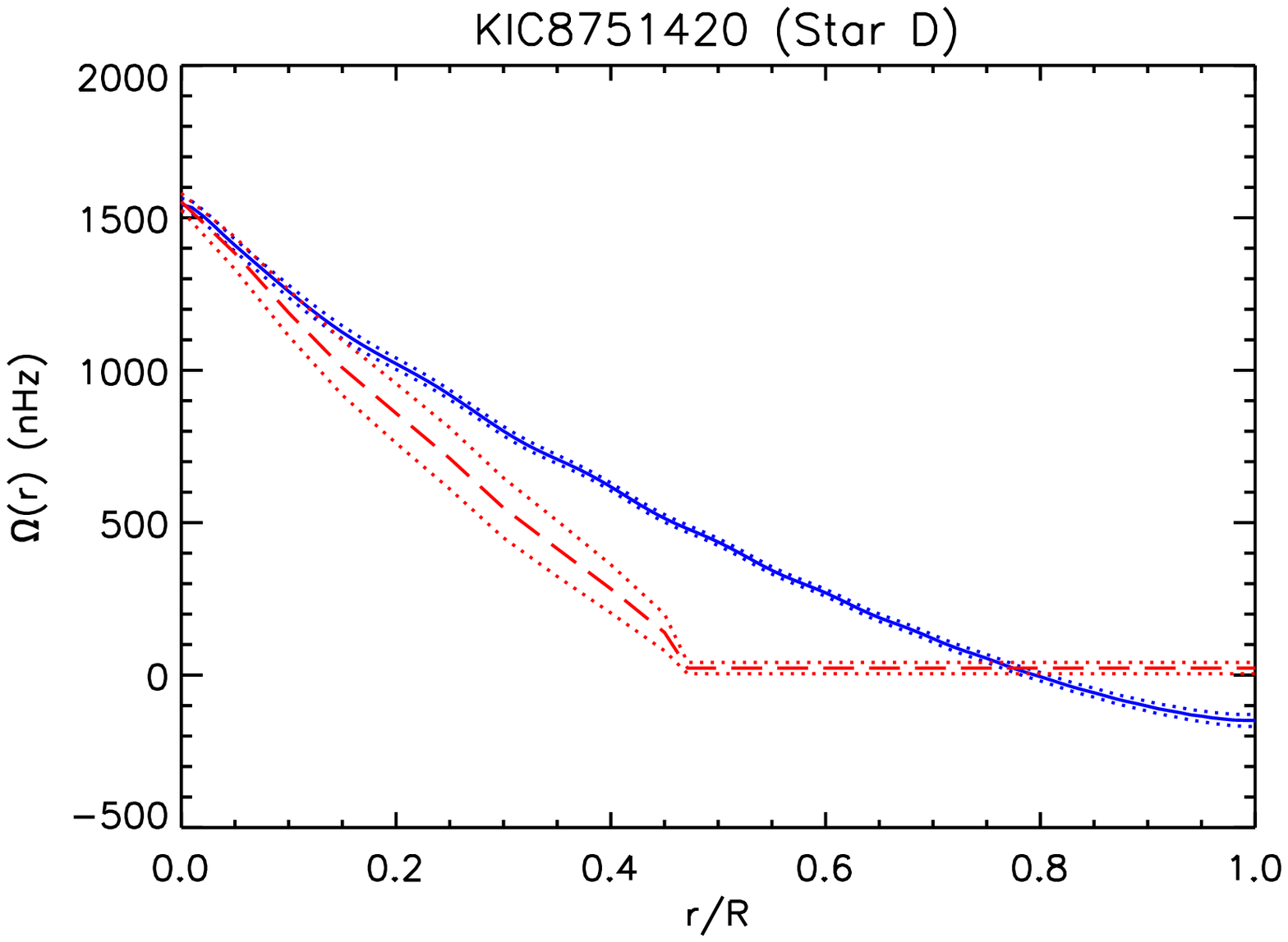}
\includegraphics[width=6cm]{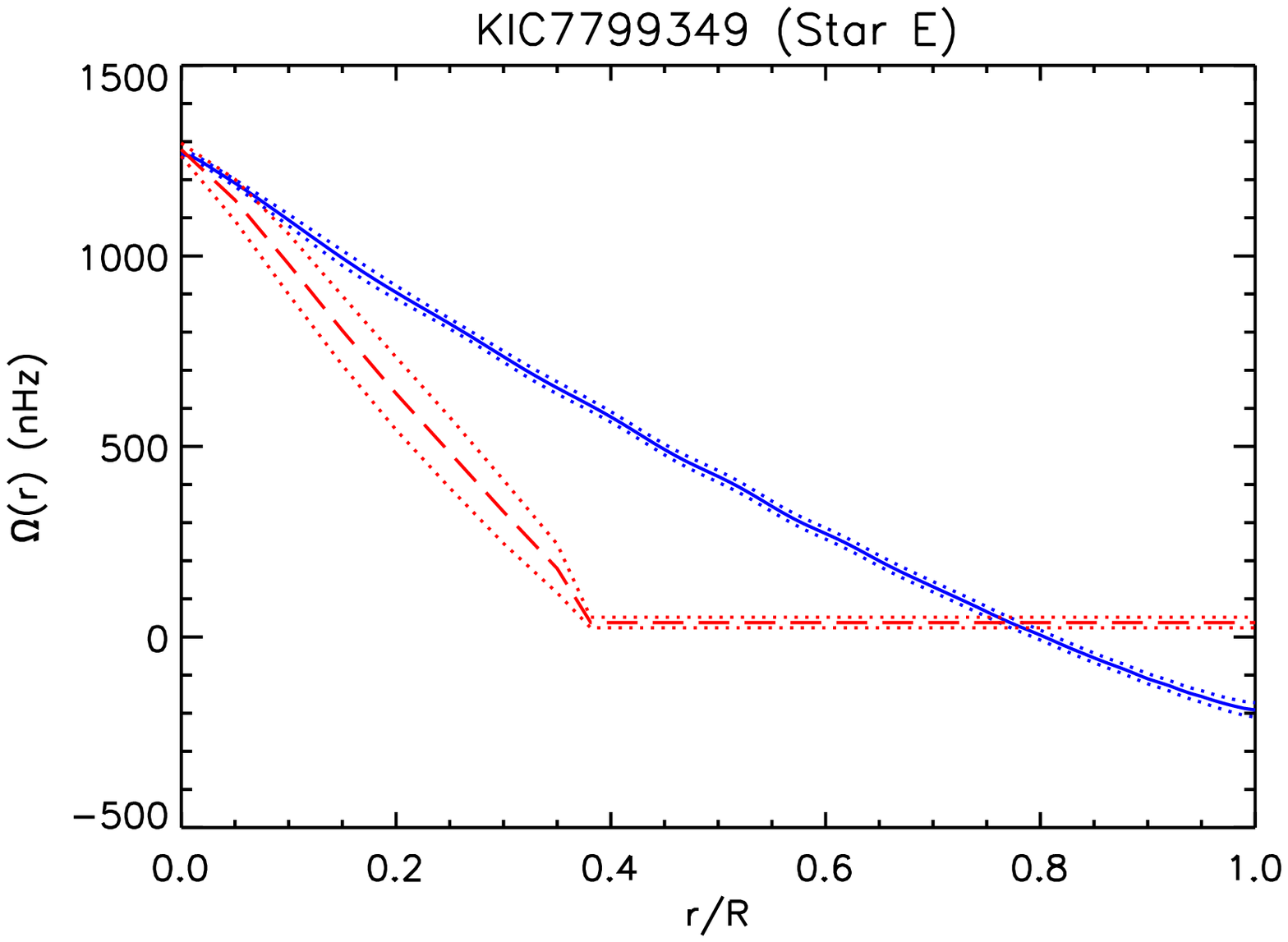}
\includegraphics[width=6cm]{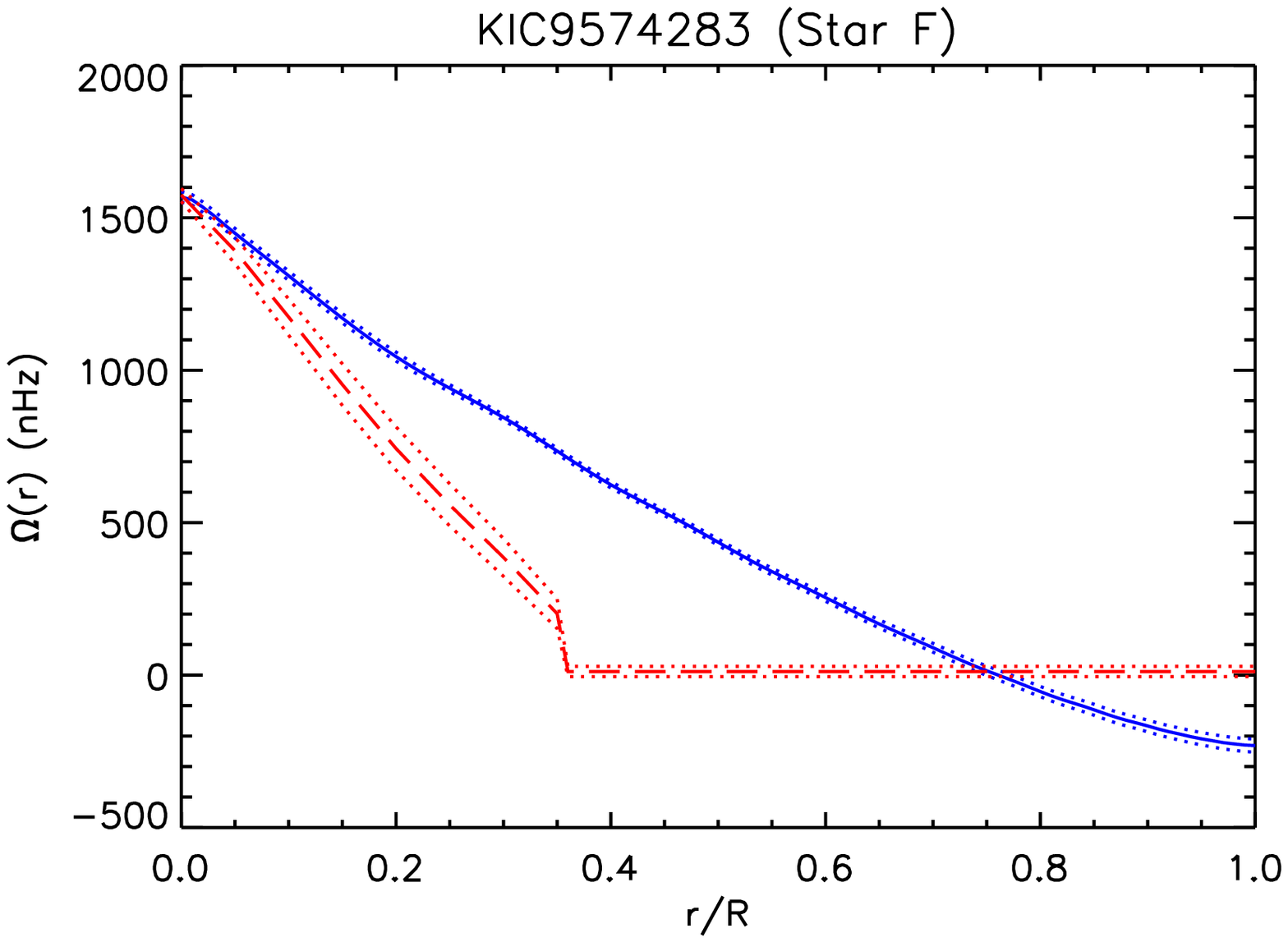}
\end{center}
\caption{Optimal rotation profiles obtained by applying the RLS method with a smoothness condition on the rotation profile on the entire star (blue solid lines) or only in the radiative interior while the convective envelope is assumed to rotate as a solid-body (red long-dashed lines). The dotted lines indicate the 1-$\sigma$ error bars for both types of inversions.
\label{fig_RLS}}
\end{figure*}

The surface averaging kernels are obtained by choosing $r_0= 1$ and searching for the coefficients $c_k(r_0)$ that minimize the function $J$. We show in Fig. \ref{fig_surface_kernels} the cumulative integral of the surface kernels $\int_0^R |\mathcal{K}(r;1)|\, dr$ for the stars of the sample. First, we note that the contribution of the g-mode cavity to the surface averaging kernels seems to be efficiently canceled. However, it is clear from Fig. \ref{fig_surface_kernels} that the surface averaging kernels provide a weighted average of the rotation rate in the whole p-mode cavity rather than an estimate of the surface rotation itself. Without knowledge of the variations in $\Omega(r)$ in the p-mode cavity, no direct information about the rotation profile in the envelope can be inferred. If we assume that the star rotates as a solid-body in the whole p-mode cavity, then the quantity $2\pi\sum_k c_k(r_0)\delta\nu_k$ provides an estimate of the envelope rotation rate. The obtained estimates are listed in Table \ref{tab_om_OLA} for the six stars. These estimates can in fact be regarded as upper limits to the rotation rate in the convective envelope. Indeed, Fig. \ref{fig_surface_kernels} shows that the radiative part of the p-mode cavity, which rotates presumably faster than the convective envelope, also contributes to the surface averaging kernels. Besides, even though the contribution from the g-mode cavity to the surface averaging kernel is small, we know that the core rotates much faster than the envelope, which leads us to overestimate the envelope rotation. This leakage from this region can be quantified by separating the contributions from the core and the envelope as follows
\begin{linenomath*}
\begin{equation}
2\pi\sum_k c_k(r_0)\delta\nu_k \approx \Omega_{\rm g} \int_{\rm g} \mathcal{K}(r;1)\,dr + \Omega_{\rm p} \int_{\rm p} \mathcal{K}(r;1)\,dr.
\label{eq_contrib_core}
\end{equation}
\end{linenomath*}
The first term of the right-hand side in Eq. \ref{eq_contrib_core} represents the contribution from the g-mode cavity and can be estimated by using the core rotation rates that were obtained in Sect. \ref{sect_OLA_core}. Taking this into account, we found that the estimates of the rotation rates in the p-mode cavity listed in Table \ref{tab_om_OLA} are reduced by only 0.6\% to 2.6\% for stars A through E. On the contrary, for star F, which is the most evolved star in our sample, the leakage from the core amounts to 26\%. Similar inversions using the \astec\ models yielded envelope rotation rate that agree with the values quoted in Table \ref{tab_om_OLA} within less than 1 $\sigma$ for all the stars.

Estimates of the ratio $\Omega\ind{p}/\Omega\ind{g}$ could be obtained by combining our estimates of the envelope rotation rates with the estimates of the core rotation rates obtained in Sect. \ref{sect_OLA_core}. The results are given in Table \ref{tab_om_OLA}. Except for star F, they are in good agreement with the values obtained with the prescription of \cite{goupil13}. The disagreement for star F probably comes from the fact that our estimate of $\Omega\ind{p}$ is overestimated due to a leakage from the core as explained above. 

\subsection{Testing the existence of strong gradients in the rotation profiles \label{sect_test_disc}}

As mentioned above, it is very difficult to obtain localized information about the rotation profile in regions other than the deepest layers, because in these zones the averaging kernels suffer from important leakage from the core and the surface. Another approach consists in trying to place constraints on the overall shape of the rotation profiles. One important question is that of the existence of discontinuities or sharp gradients in the rotation profiles. Indeed, if we assume conservation of the specific angular momentum of each layer, the evolution in the post-main sequence leads to a fast-spinning core and a slow-spinning envelope, with a sharp rotation gradient in the intermediate region, where the H-burning shell lies. If we assume on the contrary an instantaneous AM transport, the whole star rotates as a solid body. We know that the reality should be somewhere between these two limiting cases, but the shape of the rotation profile is still very uncertain. We therefore tried to test the existence of strong gradients in the rotation profiles of the stars in our sample. For this purpose, we fitted both smooth and discontinuous rotation profiles to the observed splittings.

\subsubsection{Smooth rotation profiles \label{sect_smooth}}

To ensure that the fitted rotation profiles are smooth, we used the RLS (regularized least-squares) method. It consists in searching the rotation profile that gives the closest match to the observed splittings by performing a least-squares fitting. Owing to the limited number of available splittings, which is not sufficient to reconstruct in a unique way the whole rotation profile, the problem must be regularized. The function to be minimized corresponds to the sum of the $\chi^2$ residual of the fit and a regularization function that penalizes undesirable features in the solution. Here, since we were searching for smooth rotation profiles, we imposed a smoothness condition, by taking the regularization function as the square of the norm of its first derivative. Two options were studied: we either imposed a smoothness condition to the whole rotation profile (hereafter denoted by the index 'smooth') or to the radiative interior only, assuming a solid-body rotation in the envelope (denoted by the index 'SBenv'). The balance between minimizing $\chi^2$ and having a smooth solution is controlled by a trade-off parameter $\lambda$. The value of this trade-off parameter was determined by generating artificial rotation profiles and trying to recover them for different values of $\lambda$. 

\begin{figure*}
\begin{center}
\includegraphics[width=6cm]{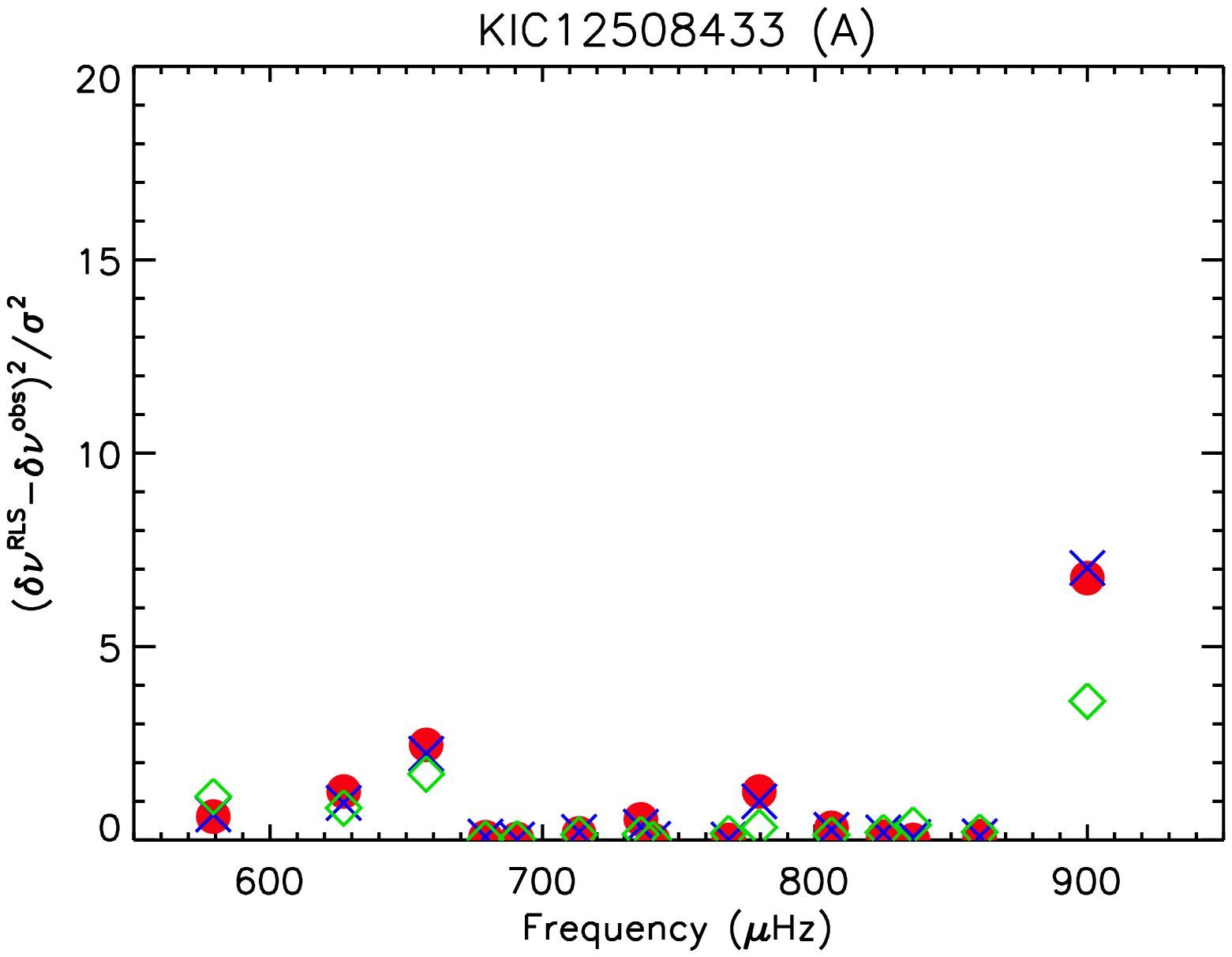}
\includegraphics[width=6cm]{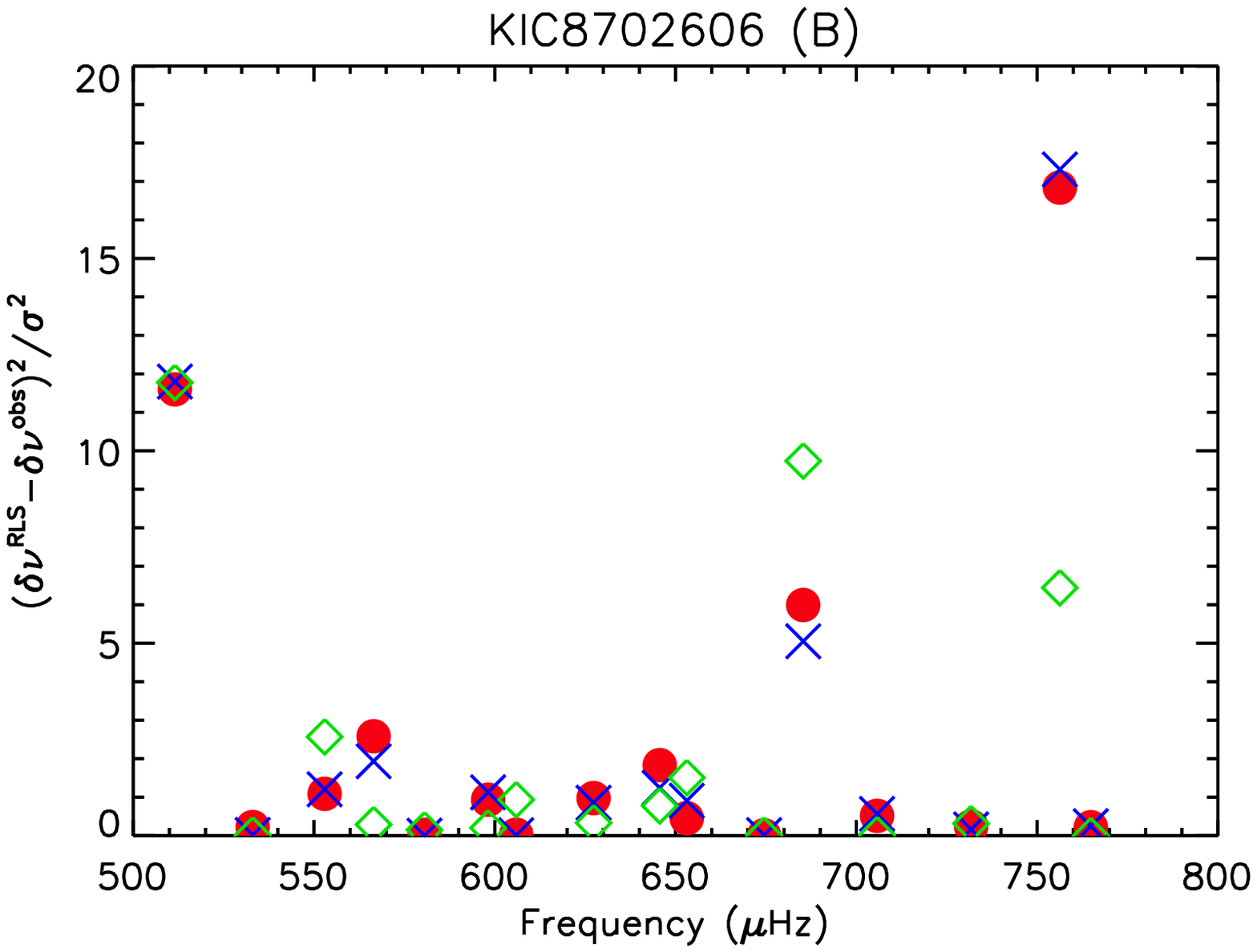}
\includegraphics[width=6cm]{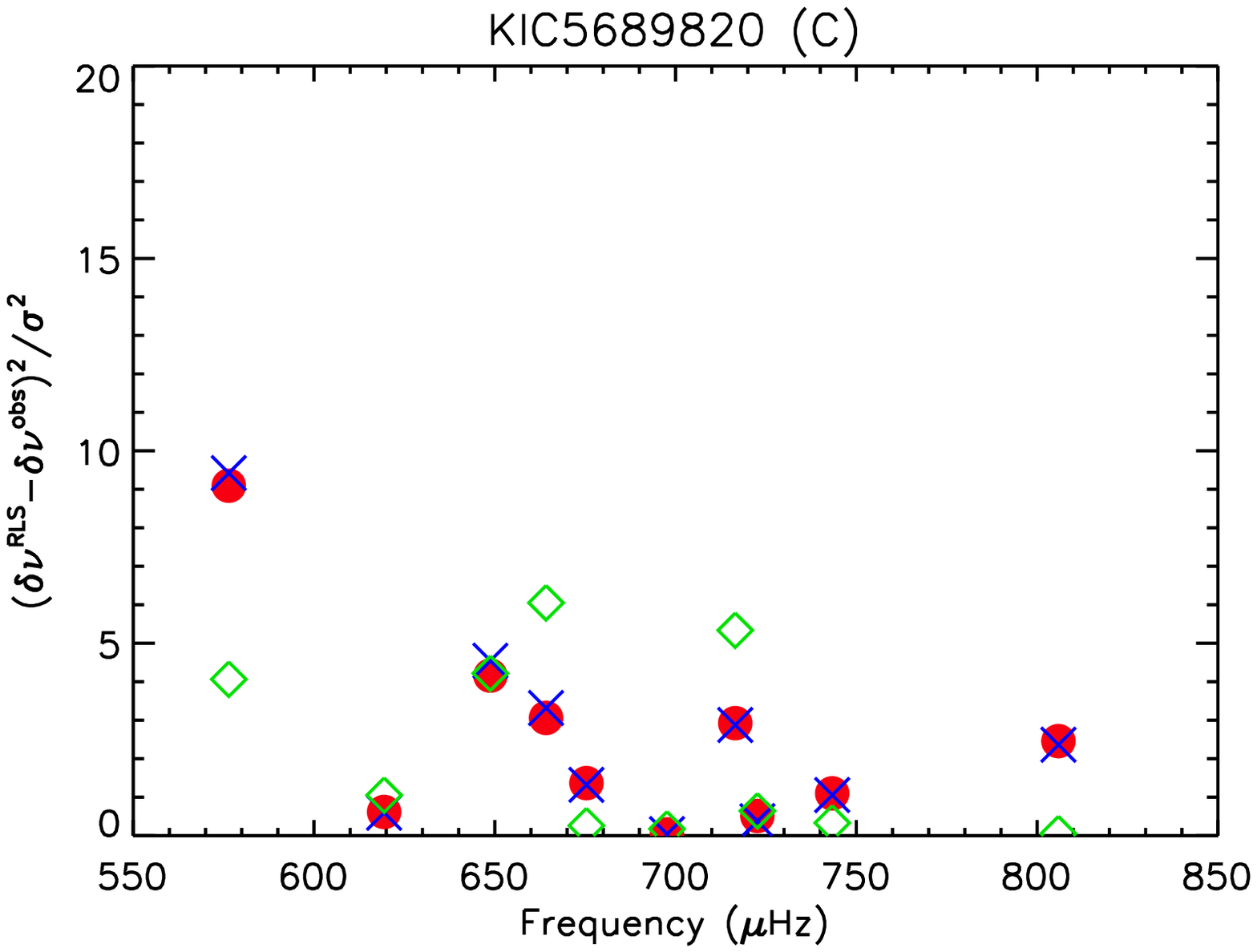}
\includegraphics[width=6cm]{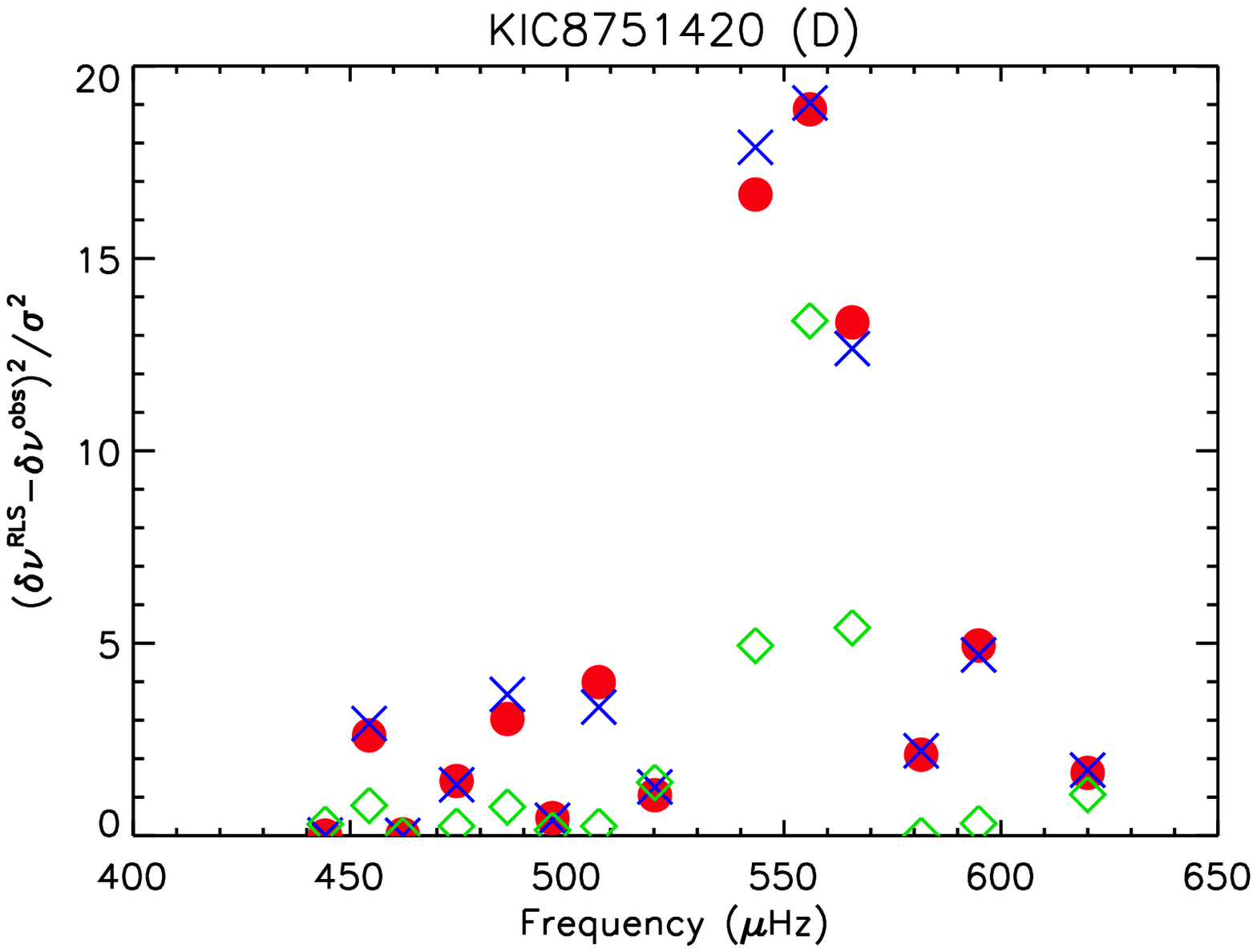}
\includegraphics[width=6cm]{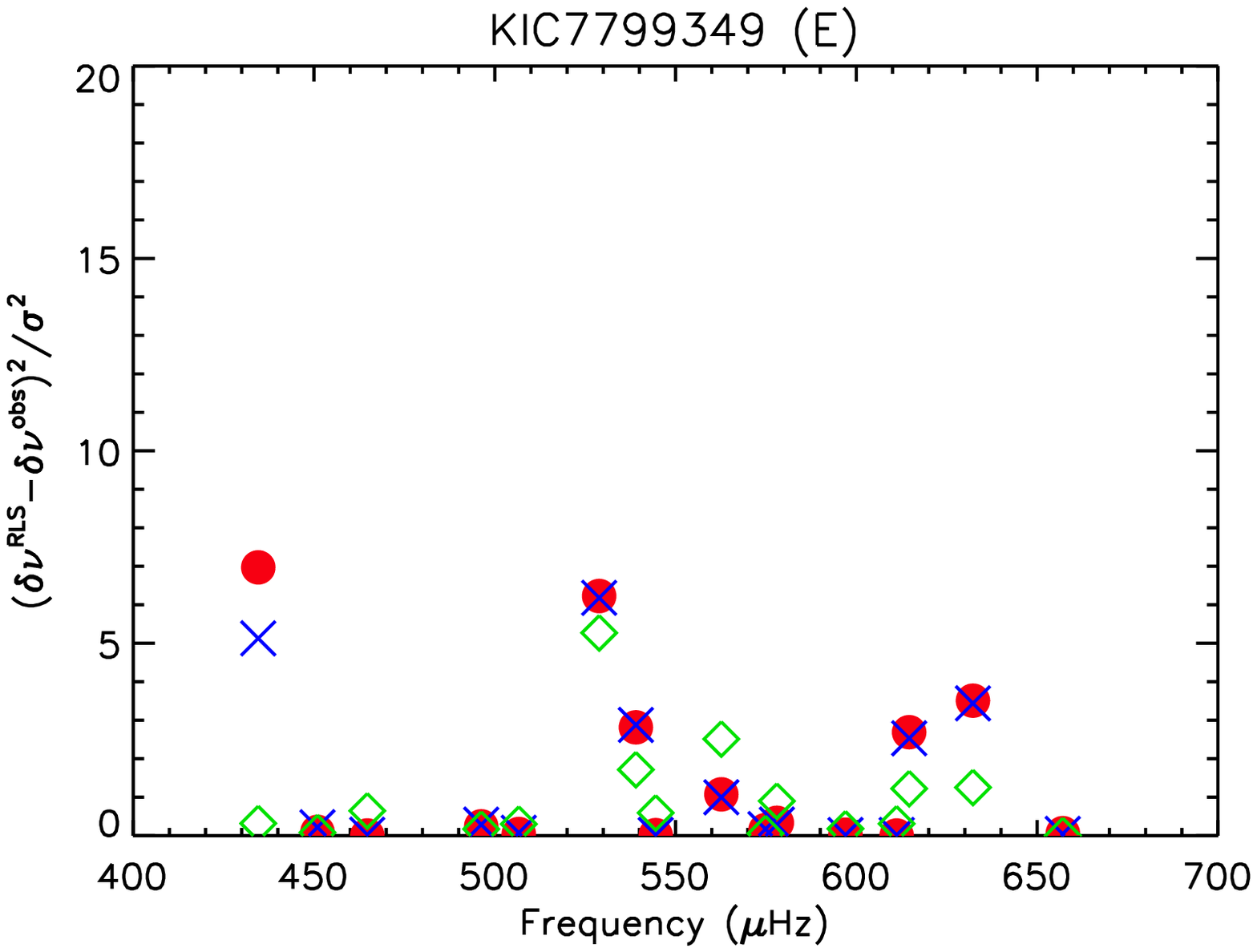}
\includegraphics[width=6cm]{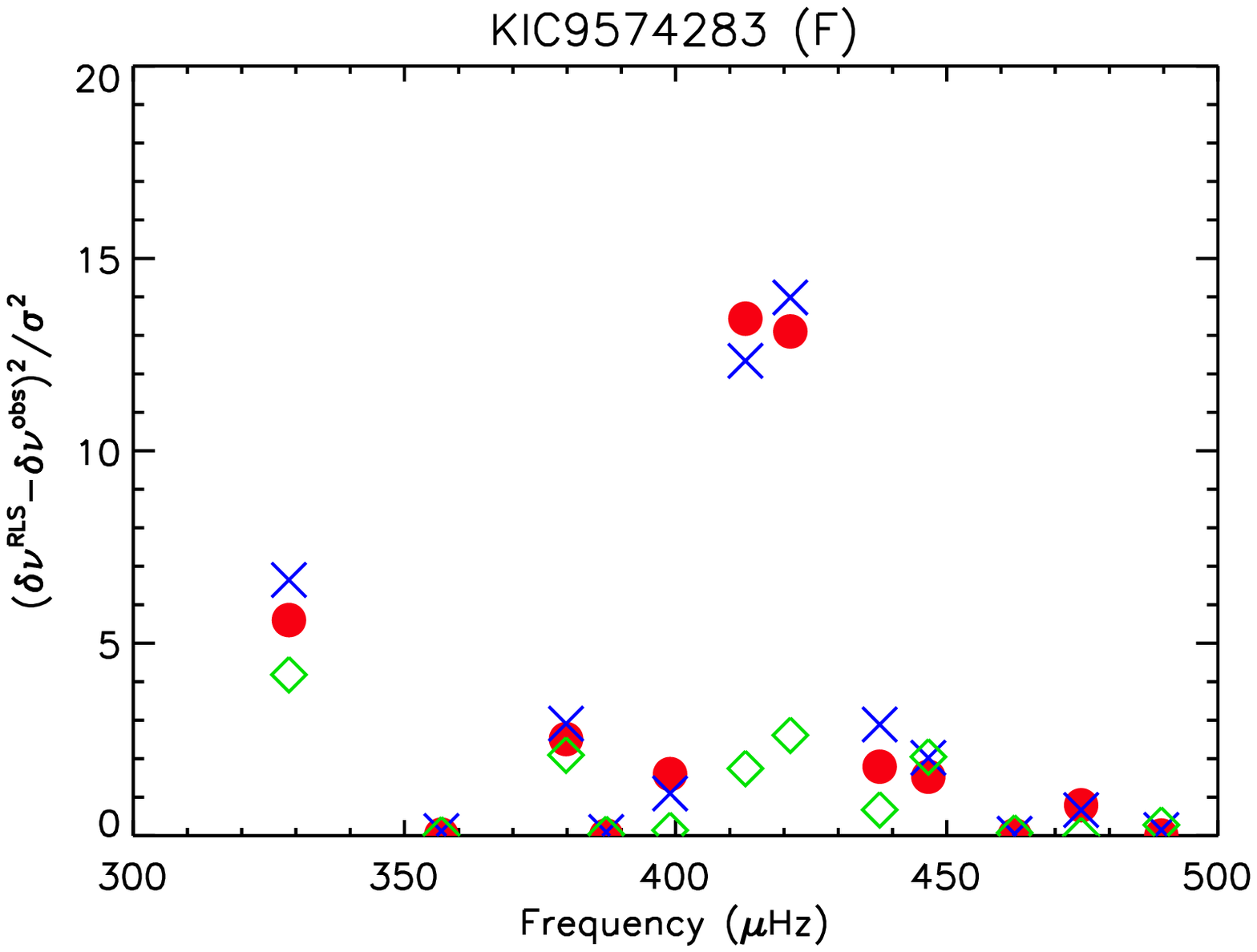}
\end{center}
\caption{Contribution of each mode to the $\chi^2$ values computed in Sect. \ref{sect_test_disc}. The symbols correspond to the splittings computed with the optimal rotation profiles that were obtained using the RLS method with a smoothness condition on the whole profile (blue crosses), the RLS method with a smoothness condition in the radiative interior only and solid-body rotation in the convective envelope (red circles), or a 2-zone model with an optimized depth of discontinuity (see Sect. \ref{sect_disc}, green diamonds).
\label{fig_contrib_chi2}}
\end{figure*}

The optimal rotation profiles that were obtained for the six stars are shown in Fig. \ref{fig_RLS}. In the case where the rotation profile is required to be smooth throughout the star (blue solid lines), the optimal profiles vary almost linearly as a function of the radius. For stars C, D, E, and F, the surface is found to be counter-rotating. However, we showed in Sect. \ref{sect_OLA_env} that with our data, we can only obtain information about the average rotation in the envelope. Since the shape of the rotation profile cannot be determined in the envelope, it is clear that the smoothness condition dominates and imposes a linear behavior for the recovered profile in this region, thus causing the surface to be counter-rotating. The optimal profiles in the case where solid-body rotation is assumed in the convective envelope (red long-dashed lines in Fig. \ref{fig_RLS}) produce rotational splittings that are very similar to those of the previous case, which appears clearly in Fig. \ref{fig_zeta_split}.

The agreement between the observed splittings and the ones produced by the optimal smooth profiles (noted $\delta\nu^{\rm RLS}$) was estimated by computing the $\chi^2$ of the residual of the fits
\begin{linenomath*}
\begin{equation}
\chi^2 = \sum_{k=1}^M \left( \frac{\delta\nu_k^{\rm RLS}-\delta\nu_k^{\rm obs}}{\sigma_{\delta\nu_k}}\right)^2.
\label{eq_chi2_RLS}
\end{equation}
\end{linenomath*}
Since our aim is to interpret the obtained values of $\chi^2$ in a statistical sense, special care needs to be given to its normalization. It is customary to normalize the $\chi^2$ by its expected value, which yields a \textit{reduced} $\chi^2$ of expected value unity. The expected value of $\chi^2$ is equal to its number of degrees of freedom. For ordinary least-squares fit, it corresponds to $M-\mu$, where $\mu$ is the number of fitted parameters. For regularized least-squares fit, it is more delicate and requires computing an \textit{effective number of degrees of freedom}, which obviously depends on the regularization parameter $\lambda$. By following \cite{hastie90}, we estimated the effective number of degrees of freedom of the RLS fits for the six studied stars, given the chosen regularization parameter $\lambda$. Details of the calculation are presented in Appendix \ref{app_dof}. We obtained values around $M-2.2$ for all the stars. This is hardly a surprise since the inverted rotation profiles are close to straight lines (see Fig. \ref{fig_RLS}) and fitting the observed splittings to a straight line would give $\mu=2$, and thus $M-2$ degrees of freedom. We thus defined the reduced $\chi^2$ as $\chi^2\ind{red}\equiv\chi^2/(M-2.2)$.

The obtained $\chi^2\ind{red}$ are given in Table \ref{tab_chi2}. First we observe that the case where a smooth profile is assumed throughout the star produces $\chi^2\ind{red}$ values that are very similar to the case where solid-body rotation is assumed in the envelope. For stars A and E, the observed splittings are very well reproduced by smooth profiles, with reduced $\chi^2$ around 1. For stars B and C, the reduced $\chi^2$ is around 3, indicating a less good match with the observations. This could indicate that the rotation profile is not smooth as supposed. However, as can be seen in Fig. \ref{fig_contrib_chi2}, the modes that contribute the most to the value of $\chi^2$ for these two stars are located at the edges of the frequency range of the oscillations, where the signal-to-noise ratio is the lowest. We can therefore not exclude that this mismatch is caused by wrong splitting estimates. For stars D and F, the reduced $\chi^2$ is even higher, reaching values of about 6.0 and 4.4, respectively. For these stars, the splittings of several modes differ from the observed splittings by 3 to 5 $\sigma$. Besides, Fig. \ref{fig_contrib_chi2} shows that the modes that contribute the most to the $\chi^2$ value lie around $\nu\ind{max}$ so it is very unlikely that this disagreement might be due to a wrong determination of the splittings.

\begin{figure*}
\begin{center}
\includegraphics[width=6cm]{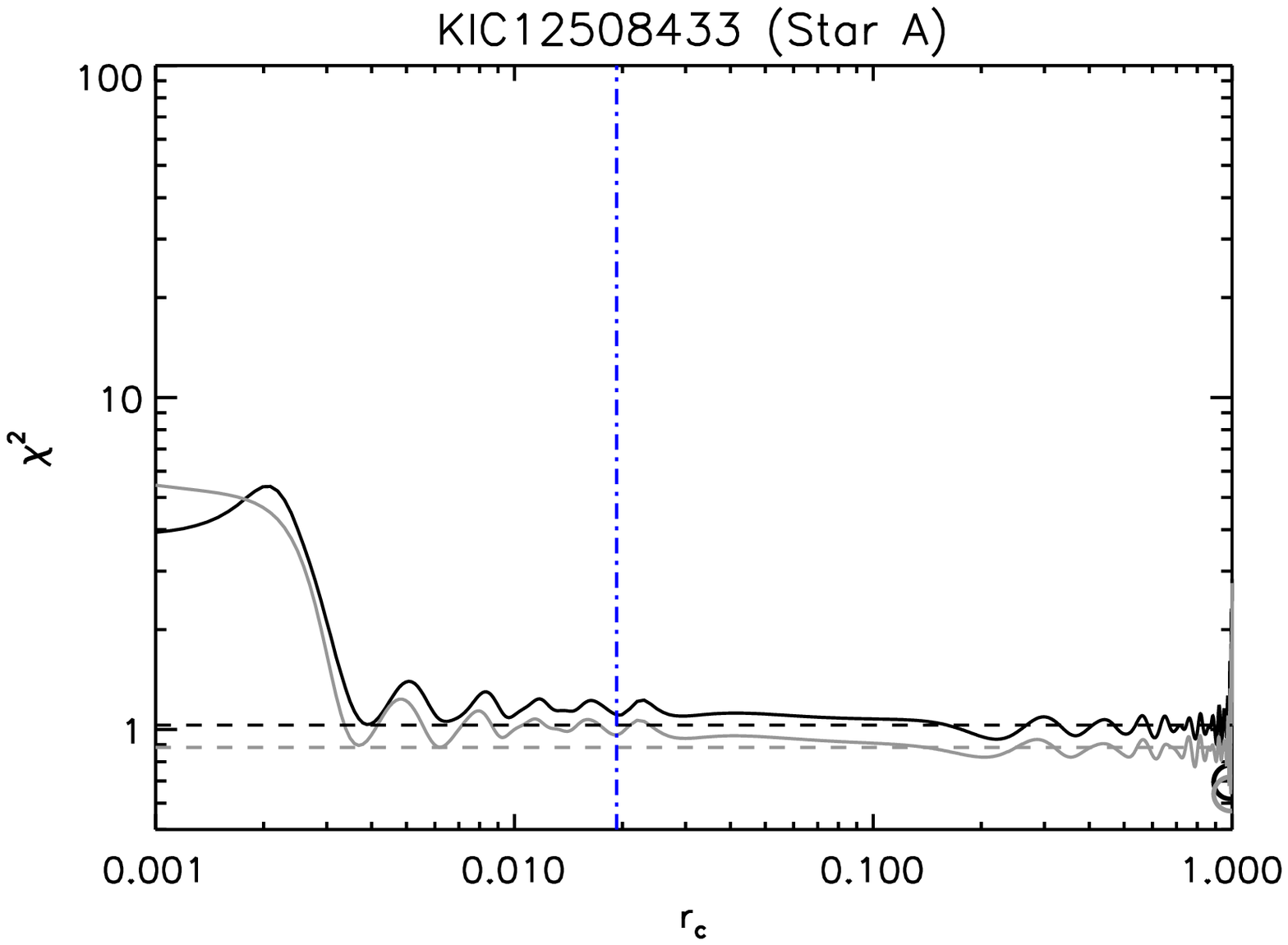}
\includegraphics[width=6cm]{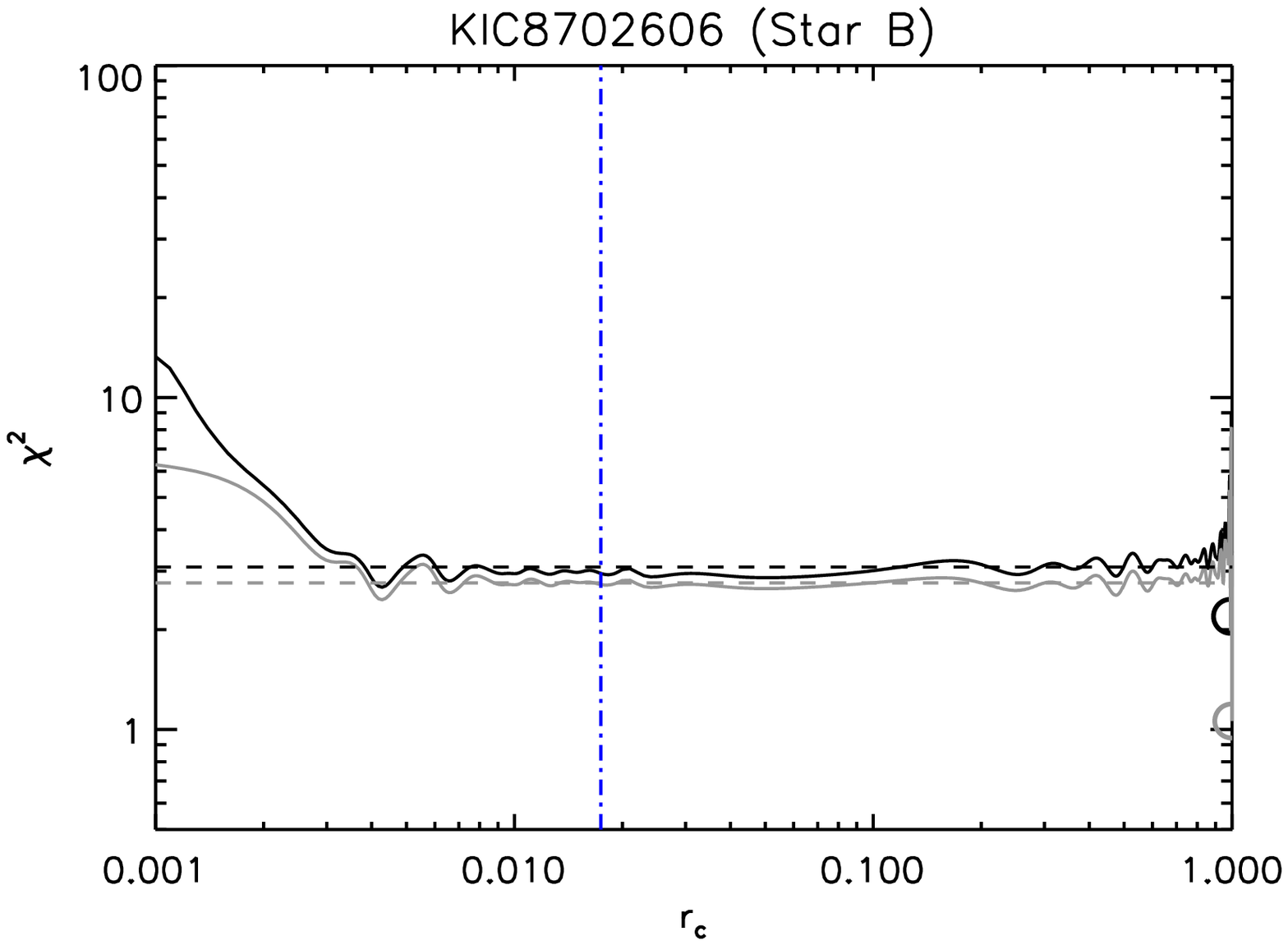}
\includegraphics[width=6cm]{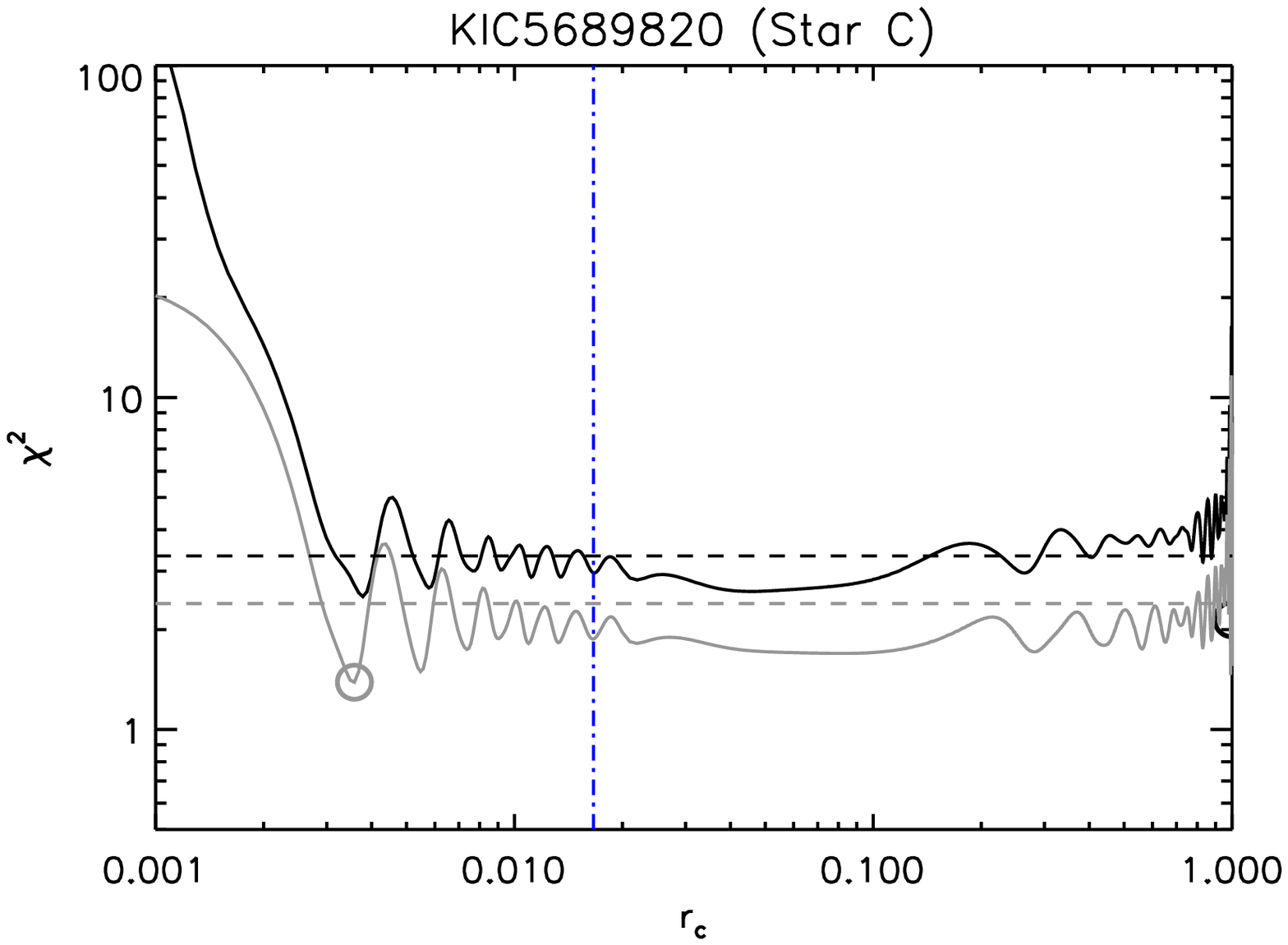}
\includegraphics[width=6cm]{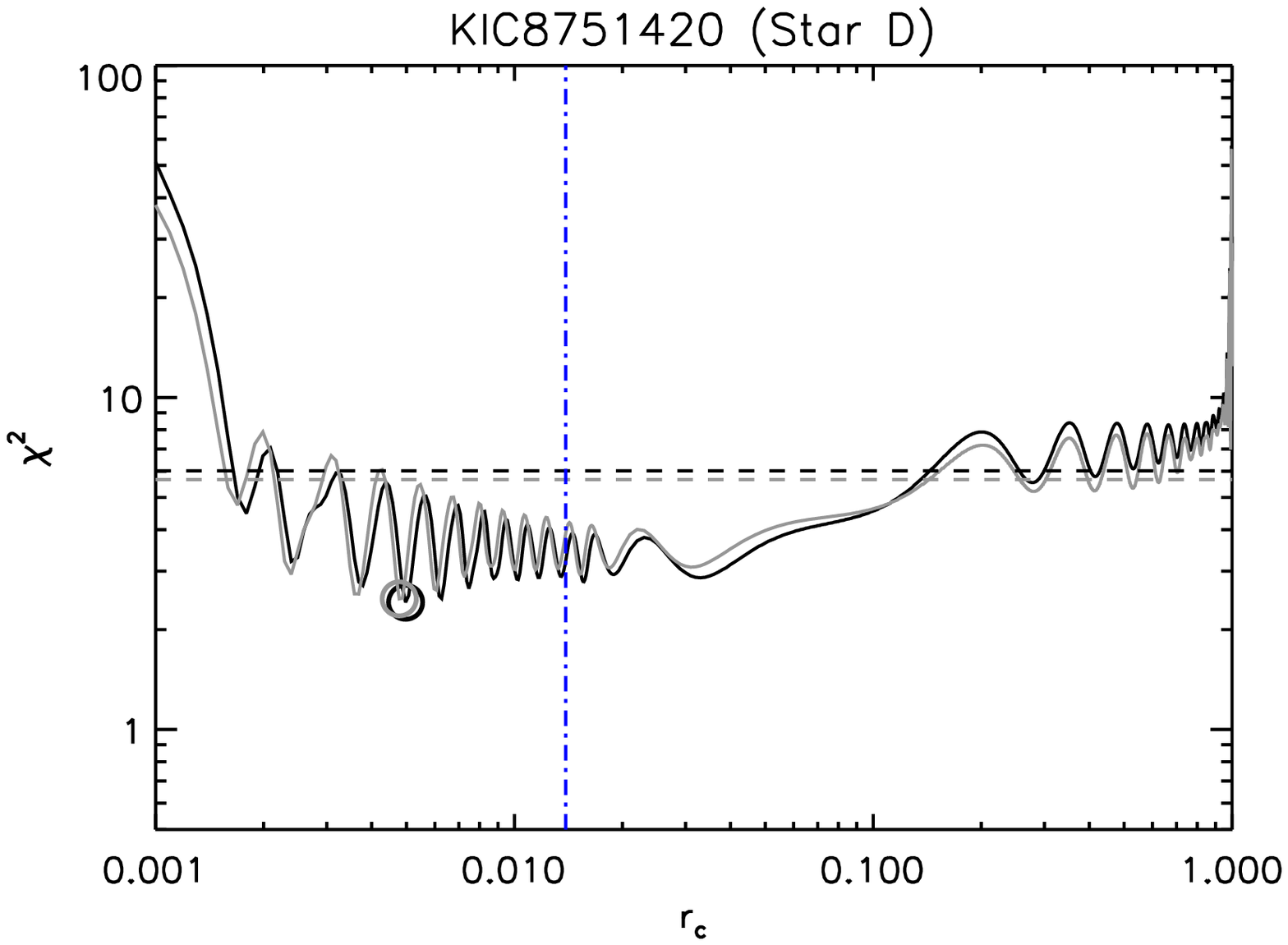}
\includegraphics[width=6cm]{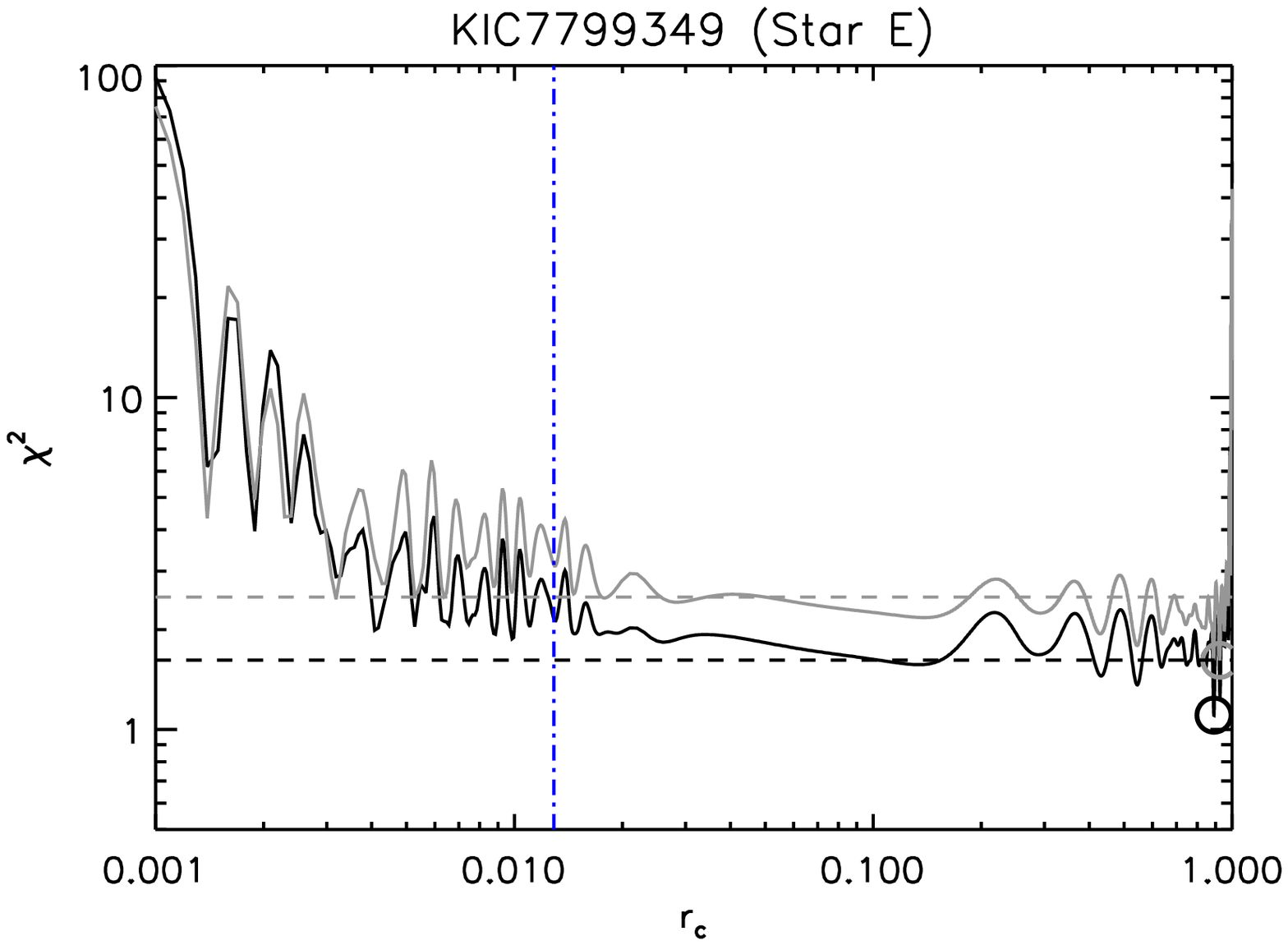}
\includegraphics[width=6cm]{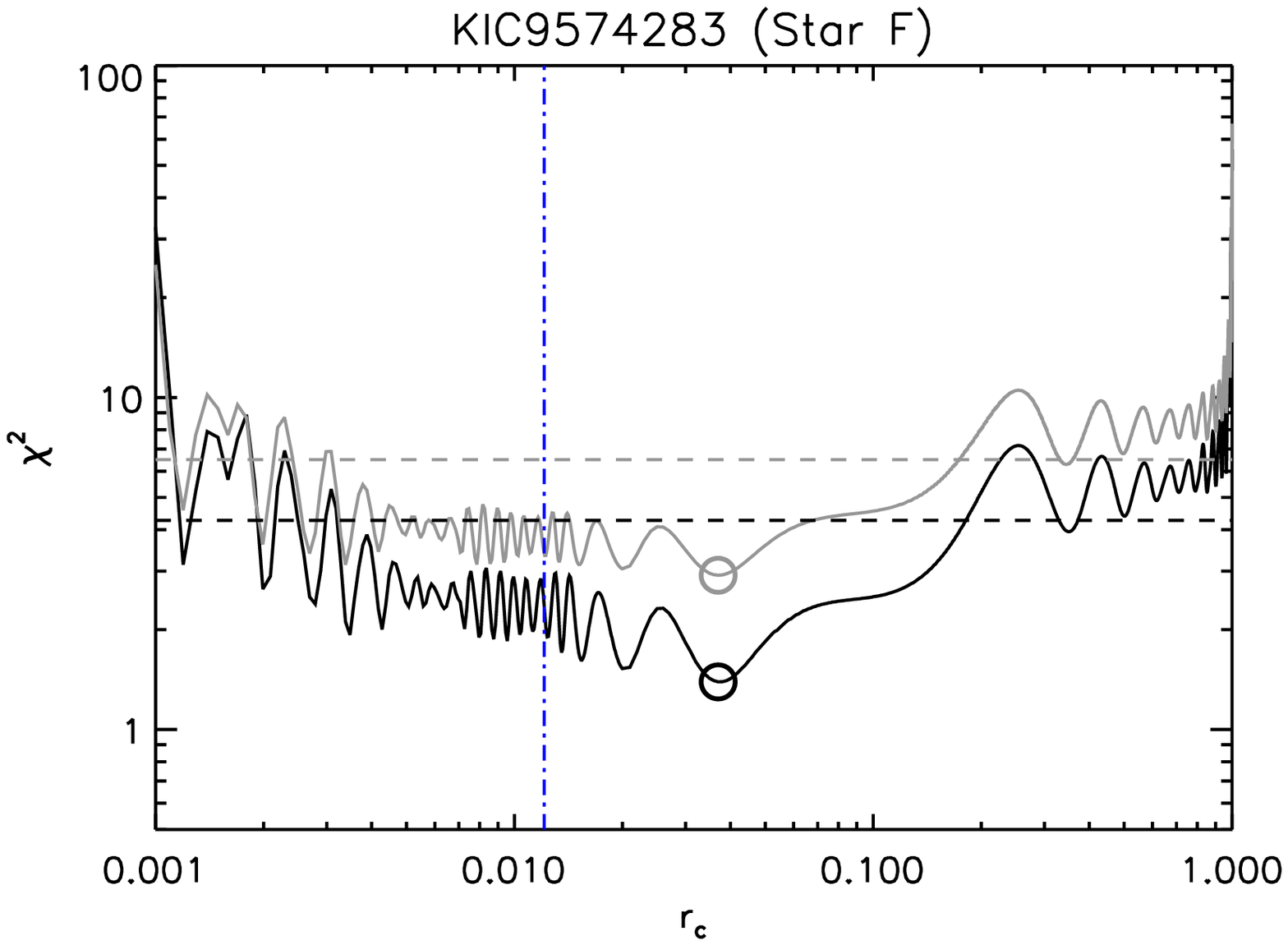}
\end{center}
\caption{Variations in the reduced $\chi^2$ as a function of the radius $r\ind{c}$ of the assumed discontinuity. The black curves correspond to the inversions performed with the optimal models of \cesam\ and the grey curves, to the inversions performed with the optimal models of \astec. The circles indicate the minimum value of the $\chi^2$ function. The blue vertical dot-dashed line indicates the location of the H-burning shell for the optimal models.
\label{fig_chi2_rc}}
\end{figure*}



\begin{table}
\caption{Values of the reduced $\chi^2$ obtained for the optimal smooth rotation profiles (Sect. \ref{sect_smooth}) and for the optimal discontinuous rotation profiles (Sect. \ref{sect_disc}).}
\begin{center}
\begin{tabular}{l c c c c c }
\hline \hline
\T \B Star & \multicolumn{2}{c}{Smooth} & & \multicolumn{2}{c}{Discontinuous} \\ 
\cline{2-3}
\cline{5-6}
\T \B & $\chi^2\ind{smooth}$ & $\chi^2\ind{SBenv}$ & & $\chi^2\ind{disc}$ & $r\ind{min}/R_\star$  \\
\hline
\T A (KIC12508433) & 1.0 & 1.0 & & 0.7 & 0.991   \\
B (KIC8702606)  & 3.1 & 3.1 & & 2.2 & 0.990  \\
C (KIC5689820) & 3.3 & 3.2 & & 2.1 & 0.998   \\
D (KIC8751420)  & 6.0 & 5.9 & & 2.4 & 0.005 \\
E (KIC7799349)  & 1.6 & 1.7 & & 1.1 & 0.889  \\
\B F (KIC9574283)  & 4.4 & 4.4 & & 1.4 & 0.037  \\
\hline
\end{tabular}
\\
\end{center}
\label{tab_chi2}
\end{table}%

\subsubsection{Discontinuous profiles \label{sect_disc}}

We then tried to fit discontinuous rotation profiles to the data. We here modeled discontinuous profiles in the crudest way, by splitting the star into two zones that are assumed to rotate as solid bodies, i.e. profiles of the type
\begin{eqnarray}
\Omega(r) = \Omega_1 &  \hbox{if} \; r\leqslant r\ind{c}, \;\; \hbox{and}\nonumber \\
\Omega(r) = \Omega_2 &  \hbox{if} \; r> r\ind{c}.
\end{eqnarray}
The radius $r\ind{c}$ that separates these zones is considered as a free parameter. We varied the radius $r\ind{c}$ between 0 and 1 and each time, we determined the rotation rates $\Omega_1$ and $\Omega_2$ that fit the observations at closest. We thus minimized the reduced $\chi^2$ defined as
\begin{linenomath*}
\begin{equation}
\chi^2\ind{red} = \frac{1}{M-2} \sum_{k=1}^M \left[\frac{\delta\nu_k\ex{obs}-\Omega_1A_k(r\ind{c})-\Omega_2 B_k(r\ind{c})}{\sigma_k\ex{obs}}\right]^2,
\label{eq_chi2_2zone}
\end{equation}
\end{linenomath*}
where $A_k(r\ind{c})\equiv\int_0^{r\ind{c}} K_k(r)\,\hbox{d}r$ and $B_k(r\ind{c})\equiv\int_{r\ind{c}}^R K_k(r)\,\hbox{d}r$. Here, there are two fitted parameters and the $\chi^2$ is thus normalized by $M-2$. In the following, the minimum value of the $\chi^2$ function is noted $\chi^2\ind{disc}$ and the corresponding depth of discontinuity, $r\ind{disc}$. The results are presented in Fig. \ref{fig_chi2_rc}. For stars A, B, C, and E, the $\chi^2$ function varies little as a function of $r\ind{c}$, except at very low values of $r\ind{c}$ ($<0.002\, R_\star$) where it sharply increases. The minimum $\chi^2$ is located very close to the surface for stars A, B, and C ($r\ind{disc}\geqslant0.99\,R_\star$), and it corresponds to a rotation profile with a thin counter-rotating layer at the surface. D12 have shown that these solutions are in fact most likely spurious, and this will be further discussed in Sect. \ref{sect_inv_simu}. For stars D and F, the minimum of the $\chi^2$ function is much more pronounced than for the other stars. For these two stars, discontinuities below a depth of about 0.04 $R_\star$ yield $\chi^2$ values that are much lower than discontinuities in other regions of the star. The global minimum of the $\chi^2$ function is located at $r\ind{disc} = 0.005\,R_\star$ for star D, and $r\ind{disc} = 0.037\,R_\star$ for star F, but other depths of discontinuities below 0.04 $R_\star$ also yield very similar values of $\chi^2$. 
The values of $\chi^2\ind{disc}$ are given in Table \ref{tab_chi2} for each star. It is striking that for stars D and F, the optimal discontinuous rotation profiles yield reduced $\chi^2$ of 2.4 and 1.4, respectively, which are much smaller than the reduced $\chi^2$ obtained with smooth rotation profiles ($\chi^2\ind{smooth}=6.0$ and 4.4 for stars D and F, see Table \ref{tab_chi2}). The question of the significance of this $\chi^2$ difference is addressed in \ref{sect_significance}. For this purpose, we introduce the quantity $\Delta\chi^2\equiv\chi^2\ind{smooth}-\chi^2\ind{disc}$. We thus have $\Delta\chi^2=3.6$ for star D and $\Delta\chi^2=3.1$ for star F. 

We observe that regardless of the chosen depth of discontinuity $r\ind{c}$, the optimal values of $\Omega_1$ and $\Omega_2$ are always in good agreement with the mean rotation rate in the g-mode cavity that was obtained with the OLA method in Sect. \ref{sect_OLA_core}. We checked this by computing $\Omega_1 \int_0^{r\ind{c}} \mathcal{K}(r;0)\,dr + \Omega_2 \int_{r\ind{c}}^{R_\star} \mathcal{K}(r;0)\,dr$ and comparing the result with the values of $\Omega\ind{g}$ listed in Table \ref{tab_om_OLA}. It is thus understandable that for depths of discontinuity smaller than the outer turning point of the g-mode cavity, the optimal value of $\Omega_1$ increases and takes values that can become much larger than $\Omega\ind{g}$. For instance for star D, the optimal depth of discontinuity $r\ind{disc} = 0.005\,R_\star$ corresponds to $\Omega_1=5254\pm92$ nHz. If the rotation profile is indeed discontinuous in stars D and F, it is therefore possible that the rotation rate in the deeper core could be significantly larger than the average rotation rate in the g-mode cavity given in Table \ref{tab_om_OLA}.

To study the influence of the choice of the optimal stellar model on the results, we also performed the same inversions as described above using the best model found with the \astec\ code instead of the \cesam\ models. The $\chi^2$ that were obtained are plotted as a function of the $r\ind{c}$ in Fig. \ref{fig_chi2_rc} (gray lines). Apart from the absolute values of $\chi^2$, which vary a little compared to the case where \cesam\ models are used, the results are very similar. In particular, the radii $r\ind{disc}$ where the $\chi^2$ functions are minimal are almost unchanged. This justifies that the results on the internal rotation of the selected stars do not critically depend on the choice of the optimal stellar model.


\subsection{Significance of the results \label{sect_significance}}

The results obtained in Sect. \ref{sect_test_disc} suggest that there might exist a deep discontinuity in the rotation profiles of stars D and F. To evaluate the significance of these results, we ran a series of simulations with artificial input rotation profiles to determine whether or not we can distinguish between a smooth and a discontinuous profile with a deep discontinuity for the six stars of the sample. These tests are presented in Appendix \ref{sect_inv_simu} and we here summarize the obtained results. 

We found that stars D and F only offer good chances of distinguishing between the two cases for the following reasons:
\begin{enumerate}
\item If we assume a discontinuous input profile with a deep discontinuity (at $0.04\,R_\star$), the probability of correctly recovering this depth of discontinuity is high (90\% for star D and 97\% for star F).
\item If we assume a smooth input profile, the probability of mistaking it with a discontinuous profile with a deep discontinuity is low (4\% for star D and 2\% for star F).
\item Assuming a smooth or discontinuous input profile yields distributions of $\Delta\chi^2$ that are clearly different from each other (see Fig. \ref{fig_histo_delta_chi2}).
\end{enumerate}
By comparison, for stars A, B, C, and E, the chances of correctly recovering a deep discontinuity are relatively low (between 28\% and 60\%), and the distributions of $\Delta\chi^2$ whether we assume a smooth or discontinuous input profile are almost overlapping (see Fig. \ref{fig_histo_delta_chi2}). This shows that the splittings that are available do not allow us to significantly distinguish between a smooth and a discontinuous rotation profile for these stars.

We now use the results of our simulations to interpret the observations. In Sect. \ref{sect_disc}, we found that a deep discontinuity yields the smallest reduced $\chi^2$ for stars D and F, which has little chance of occurring if the rotation profile is smooth, as we showed in Appendix \ref{sect_inv_simu}. Also, we obtained $\Delta\chi^2=3.6$ for star D and $\Delta\chi^2=3.1$ for star F. These values are much more compatible with the case of a discontinuous profile than with the case of a smooth profile (Fig. \ref{fig_histo_delta_chi2}). We thus conclude that the improvement in the $\chi^2$ when considering a discontinuous rotation profile is statistically significant. We note that even with discontinuous profiles, the reduced $\chi^2$ that we obtained for stars D and F is higher than 1. This is very likely because the actual rotation profile of the star is more complex than the simple two-zone model that we have assumed. However, for completeness, we also considered the possibility that it might come from an underestimation of the error bars of the observed splittings. We computed a correcting factor to the errors that would bring the reduced $\chi^2$ to 1 and we repeated the inversions performed in Sect. \ref{sect_test_disc}. We found that a discontinuous rotation profile still reproduces the observed splittings significantly better than a smooth profile for both stars.

For stars A, B, C, and E, no evidence for a discontinuous rotation profile was found in the observations. However, this does not mean that the profile is smooth because our simulations showed that it is impossible to discriminate between the two types of profiles. This raises the question of why stars D and F are more favorable to the detection of discontinuities in the rotation profiles. Answering this question is out of the scope of the present paper; however several possible explanations can already be put forward. First of all, stars D and F are the ones that have the fastest core rotation (see Table \ref{tab_om_OLA}). By performing statistical tests similar to those presented in Appendix \ref{sect_inv_simu} for input profiles with increasing core rotations, we indeed found that the probability of detecting a discontinuity increases with the core rotation. Yet, this does not explain why discontinuities cannot be detected for star E, which has a core rotation that is comparable to those of stars D and F. This could come from the precision of the splitting estimates. Indeed, we saw from Fig. \ref{fig_diff_split_875} that the splittings of several modes are crucial to distinguish between smooth and discontinuous profiles. The precision with which the splittings of these modes can be determined from the observations is therefore decisive. 




\section{Discussion and conclusion \label{sect_discussion}}

We selected a subsample of six subgiants or early red giants observed with \kepler\ with the objective to obtain constraints on the radial dependence of their rotation profile. For this purpose, spectroscopic estimates of their surface parameters were obtained, either from the literature of by performing ground observations. The \kepler\ lightcurves of the six stars were analyzed, enabling us to determine the frequencies and rotational splittings of 12 to 18 mixed modes of degree $l=1$ or 2 with a very high level of precision (uncertainties of the order of 10 nHz). We then performed a seismic modeling of the 6 targets and obtained stellar models that reproduce well both the observed atmospheric parameters and the frequencies of the observed modes. By using these models along with our estimates of the rotational splittings, we performed inversions to probe the rotation profiles of the selected targets.

\begin{figure}
\begin{center}
\includegraphics[width=9cm]{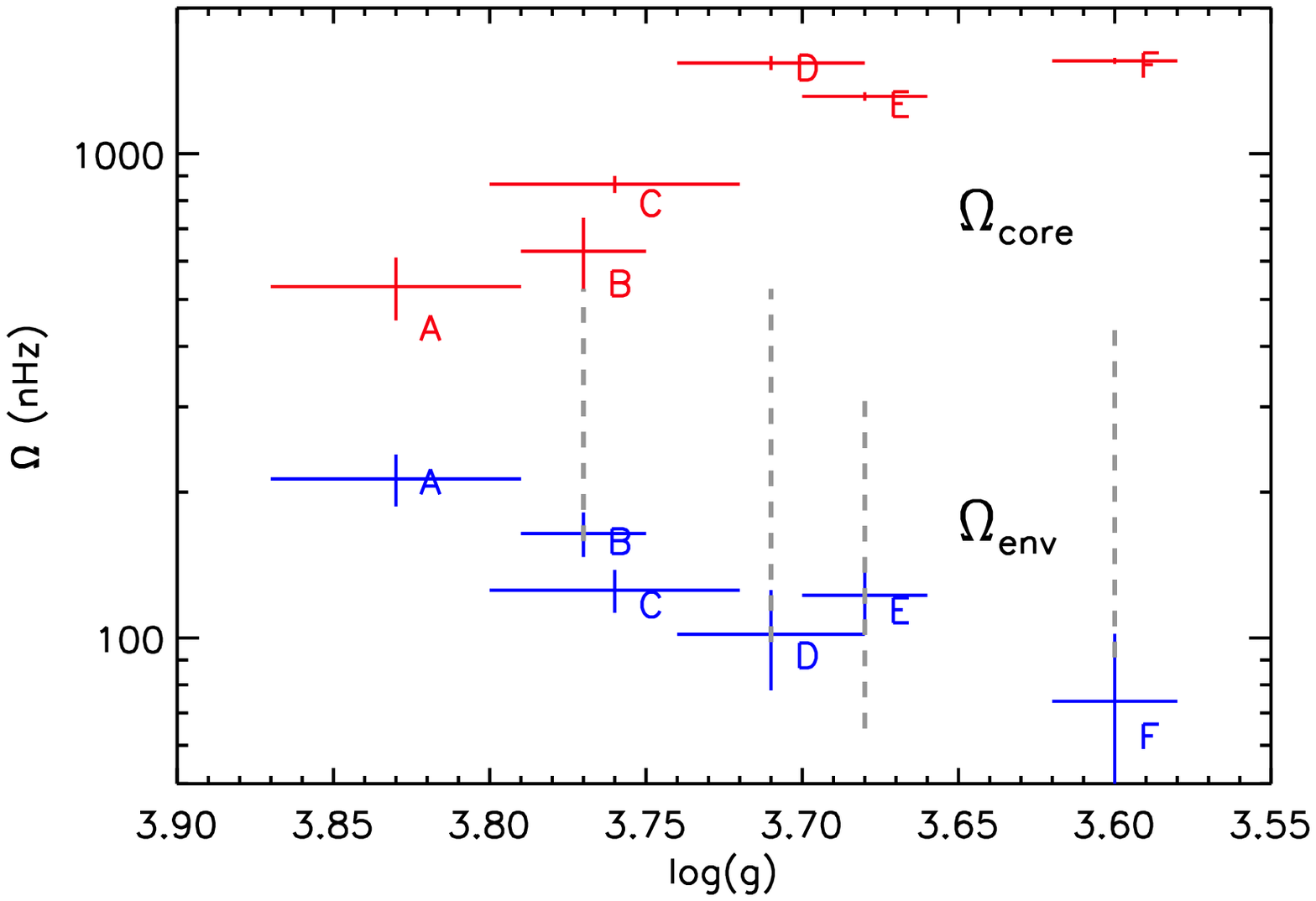}
\end{center}
\caption{Core (red symbols) and envelope (blue symbols) rotation rates obtained with the OLA method (see Sect. \ref{sect_OLA_core} and \ref{sect_OLA_env}) plotted as a function of the surface gravity. The letter corresponding to each star is specified. The gray dashed lines correspond to the range of surface rotation rates predicted by \cite{vansaders13} for the stars that lie in the range of parameters that they considered. The horizontal and vertical lines indicate 1-$\sigma$ error-bars. 
\label{fig_om_logg}}
\end{figure}

\begin{figure}
\begin{center}
\includegraphics[width=9cm]{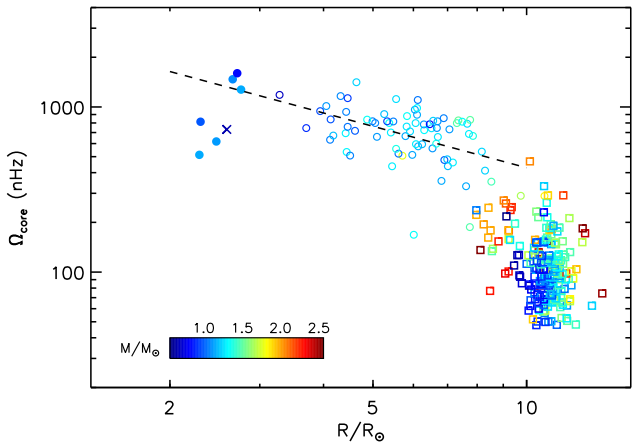}
\end{center}
\caption{Core rotation rate as a function of the stellar radius. The open symbols correspond to the stars that were studied by \cite{mosser12b} (circles: RGB stars, squares: clump stars). The filled symbols indicate the stars that were studied in this paper and the cross corresponds to the young giant KIC7341231 studied by D12.
\label{fig_pcore_radius}}
\end{figure}

By using the OLA (Optimally Localized Average) method, we were able to obtain estimates of the average rotation in the g-mode cavity (which corresponds roughly to the innermost 2\% of the stellar radius) for the six stars of our sample (Sect. \ref{sect_OLA_core}). 
It is interesting to remark that the mean core rotation rate appears to be correlated with the evolutionary status. The two stars that are the least evolved and were identified as subgiants on their way to the RGB (stars A and B) are the ones that have the slowest cores. To further illustrate this point, we plotted in Fig. \ref{fig_om_logg}, the estimated core rotation rates as a function of the surface gravities that were obtained from seismic global parameters in Sect. \ref{sect_selection}. There is a clear trend, which suggests that the core spins up as the star evolves. At first glance, this result seems at odds with the conclusions of \cite{mosser12b}, who reached the conclusion that the core of red giant stars spins down as they climb the RGB. However, the authors studied stars that are more evolved than the ones in our sample. Fig. \ref{fig_pcore_radius} reproduces Fig. 9 of \cite{mosser12b}, where we added the core rotation rates of the stars that were studied here (filled circles in the plot), as well as the rotation rate of KIC7341231 obtained by D12 (cross). Our results suggest that the core of subgiant stars spins up until the base of the RGB and subsequently spins down due to an efficient transport of AM from the core to the envelope whose origin is still unknown. This suggests that during the subgiant phase, the AM transport from the core to the envelope is not efficient enough to counterbalance the core contraction, which results in a spin-up of the core in this phase. This result, if confirmed, can be used to place constraints on the mechanisms of AM transport that operate in this phase. The confirmation of this result will require the measurement of the core rotation for more subgiant stars, which is difficult since \kepler\ observed less of these targets\footnote{Subgiants are intrinsically less bright than red giants. Besides, to perform a seismic study of these stars, short-cadence data are required, which limits the number of targets.} and besides their modes are wider, making it more difficult to estimate the rotational splittings.

We were also able to build averaging kernels that almost erase the contribution from the core, and thus obtained estimates of the rotation rate in the convective envelope for the stars of the sample (see Sect. \ref{sect_OLA_env}). We showed that except for star F, these estimates are nearly insensitive to the core rotation. However, if the radiative layers below the envelope spin much faster than the envelope, they could lead us to overestimate the envelope rotation rate. These estimates must therefore be regarded as upper limits of the envelope rotation rate. They are plotted as a function of the surface gravity in Fig. \ref{fig_om_logg}. Once again, a clear trend appears. It suggests that the convective envelope is spun down in the subgiant phase. This is hardly a surprise since the envelope rapidly expends in this phase. However, we know that AM must be redistributed from the core to the envelope to account for the relatively small core rotation rates that are observed (\citealt{eggenberger12}, \citealt{marques13}, \citealt{ceillier13}). Our results show that the AM that is gained by the envelope is not sufficient to spin it up. Our estimates of the envelope rotation rates can therefore be used to place constraints on the transport of AM in the subgiant phase. \cite{vansaders13} recently built a simplified model to predict the rotation periods of stars across the HR diagram by taking into account the loss of AM through a magnetized wind when relevant. They assumed rigid rotation at all times (i.e. instantaneous internal transport of AM), which is clearly not the case for the stars of our study, but they argued that their results are weakly sensitive on this hypothesis. Fig. \ref{fig_om_logg} shows the ranges of surface rotation rates predicted by \cite{vansaders13} for the stars of our sample that lie in the range of parameters that they considered (stars B, D, E, and F). They are in broad agreement with the estimates found in this study, but systematically larger, which is consistent with  the fact that internal transport of AM is likely less efficient than they assumed. It would be interesting to compare the observed envelope rotation rates to those predicted by models that include an additional viscosity to simulate the unknown physical process of AM transport (\citealt{eggenberger12}).

Finally, we found indications that the rotation profiles of two of the stars that we studied (stars D and F) might have a sharp gradient in the core. For these stars, the observed splittings are significantly better reproduced by a discontinuous rotation profile with a deep discontinuity (located at 0.4\% of the stellar radius for star D and 3.9\% of the radius for star F) than by a rotation profile that decreases smoothly from the core to the surface. The significance of this result was established by performing simulations with our optimal models and by assuming either discontinuous or smooth input rotation profiles. For the four other stars, no observational evidence of a discontinuous rotation profile was found. However, our statistical tests showed that with the current data, it is in fact impossible to distinguish between a smooth and a discontinuous rotation profile for these stars, which means that discontinuities could exist. Interestingly, the depth that was found most probable for the discontinuity in the rotation profiles of stars D and F is close to the location of the H-burning shell, which is located between 1\% and 2\% of the stellar radius for these stars (see Fig. \ref{fig_chi2_rc}). This suggests that the overall shape of the rotation profile might in fact be similar to the one that is predicted by current theoretical models that include rotationally-induced AM transport (e.g. \citealt{marques13}). Of course there remains the issue that the core rotation rates that these models predict are larger than the observed ones by several orders of magnitude. 

\begin{figure}
\begin{center}
\includegraphics[width=9cm]{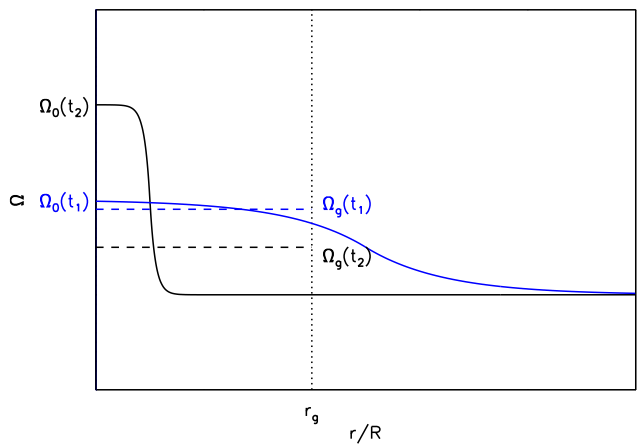}
\end{center}
\caption{Schematic rotation profiles for a star in which a rotation gradient is building up in the g-mode cavity (delimited by the vertical dotted line) between times $t_1$ (black curves) and $t_2$ (blue curves). The dashed lines indicate the mean rotation rate in the g-mode cavity $\Omega\ind{g}$ and the rotation rate in the center is denote $\Omega_0$.
\label{fig_schema_grad}}
\end{figure}

The existence of a sharp gradient within the g-mode cavity of young giants would complicate the interpretation of the average rotation velocity in the g-mode cavity $\Omega\ind{g}$. Indeed, if it is the case, the layers below this jump could have rotation rates that are much larger than our estimates of $\Omega\ind{g}$, which are listed in Table \ref{tab_om_OLA}. 
 As a result, any conclusion on the spin-up or spin-down of the most central layers depends on the evolution of the rotation gradient in the g-mode cavity. Fig. \ref{fig_schema_grad} illustrates this point schematically. We assumed rotation profiles for an imaginary star in which a rotation gradient would be building up between times $t_1$ and $t_2$. If we only have access to $\Omega\ind{g}$, we would conclude that the core is spinning down because $\Omega\ind{g}(t_2)<\Omega\ind{g}(t_1)$. But in fact, Fig. \ref{fig_schema_grad} shows that the core spins up ($\Omega_0(t_2)>\Omega_0(t_1)$) owing to the sharp gradient that is building up in the core.

Also, if  the existence of sharp gradients in the rotation profiles of early red giants can be confirmed, it would bring very important constraints on the mechanisms that transport AM in these stars. In particular, it seems incompatible with AM transport through a deep fossil magnetic field, as advocated e.g. by \cite{gough98}, because it would likely erase such sharp features. Indeed, differential rotation is expected to be damped along the poloidal filed lines (\citealt{garaud08}, \citealt{stugarek11}). On the other hand, internal gravity waves, which are also able to transport AM and have been reported to be efficient during advanced phases of stellar evolution (\citealt{talon08}), can give birth to localized weak gradients in the rotation profile caused by the extraction and deposit of AM (e.g. \citealt{talon05}). Sharp rotation gradients could also potentially trigger magnetohydrodynamic instabilities, such as the magnetorotational instability (MRI) that could in turn transport AM (\citealt{balbus94}, \citealt{arlt03}, \citealt{menou04}, \citealt{menou06}). 

To conclude, we insist that the precision with which rotational splittings can be determined from the observations is crucial to study the shape of the internal rotation profiles of young red giants. This is yet another motivation to try to obtain long time series for asteroseismic measurements, such as those obtained with \kepler.

\begin{acknowledgements}
S.D. thanks W. Ball for his comments on this paper and acknowledges support from the PNPS under the grant "Rotation interne et magn\'etisme des sous-g\'eantes et g\'eantes Kepler".
Funding for the Stellar Astrophysics Centre is provided by The Danish National Research Foundation (Grant DNRF106). The research is supported by the ASTERISK project (ASTERoseismic Investigations with SONG and Kepler) funded by the European Research Council (Grant agreement no.: 267864). 
A.O.T acknowledges support from Sonderforschungsbereich SFB 881 "The Milky Way System" (subproject A5) of the German Research Foundation (DFG). 
J.M.-\.Z acknowledges the Polish MNiSW grant N N203 405139. 
L.G. and J.S. acknowledge research funding by the Deutsche Forschungsgemeinschaft (DFG) under grant SFB 963/1 "Astrophysical flow instabilities and turbulence" (Project A18). 
T.L.C., G.R.D. and W.J.C. acknowledge financial support from the UK Science and Technology Facilities Council. 

\end{acknowledgements}

\bibliographystyle{aa.bst} 
\bibliography{biblio} 

\begin{thebibliography}{102}
\expandafter\ifx\csname natexlab\endcsname\relax\def\natexlab#1{#1}\fi

\bibitem[{{Aerts} {et~al.}(2010){Aerts}, {Christensen-Dalsgaard}, \&
  {Kurtz}}]{asteroseismology}
{Aerts}, C., {Christensen-Dalsgaard}, J., \& {Kurtz}, D.~W. 2010,
  {Asteroseismology}

\bibitem[{{Aigrain} {et~al.}(2004){Aigrain}, {Favata}, \&
  {Gilmore}}]{aigrain04}
{Aigrain}, S., {Favata}, F., \& {Gilmore}, G. 2004, in ESA Special Publication,
  Vol. 538, Stellar Structure and Habitable Planet Finding, ed. F.~{Favata},
  S.~{Aigrain}, \& A.~{Wilson}, 215--224

\bibitem[{{Aizenman} {et~al.}(1977){Aizenman}, {Smeyers}, \&
  {Weigert}}]{aizenman77}
{Aizenman}, M., {Smeyers}, P., \& {Weigert}, A. 1977, \aap, 58, 41

\bibitem[{{An} {et~al.}(2009){An}, {Pinsonneault}, {Masseron}, {Delahaye},
  {Johnson}, {Terndrup}, {Beers}, {Ivans}, \& {Ivezi{\'c}}}]{An09}
{An}, D., {Pinsonneault}, M.~H., {Masseron}, T., {et~al.} 2009, \apj, 700, 523

\bibitem[{{Anderson} {et~al.}(1990){Anderson}, {Duvall}, \&
  {Jefferies}}]{anderson90}
{Anderson}, E.~R., {Duvall}, Jr., T.~L., \& {Jefferies}, S.~M. 1990, \apj, 364,
  699

\bibitem[{{Angulo} {et~al.}(1999){Angulo}, {Arnould}, {Rayet}, {Descouvemont},
  {Baye}, {Leclercq-Willain}, {Coc}, {Barhoumi}, {Aguer}, {Rolfs}, {Kunz},
  {Hammer}, {Mayer}, {Paradellis}, {Kossionides}, {Chronidou}, {Spyrou},
  {degl'Innocenti}, {Fiorentini}, {Ricci}, {Zavatarelli}, {Providencia},
  {Wolters}, {Soares}, {Grama}, {Rahighi}, {Shotter}, \& {Lamehi
  Rachti}}]{angulo99}
{Angulo}, C., {Arnould}, M., {Rayet}, M., {et~al.} 1999, Nuclear Physics A,
  656, 3

\bibitem[{{Appourchaux} {et~al.}(2012){Appourchaux}, {Benomar}, {Gruberbauer},
  {Chaplin}, {Garc{\'{\i}}a}, {Handberg}, {Verner}, {Antia}, {Campante},
  {Davies}, {Deheuvels}, {Hekker}, {Howe}, {Salabert}, {Bedding}, {White},
  {Houdek}, {Silva Aguirre}, {Elsworth}, {van Cleve}, {Clarke}, {Hall}, \&
  {Kjeldsen}}]{appourchaux12}
{Appourchaux}, T., {Benomar}, O., {Gruberbauer}, M., {et~al.} 2012, \aap, 537,
  A134

\bibitem[{{Appourchaux} {et~al.}(2008){Appourchaux}, {Michel}, {Auvergne},
  {Baglin}, {Toutain}, {Baudin}, {Benomar}, {Chaplin}, {Deheuvels}, {Samadi},
  {Verner}, {Boumier}, {Garc{\'{\i}}a}, {Mosser}, {Hulot}, {Ballot}, {Barban},
  {Elsworth}, {Jim{\'e}nez-Reyes}, {Kjeldsen}, {R{\'e}gulo}, \&
  {Roxburgh}}]{appourchaux08}
{Appourchaux}, T., {Michel}, E., {Auvergne}, M., {et~al.} 2008, \aap, 488, 705

\bibitem[{{Arlt} {et~al.}(2003){Arlt}, {Hollerbach}, \& {R{\"u}diger}}]{arlt03}
{Arlt}, R., {Hollerbach}, R., \& {R{\"u}diger}, G. 2003, \aap, 401, 1087

\bibitem[{{Backus} \& {Gilbert}(1968)}]{backus68}
{Backus}, G. \& {Gilbert}, F. 1968, Geophysical Journal International, 16, 169

\bibitem[{{Baglin} {et~al.}(2006){Baglin}, {Auvergne}, {Boisnard}, {Lam-Trong},
  {Barge}, {Catala}, {Deleuil}, {Michel}, \& {Weiss}}]{baglin06b}
{Baglin}, A., {Auvergne}, M., {Boisnard}, L., {et~al.} 2006, in COSPAR, Plenary
  Meeting, Vol.~36, 36th COSPAR Scientific Assembly, 3749

\bibitem[{{Bai} \& {Sturrock}(1993)}]{bai93}
{Bai}, T. \& {Sturrock}, P.~A. 1993, \apj, 409, 476

\bibitem[{{Balbus} \& {Hawley}(1994)}]{balbus94}
{Balbus}, S.~A. \& {Hawley}, J.~F. 1994, \mnras, 266, 769

\bibitem[{{Ballot} {et~al.}(2011){Ballot}, {Barban}, \& {van't
  Veer-Menneret}}]{ballot11}
{Ballot}, J., {Barban}, C., \& {van't Veer-Menneret}, C. 2011, \aap, 531, A124

\bibitem[{{Ballot} {et~al.}(2006){Ballot}, {Garc{\'{\i}}a}, \&
  {Lambert}}]{ballot06}
{Ballot}, J., {Garc{\'{\i}}a}, R.~A., \& {Lambert}, P. 2006, \mnras, 369, 1281

\bibitem[{{Beck} {et~al.}(2011){Beck}, {Bedding}, {Mosser}, {Stello}, {Garcia},
  {Kallinger}, {Hekker}, {Elsworth}, {Frandsen}, {Carrier}, {De Ridder},
  {Aerts}, {White}, {Huber}, {Dupret}, {Montalb{\'a}n}, {Miglio}, {Noels},
  {Chaplin}, {Kjeldsen}, {Christensen-Dalsgaard}, {Gilliland}, {Brown},
  {Kawaler}, {Mathur}, \& {Jenkins}}]{beck11}
{Beck}, P.~G., {Bedding}, T.~R., {Mosser}, B., {et~al.} 2011, Science, 332, 205

\bibitem[{{Beck} {et~al.}(2012){Beck}, {Montalban}, {Kallinger}, {De Ridder},
  {Aerts}, {Garc{\'{\i}}a}, {Hekker}, {Dupret}, {Mosser}, {Eggenberger},
  {Stello}, {Elsworth}, {Frandsen}, {Carrier}, {Hillen}, {Gruberbauer},
  {Christensen-Dalsgaard}, {Miglio}, {Valentini}, {Bedding}, {Kjeldsen},
  {Girouard}, {Hall}, \& {Ibrahim}}]{beck12}
{Beck}, P.~G., {Montalban}, J., {Kallinger}, T., {et~al.} 2012, \nat, 481, 55

\bibitem[{{Bedding} {et~al.}(2011){Bedding}, {Mosser}, {Huber},
  {Montalb{\'a}n}, {Beck}, {Christensen-Dalsgaard}, {Elsworth},
  {Garc{\'{\i}}a}, {Miglio}, {Stello}, {White}, {De Ridder}, {Hekker}, {Aerts},
  {Barban}, {Belkacem}, {Broomhall}, {Brown}, {Buzasi}, {Carrier}, {Chaplin},
  {di Mauro}, {Dupret}, {Frandsen}, {Gilliland}, {Goupil}, {Jenkins},
  {Kallinger}, {Kawaler}, {Kjeldsen}, {Mathur}, {Noels}, {Aguirre}, \&
  {Ventura}}]{bedding11}
{Bedding}, T.~R., {Mosser}, B., {Huber}, D., {et~al.} 2011, \nat, 471, 608

\bibitem[{{Belkacem} {et~al.}(2011){Belkacem}, {Goupil}, {Dupret}, {Samadi},
  {Baudin}, {Noels}, \& {Mosser}}]{belkacem11}
{Belkacem}, K., {Goupil}, M.~J., {Dupret}, M.~A., {et~al.} 2011, \aap, 530,
  A142

\bibitem[{{Benomar} {et~al.}(2009){Benomar}, {Appourchaux}, \&
  {Baudin}}]{benomar09a}
{Benomar}, O., {Appourchaux}, T., \& {Baudin}, F. 2009, \aap, 506, 15

\bibitem[{{Benomar} {et~al.}(2013){Benomar}, {Bedding}, {Mosser}, {Stello},
  {Belkacem}, {Garcia}, {White}, {Kuehn}, {Deheuvels}, \&
  {Christensen-Dalsgaard}}]{benomar13}
{Benomar}, O., {Bedding}, T.~R., {Mosser}, B., {et~al.} 2013, \apj, 767, 158

\bibitem[{{Benomar} {et~al.}(2012){Benomar}, {Bedding}, {Stello}, {Deheuvels},
  {White}, \& {Christensen-Dalsgaard}}]{benomar12}
{Benomar}, O., {Bedding}, T.~R., {Stello}, D., {et~al.} 2012, \apjl, 745, L33

\bibitem[{{B{\"o}hm-Vitense}(1958)}]{bohm58}
{B{\"o}hm-Vitense}, E. 1958, \zap, 46, 108

\bibitem[{{Borucki} {et~al.}(2010){Borucki}, {Koch}, {Basri}, {Batalha},
  {Brown}, {Caldwell}, {Caldwell}, {Christensen-Dalsgaard}, {Cochran},
  {DeVore}, {Dunham}, {Dupree}, {Gautier}, {Geary}, {Gilliland}, {Gould},
  {Howell}, {Jenkins}, {Kondo}, {Latham}, {Marcy}, {Meibom}, {Kjeldsen},
  {Lissauer}, {Monet}, {Morrison}, {Sasselov}, {Tarter}, {Boss}, {Brownlee},
  {Owen}, {Buzasi}, {Charbonneau}, {Doyle}, {Fortney}, {Ford}, {Holman},
  {Seager}, {Steffen}, {Welsh}, {Rowe}, {Anderson}, {Buchhave}, {Ciardi},
  {Walkowicz}, {Sherry}, {Horch}, {Isaacson}, {Everett}, {Fischer}, {Torres},
  {Johnson}, {Endl}, {MacQueen}, {Bryson}, {Dotson}, {Haas}, {Kolodziejczak},
  {Van Cleve}, {Chandrasekaran}, {Twicken}, {Quintana}, {Clarke}, {Allen},
  {Li}, {Wu}, {Tenenbaum}, {Verner}, {Bruhweiler}, {Barnes}, \&
  {Prsa}}]{borucki10}
{Borucki}, W.~J., {Koch}, D., {Basri}, G., {et~al.} 2010, Science, 327, 977

\bibitem[{{Brown} {et~al.}(1991){Brown}, {Gilliland}, {Noyes}, \&
  {Ramsey}}]{brown91}
{Brown}, T.~M., {Gilliland}, R.~L., {Noyes}, R.~W., \& {Ramsey}, L.~W. 1991,
  \apj, 368, 599

\bibitem[{{Bruntt}(2009)}]{bruntt09}
{Bruntt}, H. 2009, \aap, 506, 235

\bibitem[{{Bruntt} {et~al.}(2012){Bruntt}, {Basu}, {Smalley}, {Chaplin},
  {Verner}, {Bedding}, {Catala}, {Gazzano}, {Molenda-{\.Z}akowicz}, {Thygesen},
  {Uytterhoeven}, {Hekker}, {Huber}, {Karoff}, {Mathur}, {Mosser},
  {Appourchaux}, {Campante}, {Elsworth}, {Garc{\'{\i}}a}, {Handberg},
  {Metcalfe}, {Quirion}, {R{\'e}gulo}, {Roxburgh}, {Stello},
  {Christensen-Dalsgaard}, {Kawaler}, {Kjeldsen}, {Morris}, {Quintana}, \&
  {Sanderfer}}]{bruntt12}
{Bruntt}, H., {Basu}, S., {Smalley}, B., {et~al.} 2012, \mnras, 423, 122

\bibitem[{{Campante} {et~al.}(2011){Campante}, {Handberg}, {Mathur},
  {Appourchaux}, {Bedding}, {Chaplin}, {Garc{\'{\i}}a}, {Mosser}, {Benomar},
  {Bonanno}, {Corsaro}, {Fletcher}, {Gaulme}, {Hekker}, {Karoff}, {R{\'e}gulo},
  {Salabert}, {Verner}, {White}, {Houdek}, {Brand{\~a}o}, {Creevey}, {Do{\v
  g}an}, {Bazot}, {Christensen-Dalsgaard}, {Cunha}, {Elsworth}, {Huber},
  {Kjeldsen}, {Lundkvist}, {Molenda-{\.Z}akowicz}, {Monteiro}, {Stello},
  {Clarke}, {Girouard}, \& {Hall}}]{campante11}
{Campante}, T.~L., {Handberg}, R., {Mathur}, S., {et~al.} 2011, \aap, 534, A6

\bibitem[{{Canuto} {et~al.}(1996){Canuto}, {Goldman}, \&
  {Mazzitelli}}]{canuto96}
{Canuto}, V.~M., {Goldman}, I., \& {Mazzitelli}, I. 1996, \apj, 473, 550

\bibitem[{{Casagrande} {et~al.}(2010){Casagrande}, {Ram{\'{\i}}rez},
  {Mel{\'e}ndez}, {Bessell}, \& {Asplund}}]{casagrande10}
{Casagrande}, L., {Ram{\'{\i}}rez}, I., {Mel{\'e}ndez}, J., {Bessell}, M., \&
  {Asplund}, M. 2010, \aap, 512, A54

\bibitem[{{Ceillier} {et~al.}(2013){Ceillier}, {Eggenberger}, {Garc{\'{\i}}a},
  \& {Mathis}}]{ceillier13}
{Ceillier}, T., {Eggenberger}, P., {Garc{\'{\i}}a}, R.~A., \& {Mathis}, S.
  2013, \aap, 555, A54

\bibitem[{{Chaplin}(2013)}]{chaplin13}
{Chaplin}, W.~J. 2013, \apj, in prep

\bibitem[{{Chaplin} {et~al.}(1999){Chaplin}, {Christensen-Dalsgaard},
  {Elsworth}, {Howe}, {Isaak}, {Larsen}, {New}, {Schou}, {Thompson}, \&
  {Tomczyk}}]{chaplin99}
{Chaplin}, W.~J., {Christensen-Dalsgaard}, J., {Elsworth}, Y., {et~al.} 1999,
  \mnras, 308, 405

\bibitem[{{Charbonnel} \& {Talon}(2005)}]{charbonnel05}
{Charbonnel}, C. \& {Talon}, S. 2005, Science, 309, 2189

\bibitem[{{Christensen-Dalsgaard}(2003)}]{lecturenotesJCD}
{Christensen-Dalsgaard}, J. 2003, {Lecture Notes on Stellar Oscillations}, 5th
  edn.

\bibitem[{{Christensen-Dalsgaard}(2008{\natexlab{a}})}]{astec}
{Christensen-Dalsgaard}, J. 2008{\natexlab{a}}, \apss, 316, 13

\bibitem[{{Christensen-Dalsgaard}(2008{\natexlab{b}})}]{adipls}
{Christensen-Dalsgaard}, J. 2008{\natexlab{b}}, \apss, 316, 113

\bibitem[{{Deheuvels} {et~al.}(2012){Deheuvels}, {Garc{\'{\i}}a}, {Chaplin},
  {Basu}, {Antia}, {Appourchaux}, {Benomar}, {Davies}, {Elsworth}, {Gizon},
  {Goupil}, {Reese}, {Regulo}, {Schou}, {Stahn}, {Casagrande},
  {Christensen-Dalsgaard}, {Fischer}, {Hekker}, {Kjeldsen}, {Mathur}, {Mosser},
  {Pinsonneault}, {Valenti}, {Christiansen}, {Kinemuchi}, \&
  {Mullally}}]{deheuvels12}
{Deheuvels}, S., {Garc{\'{\i}}a}, R.~A., {Chaplin}, W.~J., {et~al.} 2012, \apj,
  756, 19

\bibitem[{{Deheuvels} \& {Michel}(2010)}]{rome}
{Deheuvels}, S. \& {Michel}, E. 2010, \apss, 328, 259

\bibitem[{{Deheuvels} \& {Michel}(2011)}]{deheuvels11}
{Deheuvels}, S. \& {Michel}, E. 2011, \aap, 535, A91

\bibitem[{{Deheuvels} {et~al.}(2013){Deheuvels}, {Ouazzani}, \&
  {Basu}}]{deheuvels13}
{Deheuvels}, S., {Ouazzani}, R.~M., \& {Basu}, S. 2013, submitted to \aap

\bibitem[{{Denissenkov} {et~al.}(2010){Denissenkov}, {Pinsonneault},
  {Terndrup}, \& {Newsham}}]{denissenkov10}
{Denissenkov}, P.~A., {Pinsonneault}, M., {Terndrup}, D.~M., \& {Newsham}, G.
  2010, \apj, 716, 1269

\bibitem[{{Do{\u g}an} {et~al.}(2013){Do{\u g}an}, {Metcalfe}, {Deheuvels}, {Di
  Mauro}, {Eggenberger}, {Creevey}, {Monteiro}, {Pinsonneault}, {Frasca},
  {Karoff}, {Mathur}, {Sousa}, {Brand{\~a}o}, {Campante}, {Handberg},
  {Thygesen}, {Biazzo}, {Bruntt}, {Niemczura}, {Bedding}, {Chaplin},
  {Christensen-Dalsgaard}, {Garc{\'{\i}}a}, {Molenda-{\.Z}akowicz}, {Stello},
  {Van Saders}, {Kjeldsen}, {Still}, {Thompson}, \& {Van Cleve}}]{dogan13}
{Do{\u g}an}, G., {Metcalfe}, T.~S., {Deheuvels}, S., {et~al.} 2013, \apj, 763,
  49

\bibitem[{{Drimmel} {et~al.}(2003){Drimmel}, {Cabrera-Lavers}, \&
  {L{\'o}pez-Corredoira}}]{drimmel03}
{Drimmel}, R., {Cabrera-Lavers}, A., \& {L{\'o}pez-Corredoira}, M. 2003, \aap,
  409, 205

\bibitem[{{Dupret} {et~al.}(2009){Dupret}, {Belkacem}, {Samadi}, {Montalban},
  {Moreira}, {Miglio}, {Godart}, {Ventura}, {Ludwig}, {Grigahc{\`e}ne},
  {Goupil}, {Noels}, \& {Caffau}}]{dupret09}
{Dupret}, M.-A., {Belkacem}, K., {Samadi}, R., {et~al.} 2009, \aap, 506, 57

\bibitem[{{Eff-Darwich} \& {Korzennik}(2013)}]{effdarwich13}
{Eff-Darwich}, A. \& {Korzennik}, S.~G. 2013, \solphys, 287, 43

\bibitem[{{Eggenberger} {et~al.}(2012){Eggenberger}, {Montalb{\'a}n}, \&
  {Miglio}}]{eggenberger12}
{Eggenberger}, P., {Montalb{\'a}n}, J., \& {Miglio}, A. 2012, \aap, 544, L4

\bibitem[{{Frasca} {et~al.}(2003){Frasca}, {Alcal{\'a}}, {Covino}, {Catalano},
  {Marilli}, \& {Paladino}}]{frasca03}
{Frasca}, A., {Alcal{\'a}}, J.~M., {Covino}, E., {et~al.} 2003, \aap, 405, 149

\bibitem[{{Garaud} \& {Garaud}(2008)}]{garaud08}
{Garaud}, P. \& {Garaud}, J.-D. 2008, \mnras, 391, 1239

\bibitem[{{Garc{\'{\i}}a} {et~al.}(2004){Garc{\'{\i}}a}, {Corbard}, {Chaplin},
  {Couvidat}, {Eff-Darwich}, {Jim{\'e}nez-Reyes}, {Korzennik}, {Ballot},
  {Boumier}, {Fossat}, {Henney}, {Howe}, {Lazrek}, {Lochard}, {Pall{\'e}}, \&
  {Turck-Chi{\`e}ze}}]{garcia04}
{Garc{\'{\i}}a}, R.~A., {Corbard}, T., {Chaplin}, W.~J., {et~al.} 2004,
  \solphys, 220, 269

\bibitem[{{Garc{\'{\i}}a} {et~al.}(2011){Garc{\'{\i}}a}, {Hekker}, {Stello},
  {Guti{\'e}rrez-Soto}, {Handberg}, {Huber}, {Karoff}, {Uytterhoeven},
  {Appourchaux}, {Chaplin}, {Elsworth}, {Mathur}, {Ballot},
  {Christensen-Dalsgaard}, {Gilliland}, {Houdek}, {Jenkins}, {Kjeldsen},
  {McCauliff}, {Metcalfe}, {Middour}, {Molenda-Zakowicz}, {Monteiro}, {Smith},
  \& {Thompson}}]{garcia11}
{Garc{\'{\i}}a}, R.~A., {Hekker}, S., {Stello}, D., {et~al.} 2011, \mnras, 414,
  L6

\bibitem[{{Gizon} {et~al.}(2013){Gizon}, {Ballot}, {Michel}, {Stahn},
  {Vauclair}, {Bruntt}, {Quirion}, {Benomar}, {Vauclair}, {Appourchaux},
  {Auvergne}, {Baglin}, {Barban}, {Baudin}, {Bazot}, {Campante}, {Catala},
  {Chaplin}, {Creevey}, {Deheuvels}, {Dolez}, {Elsworth}, {Garcia}, {Gaulme},
  {Mathis}, {Mathur}, {Mosser}, {Regulo}, {Roxburgh}, {Salabert}, {Samadi},
  {Sato}, {Verner}, {Hanasoge}, \& {Sreenivasan}}]{gizon13}
{Gizon}, L., {Ballot}, J., {Michel}, E., {et~al.} 2013, Proceedings of the
  National Academy of Science, 110, 13267

\bibitem[{{Gizon} \& {Solanki}(2003)}]{gizon03}
{Gizon}, L. \& {Solanki}, S.~K. 2003, \apj, 589, 1009

\bibitem[{{Gough} \& {Kosovichev}(1993)}]{gough93}
{Gough}, D.~O. \& {Kosovichev}, A.~G. 1993, in Astronomical Society of the
  Pacific Conference Series, Vol.~40, IAU Colloq. 137: Inside the Stars, ed.
  {W.~W.~Weiss \& A.~Baglin}, 566

\bibitem[{{Gough} \& {McIntyre}(1998)}]{gough98}
{Gough}, D.~O. \& {McIntyre}, M.~E. 1998, \nat, 394, 755

\bibitem[{{Goupil} {et~al.}(2013){Goupil}, {Mosser}, {Marques}, {Ouazzani},
  {Belkacem}, {Lebreton}, \& {Samadi}}]{goupil13}
{Goupil}, M.~J., {Mosser}, B., {Marques}, J.~P., {et~al.} 2013, \aap, 549, A75

\bibitem[{{Grevesse} \& {Noels}(1993)}]{grevesse93}
{Grevesse}, N. \& {Noels}, A. 1993, in Origin and Evolution of the Elements,
  ed. {N.~Prantzos, E.~Vangioni-Flam, \& M.~Casse}, 15--25

\bibitem[{{Harvey}(1985)}]{harvey85}
{Harvey}, J. 1985, {High-resolution helioseismology}, Tech. rep.

\bibitem[{{Hastie} \& {Tibshirani}(1990)}]{hastie90}
{Hastie}, T.~J. \& {Tibshirani}, R.~J. 1990, {Generalized Additive Models}

\bibitem[{{Huber} {et~al.}(2011){Huber}, {Bedding}, {Stello}, {Hekker},
  {Mathur}, {Mosser}, {Verner}, {Bonanno}, {Buzasi}, {Campante}, {Elsworth},
  {Hale}, {Kallinger}, {Silva Aguirre}, {Chaplin}, {De Ridder},
  {Garc{\'{\i}}a}, {Appourchaux}, {Frandsen}, {Houdek}, {Molenda-{\.Z}akowicz},
  {Monteiro}, {Christensen-Dalsgaard}, {Gilliland}, {Kawaler}, {Kjeldsen},
  {Broomhall}, {Corsaro}, {Salabert}, {Sanderfer}, {Seader}, \&
  {Smith}}]{huber11}
{Huber}, D., {Bedding}, T.~R., {Stello}, D., {et~al.} 2011, \apj, 743, 143

\bibitem[{{Huber et al.}(2012)}]{huber12}
{Huber et al.} 2012, en pr\'eparation

\bibitem[{{Jenkins} {et~al.}(2010){Jenkins}, {Caldwell}, {Chandrasekaran},
  {Twicken}, {Bryson}, {Quintana}, {Clarke}, {Li}, {Allen}, {Tenenbaum}, {Wu},
  {Klaus}, {Van Cleve}, {Dotson}, {Haas}, {Gilliland}, {Koch}, \&
  {Borucki}}]{jenkins10}
{Jenkins}, J.~M., {Caldwell}, D.~A., {Chandrasekaran}, H., {et~al.} 2010,
  \apjl, 713, L120

\bibitem[{{Karoff}(2012)}]{karoff12}
{Karoff}, C. 2012, \mnras, 421, 3170

\bibitem[{{Karoff} {et~al.}(2013){Karoff}, {Campante}, {Ballot}, {Kallinger},
  {Gruberbauer}, {Garc{\'{\i}}a}, {Caldwell}, {Christiansen}, \&
  {Kinemuchi}}]{karoff13}
{Karoff}, C., {Campante}, T.~L., {Ballot}, J., {et~al.} 2013, \apj, 767, 34

\bibitem[{{Kjeldsen} {et~al.}(2008){Kjeldsen}, {Bedding}, \&
  {Christensen-Dalsgaard}}]{kjeldsen08}
{Kjeldsen}, H., {Bedding}, T.~R., \& {Christensen-Dalsgaard}, J. 2008, \apjl,
  683, L175

\bibitem[{{Kjeldsen} {et~al.}(1995){Kjeldsen}, {Bedding}, {Viskum}, \&
  {Frandsen}}]{kjeldsen95b}
{Kjeldsen}, H., {Bedding}, T.~R., {Viskum}, M., \& {Frandsen}, S. 1995, \aj,
  109, 1313

\bibitem[{{Lebreton} {et~al.}(2008){Lebreton}, {Monteiro}, {Montalb{\'a}n},
  {Moya}, {Baglin}, {Christensen-Dalsgaard}, {Goupil}, {Michel}, {Provost},
  {Roxburgh}, {Scuflaire}, \& {ESTA Team}}]{lebreton08}
{Lebreton}, Y., {Monteiro}, M.~J.~P.~F.~G., {Montalb{\'a}n}, J., {et~al.} 2008,
  \apss, 316, 1

\bibitem[{{Lomb}(1976)}]{lomb76}
{Lomb}, N.~R. 1976, \apss, 39, 447

\bibitem[{{Marques} {et~al.}(2013){Marques}, {Goupil}, {Lebreton}, {Talon},
  {Palacios}, {Belkacem}, {Ouazzani}, {Mosser}, {Moya}, {Morel}, {Pichon},
  {Mathis}, {Zahn}, {Turck-Chi{\`e}ze}, \& {Nghiem}}]{marques13}
{Marques}, J.~P., {Goupil}, M.~J., {Lebreton}, Y., {et~al.} 2013, \aap, 549,
  A74

\bibitem[{{Mathis} \& {Zahn}(2004)}]{mathis04}
{Mathis}, S. \& {Zahn}, J. 2004, \aap, 425, 229

\bibitem[{{Mathur} {et~al.}(2011){Mathur}, {Handberg}, {Campante},
  {Garc{\'{\i}}a}, {Appourchaux}, {Bedding}, {Mosser}, {Chaplin}, {Ballot},
  {Benomar}, {Bonanno}, {Corsaro}, {Gaulme}, {Hekker}, {R{\'e}gulo},
  {Salabert}, {Verner}, {White}, {Brand{\~a}o}, {Creevey}, {Do{\v g}an},
  {Elsworth}, {Huber}, {Hale}, {Houdek}, {Karoff}, {Metcalfe},
  {Molenda-{\.Z}akowicz}, {Monteiro}, {Thompson}, {Christensen-Dalsgaard},
  {Gilliland}, {Kawaler}, {Kjeldsen}, {Quintana}, {Sanderfer}, \&
  {Seader}}]{mathur11}
{Mathur}, S., {Handberg}, R., {Campante}, T.~L., {et~al.} 2011, \apj, 733, 95

\bibitem[{{Menou} {et~al.}(2004){Menou}, {Balbus}, \& {Spruit}}]{menou04}
{Menou}, K., {Balbus}, S.~A., \& {Spruit}, H.~C. 2004, \apj, 607, 564

\bibitem[{{Menou} \& {Le Mer}(2006)}]{menou06}
{Menou}, K. \& {Le Mer}, J. 2006, \apj, 650, 1208

\bibitem[{{Metcalfe} {et~al.}(2010){Metcalfe}, {Monteiro}, {Thompson},
  {Molenda-{\.Z}akowicz}, {Appourchaux}, {Chaplin}, {Do{\u g}an},
  {Eggenberger}, {Bedding}, {Bruntt}, {Creevey}, {Quirion}, {Stello},
  {Bonanno}, {Silva Aguirre}, {Basu}, {Esch}, {Gai}, {Di Mauro}, {Kosovichev},
  {Kitiashvili}, {Su{\'a}rez}, {Moya}, {Piau}, {Garc{\'{\i}}a}, {Marques},
  {Frasca}, {Biazzo}, {Sousa}, {Dreizler}, {Bazot}, {Karoff}, {Frandsen},
  {Wilson}, {Brown}, {Christensen-Dalsgaard}, {Gilliland}, {Kjeldsen},
  {Campante}, {Fletcher}, {Handberg}, {R{\'e}gulo}, {Salabert}, {Schou},
  {Verner}, {Ballot}, {Broomhall}, {Elsworth}, {Hekker}, {Huber}, {Mathur},
  {New}, {Roxburgh}, {Sato}, {White}, {Borucki}, {Koch}, \&
  {Jenkins}}]{metcalfe10}
{Metcalfe}, T.~S., {Monteiro}, M.~J.~P.~F.~G., {Thompson}, M.~J., {et~al.}
  2010, \apj, 723, 1583

\bibitem[{{Molenda-{\.Z}akowicz} {et~al.}(2013){Molenda-{\.Z}akowicz}, {Sousa},
  {Frasca}, {Uytterhoeven}, {Briquet}, {Van Winckel}, {Drobek}, {Niemczura},
  {Lampens}, {Lykke}, {Bloemen}, {Gameiro}, {Jean}, {Volpi}, {Gorlova},
  {Mortier}, {Tsantaki}, \& {Raskin}}]{molenda13}
{Molenda-{\.Z}akowicz}, J., {Sousa}, S.~G., {Frasca}, A., {et~al.} 2013,
  \mnras, 434, 1422

\bibitem[{{Morel}(1997)}]{cesam}
{Morel}, P. 1997, \aaps, 124, 597

\bibitem[{{Mosser} {et~al.}(2011){Mosser}, {Barban}, {Montalb{\'a}n}, {Beck},
  {Miglio}, {Belkacem}, {Goupil}, {Hekker}, {De Ridder}, {Dupret}, {Elsworth},
  {Noels}, {Baudin}, {Michel}, {Samadi}, {Auvergne}, {Baglin}, \&
  {Catala}}]{mosser11}
{Mosser}, B., {Barban}, C., {Montalb{\'a}n}, J., {et~al.} 2011, \aap, 532, A86

\bibitem[{{Mosser} {et~al.}(2012{\natexlab{a}}){Mosser}, {Elsworth}, {Hekker},
  {Huber}, {Kallinger}, {Mathur}, {Belkacem}, {Goupil}, {Samadi}, {Barban},
  {Bedding}, {Chaplin}, {Garc{\'{\i}}a}, {Stello}, {De Ridder}, {Middour},
  {Morris}, \& {Quintana}}]{mosser12c}
{Mosser}, B., {Elsworth}, Y., {Hekker}, S., {et~al.} 2012{\natexlab{a}}, \aap,
  537, A30

\bibitem[{{Mosser} {et~al.}(2012{\natexlab{b}}){Mosser}, {Goupil}, {Belkacem},
  {Marques}, {Beck}, {Bloemen}, {De Ridder}, {Barban}, {Deheuvels}, {Elsworth},
  {Hekker}, {Kallinger}, {Ouazzani}, {Pinsonneault}, {Samadi}, {Stello},
  {Garc{\'{\i}}a}, {Klaus}, {Li}, {Mathur}, \& {Morris}}]{mosser12b}
{Mosser}, B., {Goupil}, M.~J., {Belkacem}, K., {et~al.} 2012{\natexlab{b}},
  \aap, 548, A10

\bibitem[{{Mosser} {et~al.}(2012{\natexlab{c}}){Mosser}, {Goupil}, {Belkacem},
  {Michel}, {Stello}, {Marques}, {Elsworth}, {Barban}, {Beck}, {Bedding}, {De
  Ridder}, {Garc{\'{\i}}a}, {Hekker}, {Kallinger}, {Samadi}, {Stumpe},
  {Barclay}, \& {Burke}}]{mosser12a}
{Mosser}, B., {Goupil}, M.~J., {Belkacem}, K., {et~al.} 2012{\natexlab{c}},
  \aap, 540, A143

\bibitem[{{Mosser} {et~al.}(2013){Mosser}, {Michel}, {Belkacem}, {Goupil},
  {Baglin}, {Barban}, {Provost}, {Samadi}, {Auvergne}, \& {Catala}}]{mosser13}
{Mosser}, B., {Michel}, E., {Belkacem}, K., {et~al.} 2013, \aap, 550, A126

\bibitem[{{Osaki}(1975)}]{osaki75}
{Osaki}, J. 1975, \pasj, 27, 237

\bibitem[{{Ouazzani} {et~al.}(2013){Ouazzani}, {Goupil}, {Dupret}, \&
  {Marques}}]{ouazzani13}
{Ouazzani}, R.-M., {Goupil}, M.~J., {Dupret}, M.-A., \& {Marques}, J.~P. 2013,
  \aap, 554, A80

\bibitem[{{Pinsonneault} {et~al.}(2012){Pinsonneault}, {An},
  {Molenda-{\.Z}akowicz}, {Chaplin}, {Metcalfe}, \& {Bruntt}}]{pinsonneault12}
{Pinsonneault}, M.~H., {An}, D., {Molenda-{\.Z}akowicz}, J., {et~al.} 2012,
  \apjs, 199, 30

\bibitem[{{Raskin} {et~al.}(2011){Raskin}, {van Winckel}, {Hensberge},
  {Jorissen}, {Lehmann}, {Waelkens}, {Avila}, {de Cuyper}, {Degroote},
  {Dubosson}, {Dumortier}, {Fr{\'e}mat}, {Laux}, {Michaud}, {Morren}, {Perez
  Padilla}, {Pessemier}, {Prins}, {Smolders}, {van Eck}, \&
  {Winkler}}]{raskin11}
{Raskin}, G., {van Winckel}, H., {Hensberge}, H., {et~al.} 2011, \aap, 526, A69

\bibitem[{{Scargle}(1982)}]{scargle82}
{Scargle}, J.~D. 1982, \apj, 263, 835

\bibitem[{{Schou} {et~al.}(1998){Schou}, {Antia}, {Basu}, {Bogart}, {Bush},
  {Chitre}, {Christensen-Dalsgaard}, {di Mauro}, {Dziembowski}, {Eff-Darwich},
  {Gough}, {Haber}, {Hoeksema}, {Howe}, {Korzennik}, {Kosovichev}, {Larsen},
  {Pijpers}, {Scherrer}, {Sekii}, {Tarbell}, {Title}, {Thompson}, \&
  {Toomre}}]{schou98}
{Schou}, J., {Antia}, H.~M., {Basu}, S., {et~al.} 1998, \apj, 505, 390

\bibitem[{{Scuflaire} {et~al.}(2008){Scuflaire}, {Montalb{\'a}n}, {Th{\'e}ado},
  {Bourge}, {Miglio}, {Godart}, {Thoul}, \& {Noels}}]{losc}
{Scuflaire}, R., {Montalb{\'a}n}, J., {Th{\'e}ado}, S., {et~al.} 2008, \apss,
  316, 149

\bibitem[{{Silva Aguirre} {et~al.}(2012){Silva Aguirre}, {Casagrande}, {Basu},
  {Campante}, {Chaplin}, {Huber}, {Miglio}, {Serenelli}, {Ballot}, {Bedding},
  {Christensen-Dalsgaard}, {Creevey}, {Elsworth}, {Garc{\'{\i}}a}, {Gilliland},
  {Hekker}, {Kjeldsen}, {Mathur}, {Metcalfe}, {Monteiro}, {Mosser},
  {Pinsonneault}, {Stello}, {Weiss}, {Tenenbaum}, {Twicken}, \&
  {Uddin}}]{silva12}
{Silva Aguirre}, V., {Casagrande}, L., {Basu}, S., {et~al.} 2012, \apj, 757, 99

\bibitem[{{Skrutskie} {et~al.}(2006){Skrutskie}, {Cutri}, {Stiening},
  {Weinberg}, {Schneider}, {Carpenter}, {Beichman}, {Capps}, {Chester},
  {Elias}, {Huchra}, {Liebert}, {Lonsdale}, {Monet}, {Price}, {Seitzer},
  {Jarrett}, {Kirkpatrick}, {Gizis}, {Howard}, {Evans}, {Fowler}, {Fullmer},
  {Hurt}, {Light}, {Kopan}, {Marsh}, {McCallon}, {Tam}, {Van Dyk}, \&
  {Wheelock}}]{skrutskie06}
{Skrutskie}, M.~F., {Cutri}, R.~M., {Stiening}, R., {et~al.} 2006, \aj, 131,
  1163

\bibitem[{{Sneden}(1973)}]{sneden73}
{Sneden}, C.~A. 1973, PhD thesis, THE UNIVERSITY OF TEXAS AT AUSTIN.

\bibitem[{{Sousa} {et~al.}(2007){Sousa}, {Santos}, {Israelian}, {Mayor}, \&
  {Monteiro}}]{sousa07}
{Sousa}, S.~G., {Santos}, N.~C., {Israelian}, G., {Mayor}, M., \& {Monteiro},
  M.~J.~P.~F.~G. 2007, \aap, 469, 783

\bibitem[{{Spruit}(1999)}]{spruit99}
{Spruit}, H.~C. 1999, \aap, 349, 189

\bibitem[{{Strugarek} {et~al.}(2011){Strugarek}, {Brun}, \&
  {Zahn}}]{stugarek11}
{Strugarek}, A., {Brun}, A.~S., \& {Zahn}, J.-P. 2011, \aap, 532, A34

\bibitem[{{Talon} \& {Charbonnel}(2005)}]{talon05}
{Talon}, S. \& {Charbonnel}, C. 2005, \aap, 440, 981

\bibitem[{{Talon} \& {Charbonnel}(2008)}]{talon08}
{Talon}, S. \& {Charbonnel}, C. 2008, \aap, 482, 597

\bibitem[{{Tayar} \& {Pinsonneault}(2013)}]{tayar13}
{Tayar}, J. \& {Pinsonneault}, M.~H. 2013, ArXiv e-prints

\bibitem[{{Thygesen} {et~al.}(2012){Thygesen}, {Frandsen}, {Bruntt},
  {Kallinger}, {Andersen}, {Elsworth}, {Hekker}, {Karoff}, {Stello},
  {Brogaard}, {Burke}, {Caldwell}, \& {Christiansen}}]{thygesen12}
{Thygesen}, A.~O., {Frandsen}, S., {Bruntt}, H., {et~al.} 2012, \aap, 543, A160

\bibitem[{{van Saders} \& {Pinsonneault}(2013)}]{vansaders13}
{van Saders}, J.~L. \& {Pinsonneault}, M.~H. 2013, ArXiv e-prints

\bibitem[{{V{\'a}zquez Rami{\'o}} {et~al.}(2005){V{\'a}zquez Rami{\'o}},
  {R{\'e}gulo}, \& {Roca Cort{\'e}s}}]{vazquez05}
{V{\'a}zquez Rami{\'o}}, H., {R{\'e}gulo}, C., \& {Roca Cort{\'e}s}, T. 2005,
  \aap, 443, L11

\bibitem[{{Yang}(2013)}]{yang13}
{Yang}, X.~H. 2013, in prep

\bibitem[{{Zahn}(1992)}]{zahn92}
{Zahn}, J.-P. 1992, \aap, 265, 115

\end{thebibliography}


\begin{appendix}

\section{Computation of asymptotic period spacings from stellar models \label{app_deltap}}

According to the asymptotic analysis of g modes (e.g. \citealt{lecturenotesJCD}), the frequencies of two g modes of same degree $l$ and consecutive radial order $n$ verify the following equation
\begin{linenomath*}
\begin{equation}
L\left[I(\omega_{n+1})-I(\omega_{n})\right] = \pi
\label{eq_deltap1}
\end{equation}
\end{linenomath*}
where $L\equiv\sqrt{l(l+1)}$ and
\begin{linenomath*}
\begin{equation}
I(\omega) \equiv \int_{r_a(\omega)}^{r_b(\omega)} \, \left(\frac{N_{\rm BV}^2}{\omega^2}-1\right)^{1/2}\frac{dr}{r}.
\label{eq_deltap2}
\end{equation}
\end{linenomath*}
The radii $r_a(\omega)$ and $r_b(\omega)$ correspond to the turning points of the g-mode cavity for the mode of frequency $\omega$. For red giants, one usually assumes that $\omega\ll N\ind{BV}(r)$ in the whole g-mode cavity and thus derives
\begin{linenomath*}
\begin{equation}
\Delta\Pi=\frac{2\pi^2}{L}\left(\int_{r_a(\omega_{n})}^{r_b(\omega_n)} \frac{N_{\rm BV}}{r} \,dr\right)^{-1}
\label{eq_deltap_RG}
\end{equation}
\end{linenomath*}
from Eq. \ref{eq_deltap1}.
For young giants, this approximation is not legitimate because the core is less dense than in more evolved giants and the integrand of Eq. \ref{eq_deltap2} cannot be simplified. In this case, computing the asymptotic period spacing from a model at a frequency $\omega_n$ requires an iterative method. By definition of the period spacing, we have
\begin{linenomath*}
\begin{equation}
\omega_{n+1}=\left(\frac{1}{\omega_n}+\frac{\Delta\Pi_l^{\rm mod}}{2\pi}\right)^{-1}
\label{eq_deltap3}
\end{equation}
\end{linenomath*}
This expression of $\omega_{n+1}$ is then plugged into Eq. \ref{eq_deltap1} and we use the Newton method to solve this latter equation for $\Delta\Pi_l^{\rm mod}$. The obtained values are given in Table \ref{tab_deltap1}. They differ from those given by Eq. \ref{eq_deltap_RG} by 3\% (for the most evolved targets) to 11\% (for the least evolved ones). This shows a posteriori that the usual approximation is not valid for the young red giants.

\section{Effective number of degrees of freedom for RLS fit \label{app_dof}}

Estimating the number of degrees of freedom for a RLS fit is not straightforward because this number cannot be interpreted as the dimension of a certain vector subspace as is the case for ordinary least squares fits. However, the fitted values (the rotational splittings that correspond to the inverted rotation profile $\vect{\delta\nu}\ex{RLS}$ in our case) remain linear in the observations $\vect{\delta\nu}\ex{obs}$. 
There exists a matrix $\vect{H}$, usually referred to as the \textit{hat} matrix, such that
\begin{linenomath*}
\begin{equation}
\vect{\delta\nu}\ex{RLS}=\vect{H}\,\vect{\delta\nu}\ex{obs}.
\end{equation}
\end{linenomath*}
We note that both the matrix $\vect{H}$ and the vector $\vect{\delta\nu}\ex{RLS}$ depend on the regularization parameter.

The number of degrees of freedom of the fit corresponds to the expected value of $\chi^2$, which is given by Eq. \ref{eq_chi2_RLS}. To calculate $E(\chi^2)$, we make the assumption that the RLS fit is unbiased, i.e. we assume that the "true" splittings would be recovered with this method if the observations were noise-free. This assumption is probably not completely justified, but if a bias exists, it will increase the expected value of the $\chi^2$. Therefore, by neglecting the bias, we obtain higher values of the reduced $\chi^2$ and our assumption is thus conservative. We also consider that the observed splittings are uncorrelated. In these conditions, we can use the development of \cite{hastie90}. The only difference is that in our case the variances of the data points are not all the same. To place ourselves in this particular case, we normalize the rotational splittings by the observed errors by defining 
\begin{linenomath*}
\begin{equation}
\widetilde{\delta\nu}\ex{RLS}_k\equiv\frac{\delta\nu\ex{RLS}}{\sigma_k} \;\;\; \hbox{and} \;\;\; \widetilde{\delta\nu}\ex{obs}_k\equiv\frac{\delta\nu\ex{obs}}{\sigma_k}, \; \forall k
\end{equation}
\end{linenomath*}
We thus have
\begin{linenomath*}
\begin{equation}
\vect{\widetilde{\delta\nu}}\ex{RLS}=\vect{\widetilde{H}}\,\vect{\widetilde{\delta\nu}}\ex{obs}
\end{equation}
\end{linenomath*}
where the corresponding hat matrix is such that $\widetilde{H}_{i,j}\equiv H_{i,j}\sigma_j/\sigma_i$. In this case, \cite{hastie90} have shown that the expected value of $\chi^2$ is $E(\chi^2) = M+{\rm Tr}(\vect{\widetilde{H}\widetilde{H}}^T)-2{\rm Tr}(\vect{\widetilde{H}})$. Finally we obtain
\begin{linenomath*}
\begin{equation}
E(\chi^2)=M+\sum_{i,k} H_{k,i}^2 \frac{\sigma_i^2}{\sigma_k^2}-2\sum_k H_{k,k}.
\label{eq_esperance}
\end{equation}
\end{linenomath*}


\section{Can we distinguish between a smooth and a discontinuous rotation profile in the stars of the sample? \label{sect_inv_simu}}

\begin{figure*}
\begin{center}
\includegraphics[width=6cm]{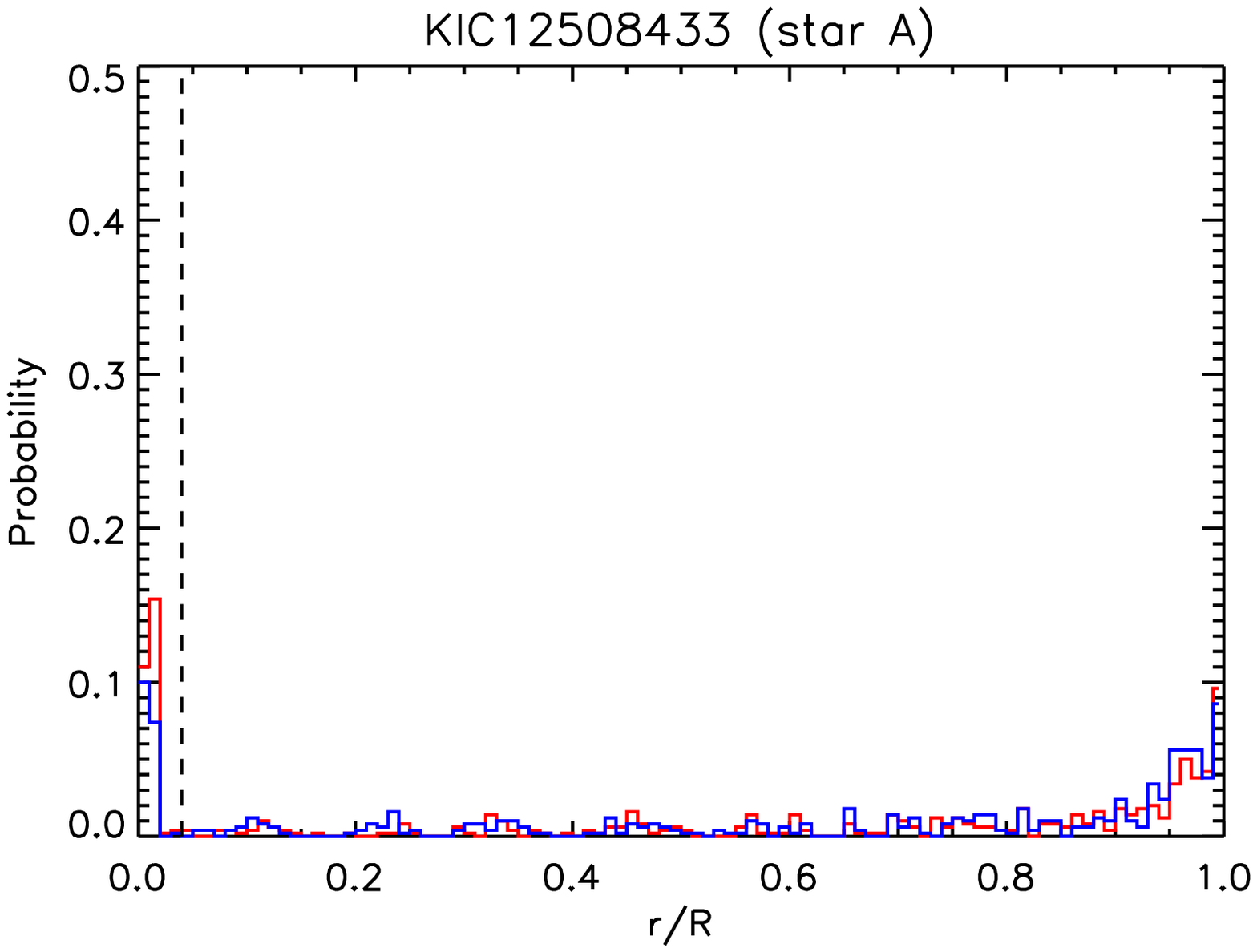}
\includegraphics[width=6cm]{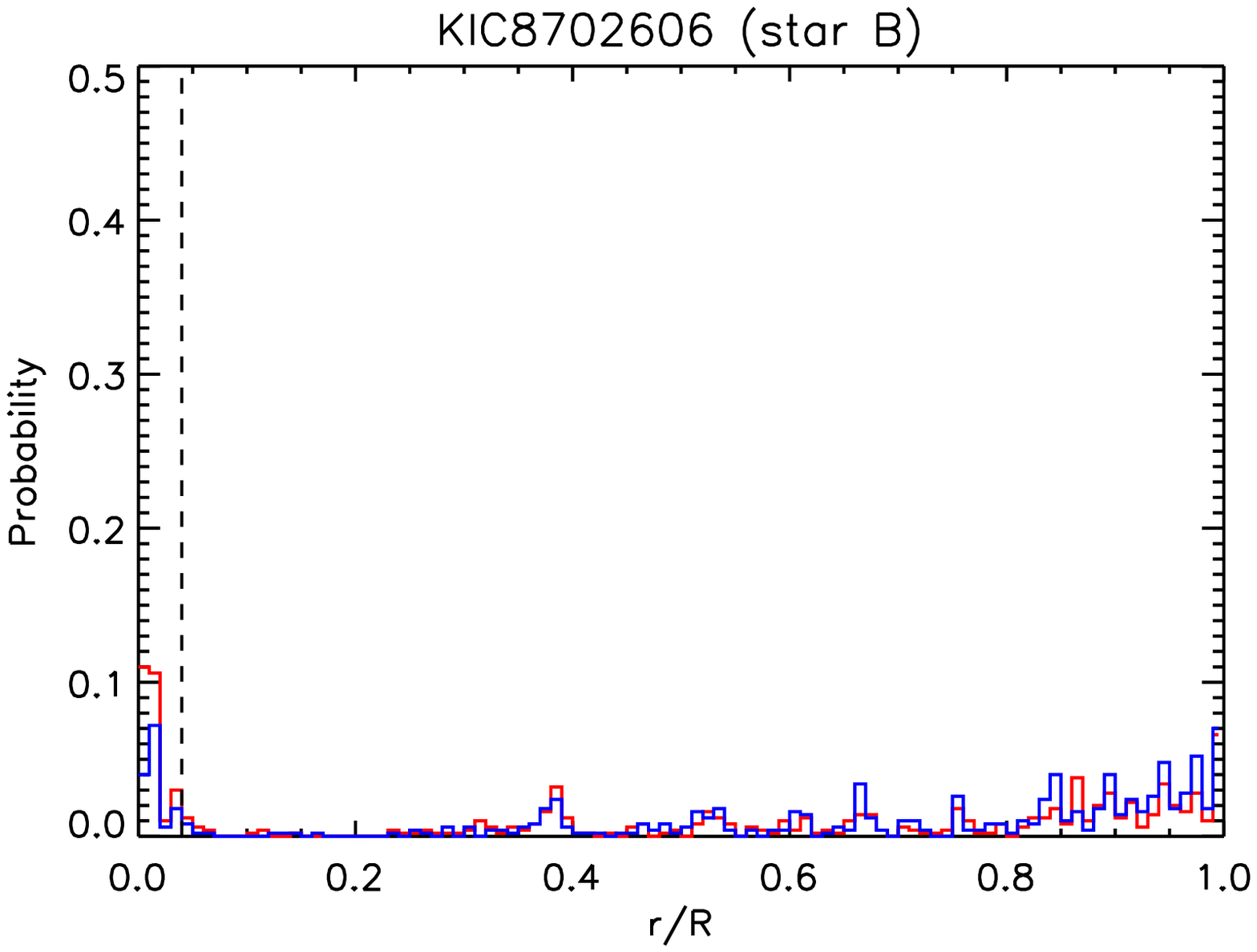}
\includegraphics[width=6cm]{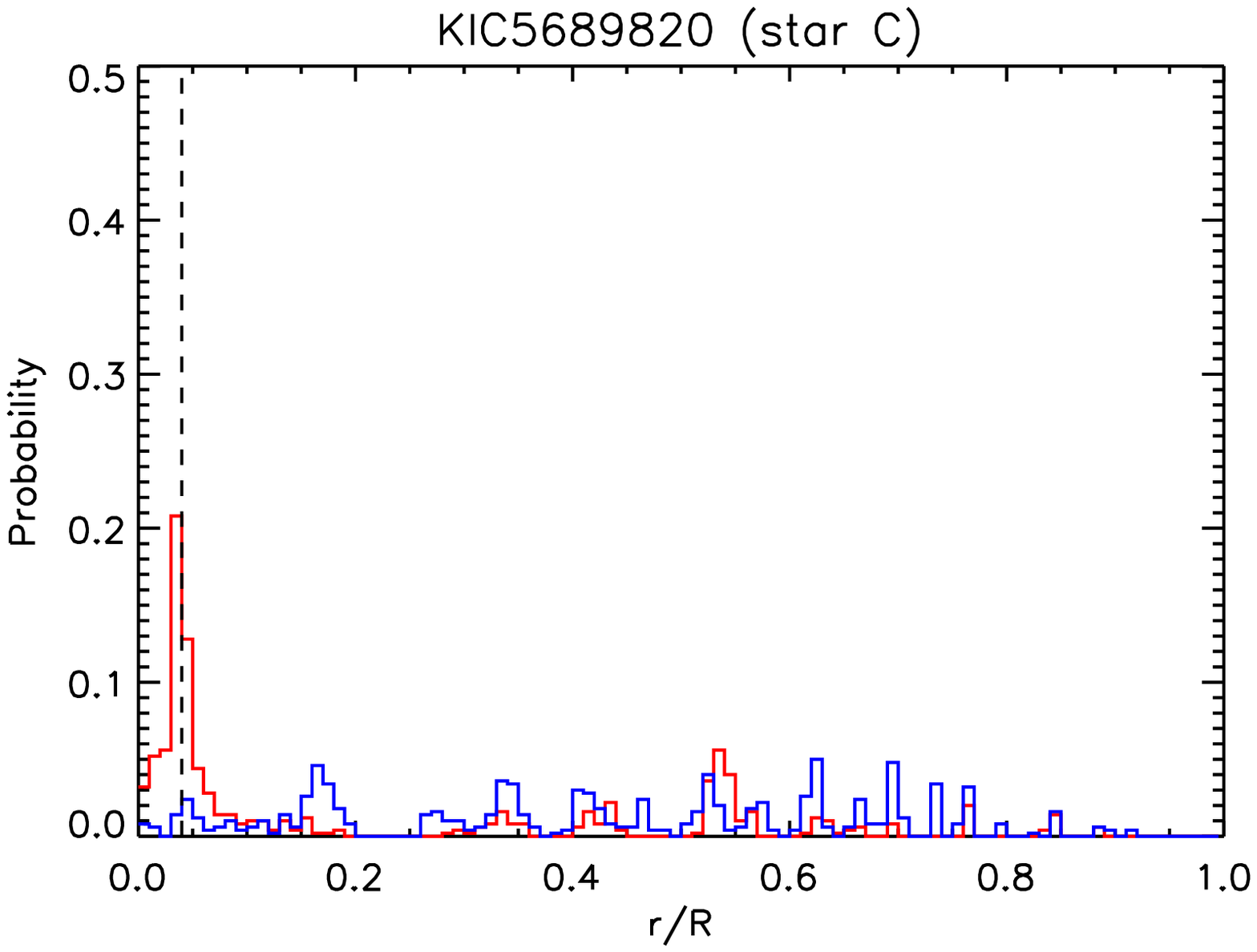}
\includegraphics[width=6cm]{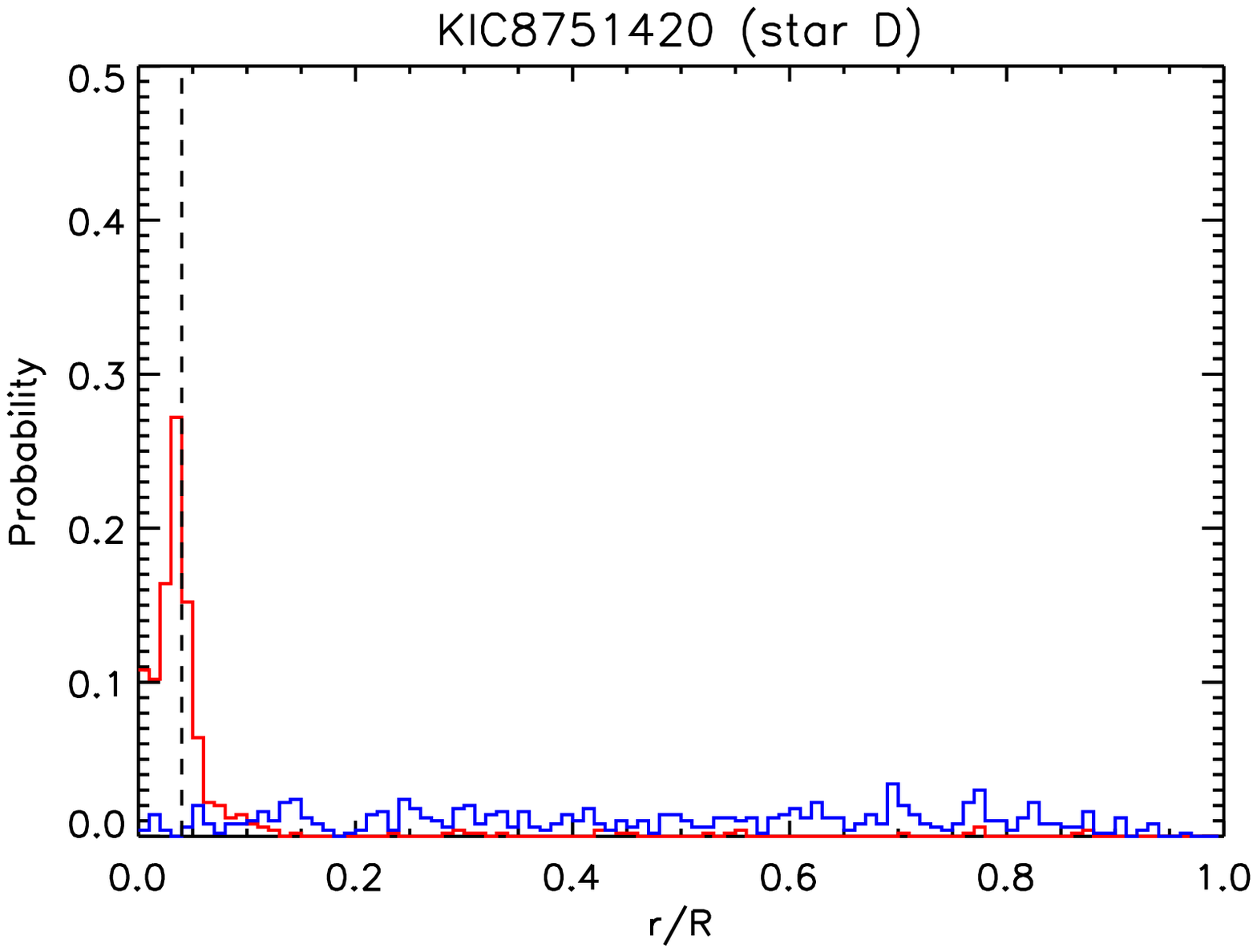}
\includegraphics[width=6cm]{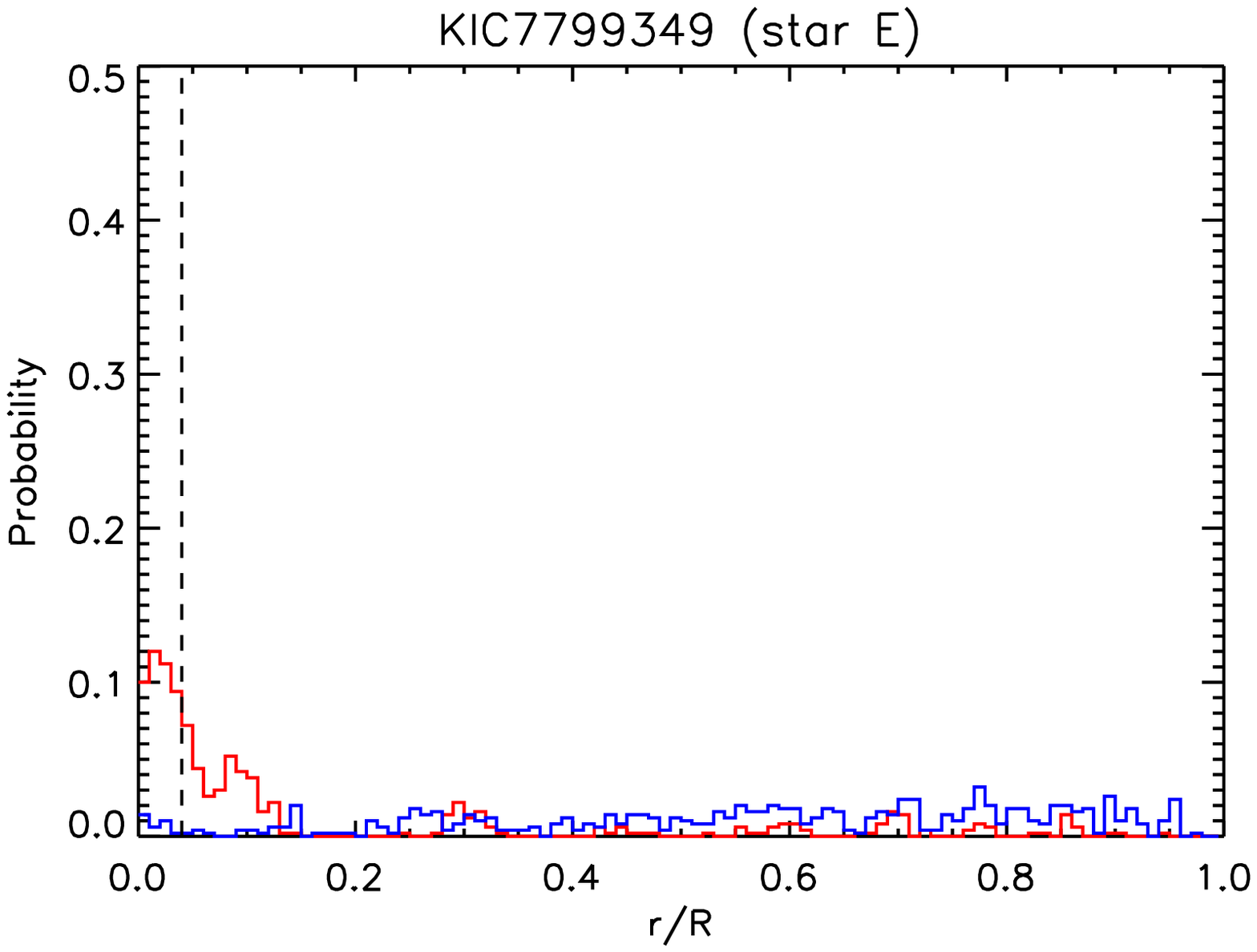}
\includegraphics[width=6cm]{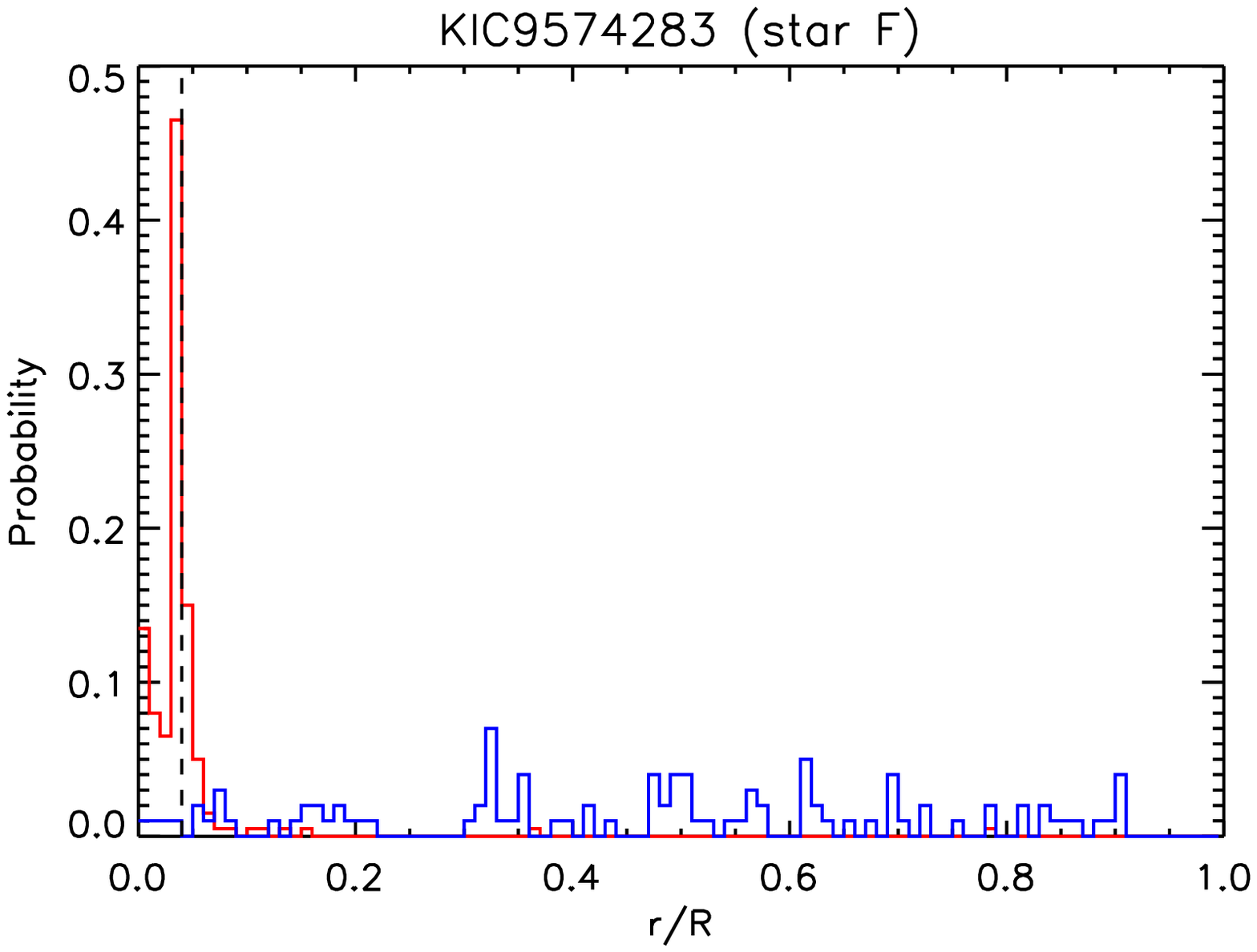}
\end{center}
\caption{Distribution of the depth $r\ind{min}$ at which a discontinuity in $\Omega(r)$ is found most probable for simulations that consider an input  rotation profile that is either smooth (blue histogram) or discontinuous at a depth $r\ind{c}\ex{th}=0.04\,R_\star$ (red histogram). The location of the true discontinuity for the latter case is indicated by the vertical dashed line.
\label{fig_rmin}}
\end{figure*}

\subsection{Simulations with a discontinuous rotation profile}

To answer this question, we performed simulations by assuming a discontinuous input rotation profile $\Omega\ex{th}(r)$. We took $r\ind{c}\ex{th}=0.04\,R_\star$ to determine whether or not a discontinuity at this depth can be detected.
For each star, the values of $\Omega_1$ and $\Omega_2$ were fixed to the optimal values that were found by assuming $r\ind{c}=0.04\,R_\star$ in Sect. \ref{sect_disc}. We then used the rotational kernels of the observed modes to compute the theoretical rotational splittings $\delta\nu_i\ex{th}$ that correspond to $\Omega\ex{th}(r)$. We used the same sets of modes as the observed ones. To simulate observed splittings, we added to the theoretical splittings $\delta\nu_i\ex{th}$ a random noise following a gaussian distribution with rms $\sigma_i\ex{obs}$. We then tried to recover the input rotation profile $\Omega\ex{th}(r)$ by performing the same inversion procedures as for the observations. The last two steps were repeated a large number of times (500 iterations per star for each inversion method), in order to study the statistics.

We first performed inversions by assuming that the rotation profile is discontinuous. We minimized the $\chi^2$ as defined by Eq. \ref{eq_chi2_2zone} for different values of $r\ind{c}$. To determine whether or not the input depth of the discontinuity $r\ind{c}\ex{th}$ can be recovered, we located the radius $r\ind{disc}$ at which the $\chi^2$ function is minimal for each iteration. Correspondingly, the minimum value of the $\chi^2$ function is further noted $\chi^2\ind{disc}$.
The distribution of $r\ind{disc}$ is shown in Fig. \ref{fig_rmin} for all the stars. We remark that for stars D and F, the depth of the discontinuity is correctly recovered in 90\% and 97\% of the cases, respectively (we arbitrarily considered that the depth of the discontinuity is recovered if $r\ind{disc}=r\ind{c}\ex{th}\pm0.04\,R_\star$), which already suggests that we should be able to detect a discontinuity in the rotation profile at a depth of $r\ind{c}\ex{th}=0.04\,R_\star$ in these stars. For the least evolved stars A and B, the success rate is much smaller (28\% for both stars). For stars C and E, the correct discontinuity is recovered about 60\% of the time.


\begin{figure*}
\begin{center}
\includegraphics[width=6cm]{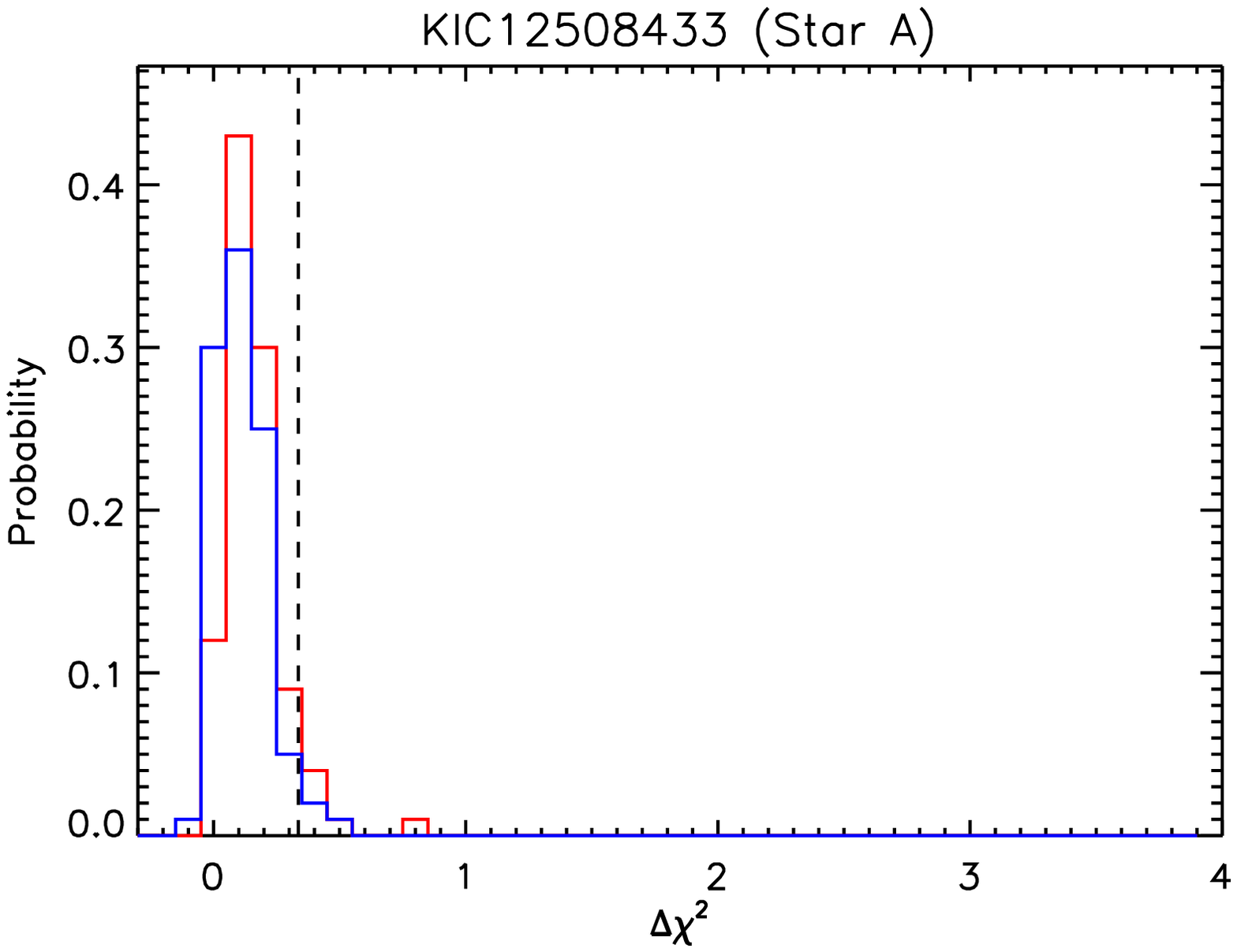}
\includegraphics[width=6cm]{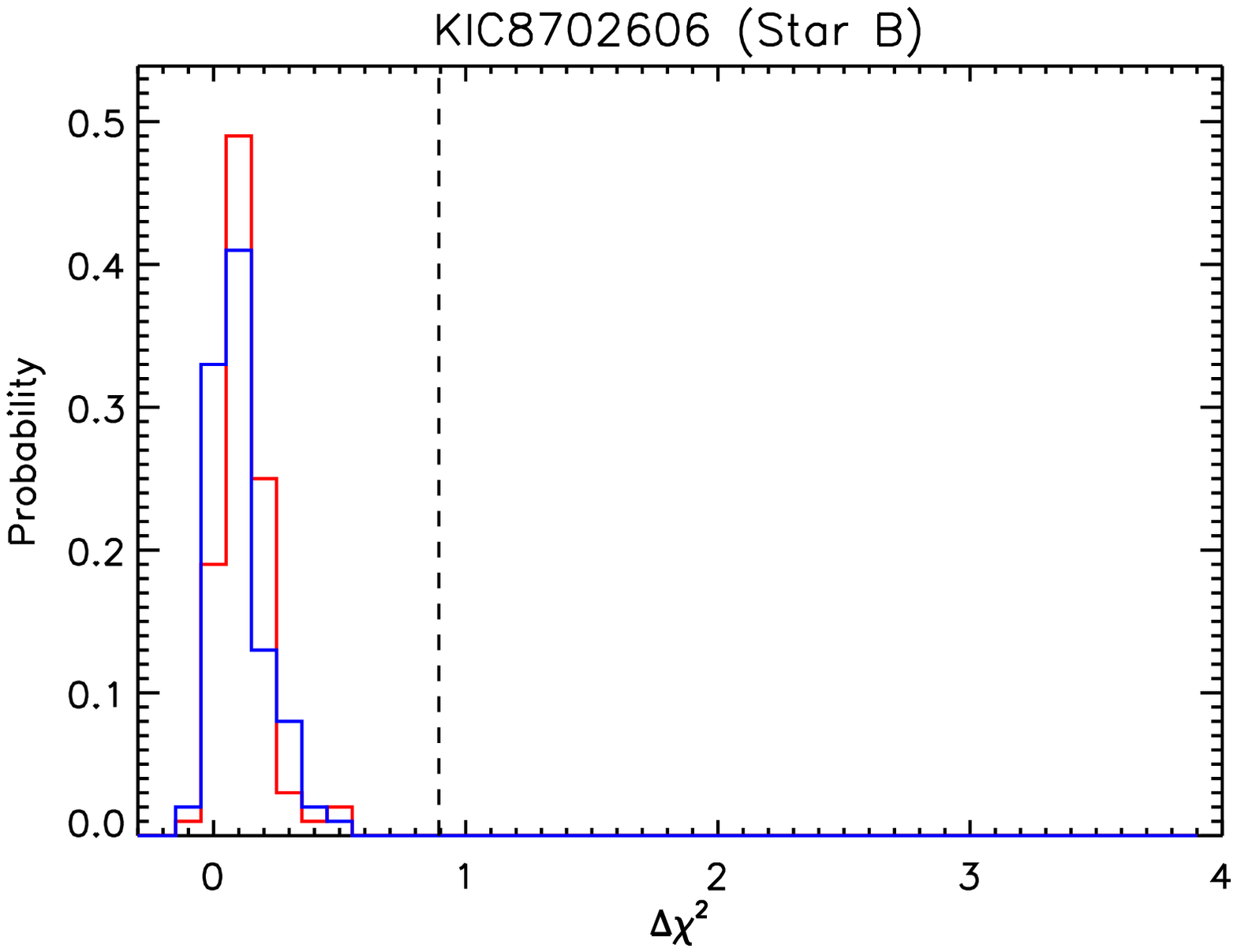}
\includegraphics[width=6cm]{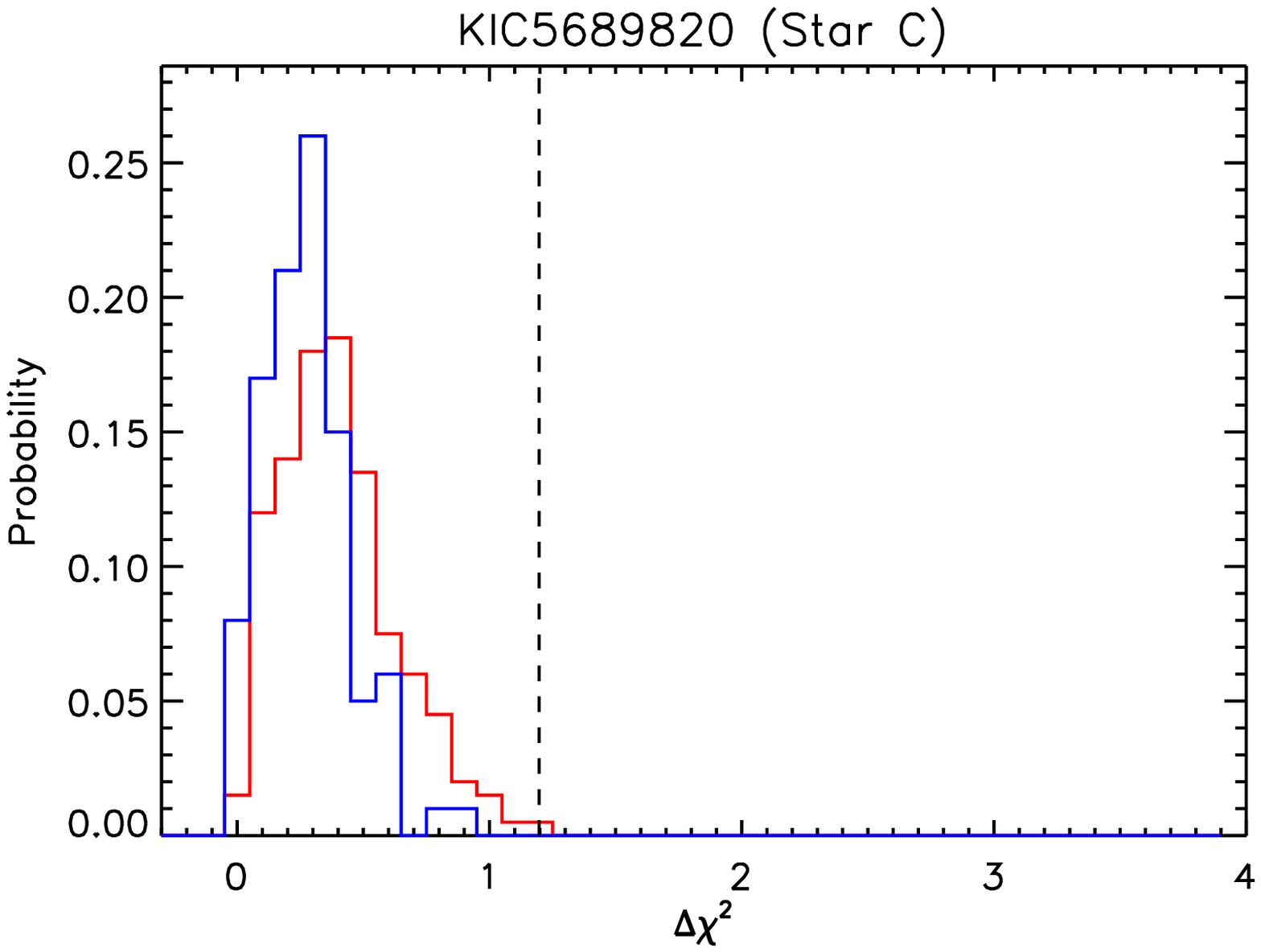}
\includegraphics[width=6cm]{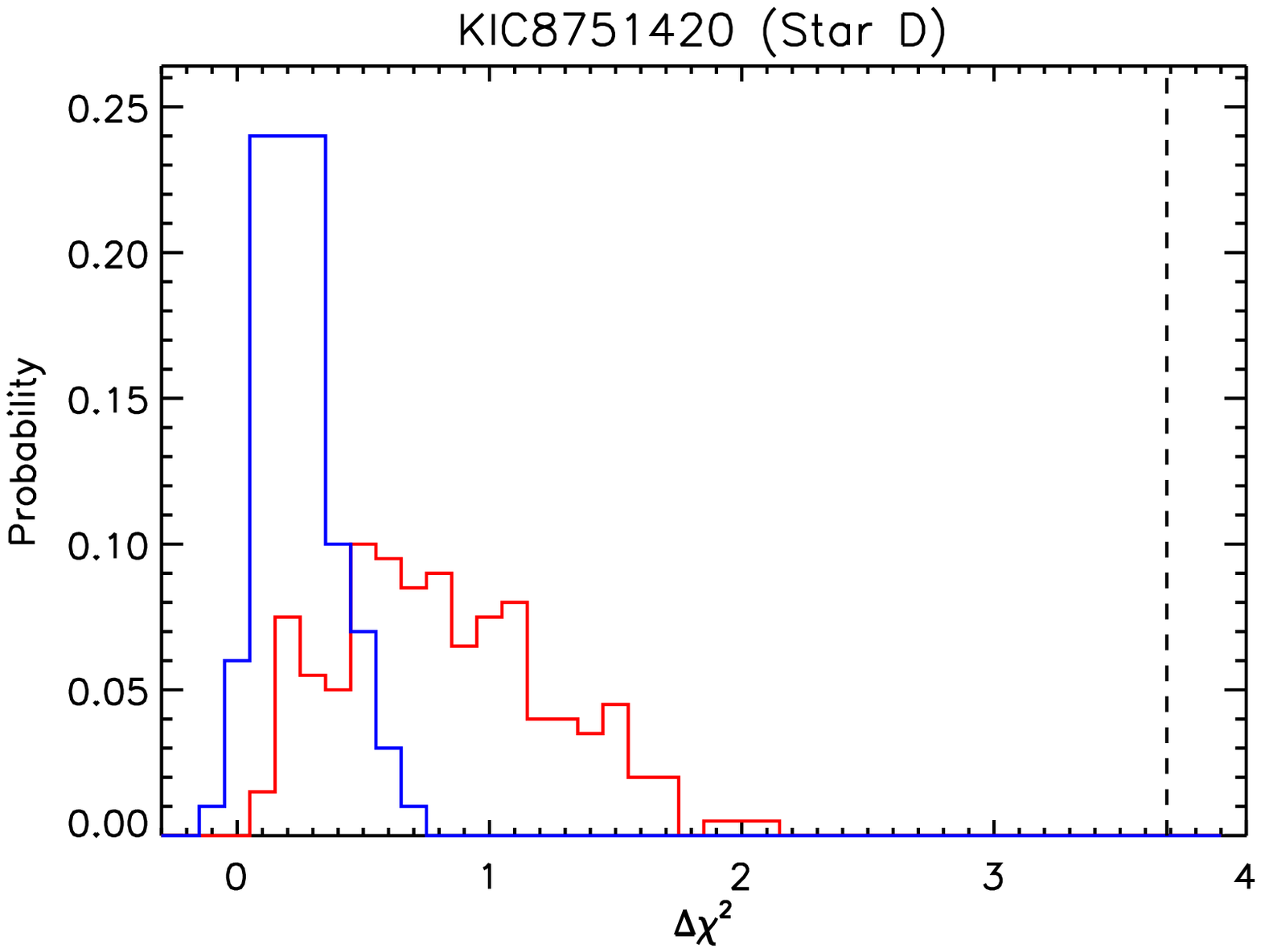}
\includegraphics[width=6cm]{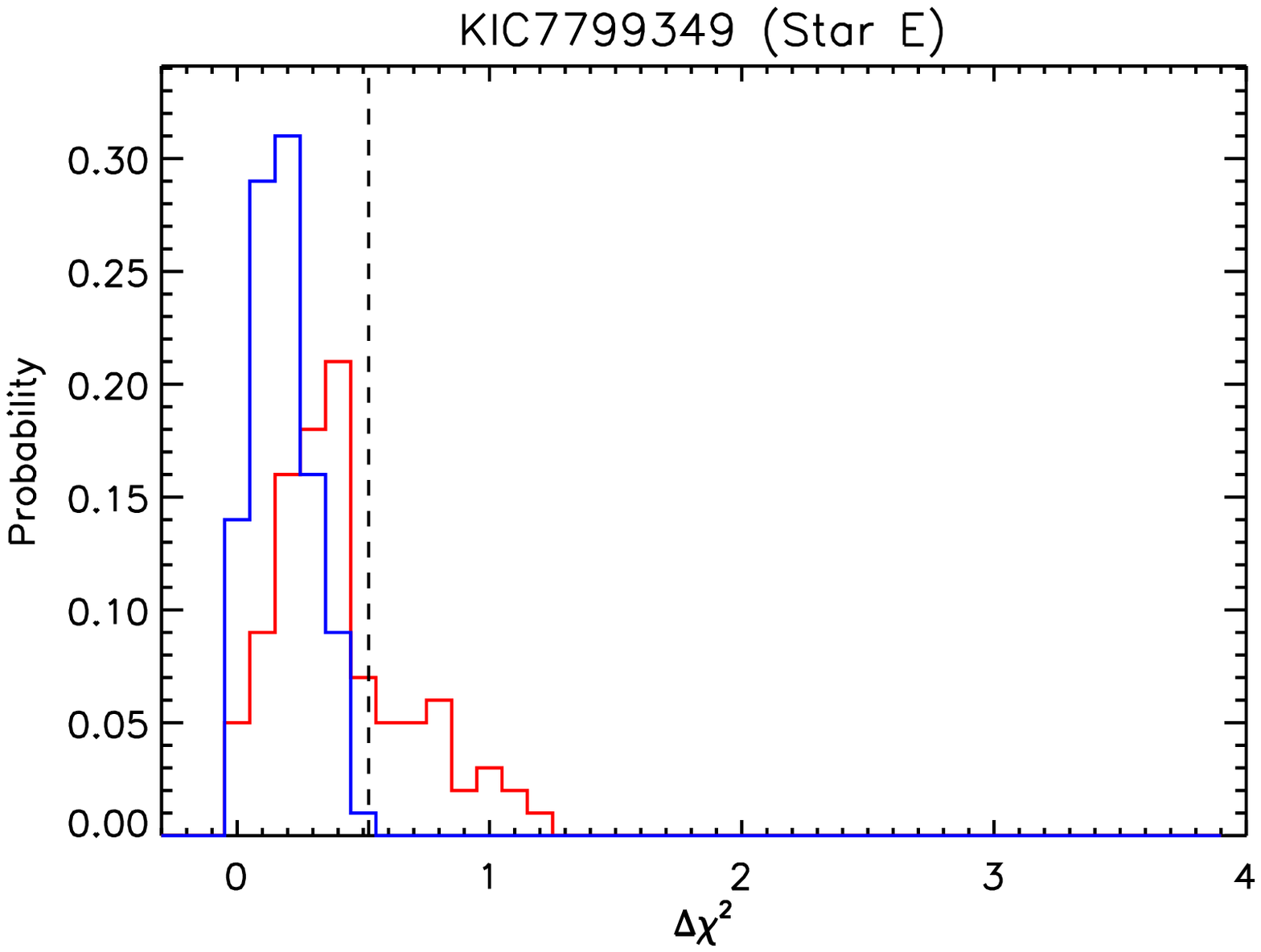}
\includegraphics[width=6cm]{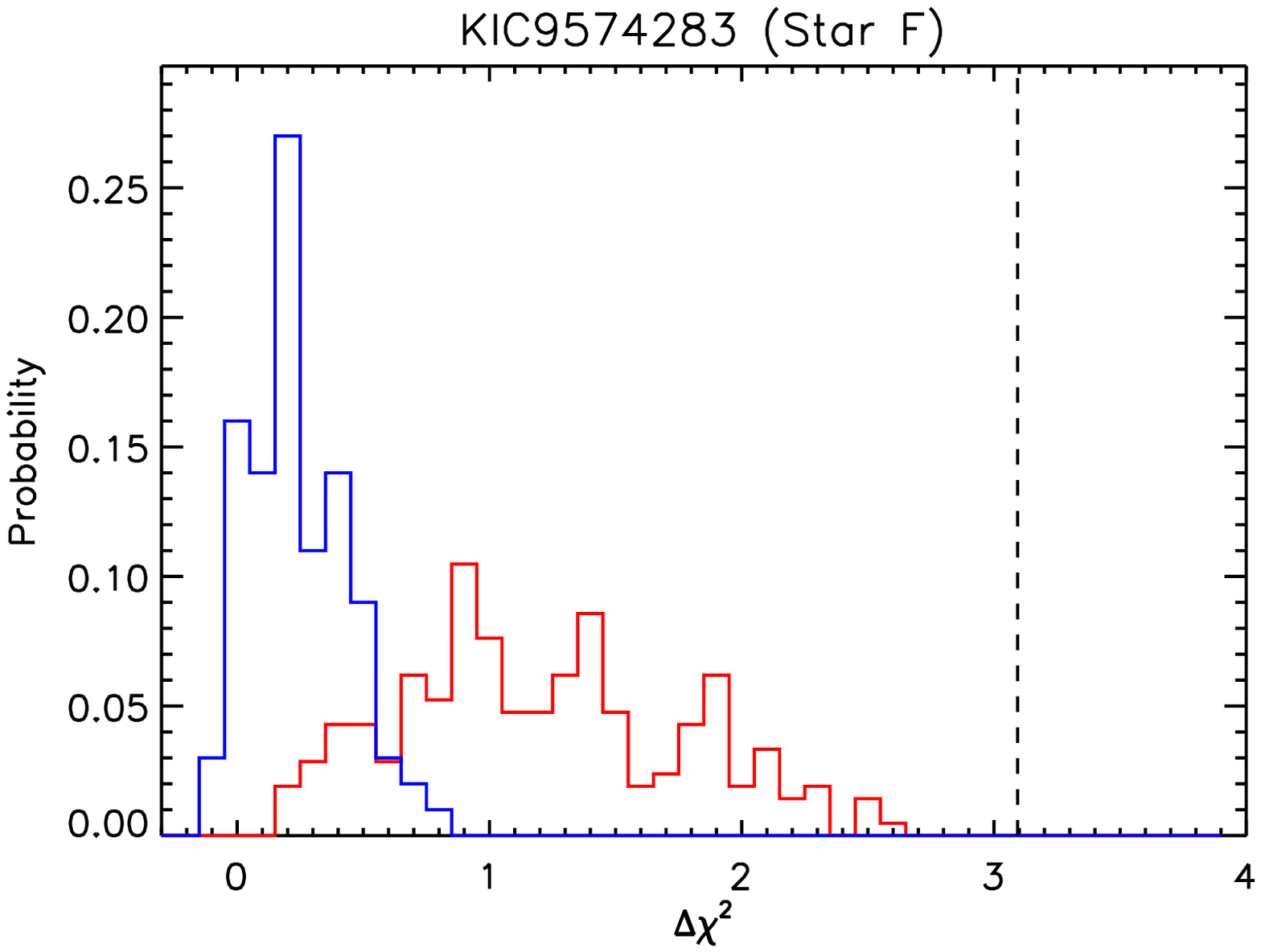}
\end{center}
\caption{Distribution of $\Delta\chi^2 \equiv \chi^2\ind{smooth}-\chi^2\ind{disc}$ (see text) for simulations that consider an input rotation profile that is either smooth (blue histograms) or discontinuous at a depth $r\ind{c}\ex{th}=0.04\,R_\star$ (red histogram). The observed values of $\Delta\chi^2$ are indicated by the vertical dashed lines for the six stars.
\label{fig_histo_delta_chi2}}
\end{figure*}

We then tried to recover the rotation profile by (wrongly) assuming that it is smooth for each of the 500 iterations. For this purpose, we used the RLS method with a smoothness condition, as was done for the observations in Sect. \ref{sect_smooth}. Of course, the resulting profiles contain no discontinuity. For each iteration, the agreement between the simulated splittings and the ones corresponding to the inverted rotation profile was estimated by computing a $\chi^2$, further referred to as $\chi^2\ind{smooth}$. We compared the values of $\chi^2\ind{smooth}$ to those of $\chi^2\ind{disc}$, which were obtained above, assuming that the profile is discontinuous. Fig. \ref{fig_histo_delta_chi2} shows the distribution of $\Delta\chi^2\equiv\chi^2\ind{smooth}-\chi^2\ind{disc}$ for each star. For stars D and F, $\chi^2\ind{smooth}$ was found larger than $\chi^2\ind{disc}$ for all the 500 iterations, with $\Delta\chi^2 = 0.8\pm0.4$ for star D and $1.1\pm0.6$ for star F. 

This set of simulations shows that for stars D and F, a discontinuity located at $r\ind{c}\ex{th}=0.04\,R_\star$ is recovered more than 90\% of the time, and the agreement with the input theoretical splittings is better when the rotation profile is assumed to be discontinuous than when it is assumed to be smooth ($\chi^2\ind{smooth}-\chi^2\ind{disc}\sim 1$). To determine whether or not the profile is indeed discontinuous in these stars, we must also show that a smooth profile has little chances of producing the same features.

%

\subsection{Simulations with a smooth rotation profile}

For this purpose, we considered an input rotation profile that is smooth. We took the optimal profile obtained with the RLS method in Sect. \ref{sect_smooth}. As before, we computed the theoretical splittings that correspond to this rotation profile, added a random noise to them, and tried to recover the input rotation profile. 500 iterations were performed for each star.

We first performed inversions by (wrongly) assuming that the rotation profile is discontinuous. For each iteration, we determined the radius $r\ind{disc}$ at which a discontinuity is most probable and we kept a record of the minimum $\chi^2$ ($\chi^2\ind{disc}$). The distributions of $r\ind{disc}$ that we obtained for all the stars are shown in Fig. \ref{fig_rmin} (blue histograms). For stars A, B, C, and E, we observe that $r\ind{disc}$ is distributed more or less randomly between 0 and 1. In particular, for stars A and B, the distribution of $r\ind{disc}$ is very similar to the one that was obtained when considering a discontinuous input profile, which confirms the impossibility to detect a discontinuity in the rotation profile for these stars. For stars D and F, some radii are more probable than others (e.g. 0.16, 0.33, 0.50, 0.60, and 0.90 $R_\star$ for star F). This means that the splittings that are produced by a smooth rotation profile are quite similar to the ones that are produced by discontinuous profiles with specific depths of discontinuity. It would therefore be hopeless to try to detect a discontinuity around these specific radii. However, smooth profiles very seldom produce rotational splittings that are similar to the ones that correspond to a discontinuous profile with $r\ind{c}$ as deep as $0.04 \,R_\star$. Indeed, for stars D and F, a value of $r\ind{disc}$ below 0.08 $R_\star$ was obtained for only 4\% and 2\% of the iterations, respectively. 

\begin{figure}
\begin{center}
\includegraphics[width=9cm]{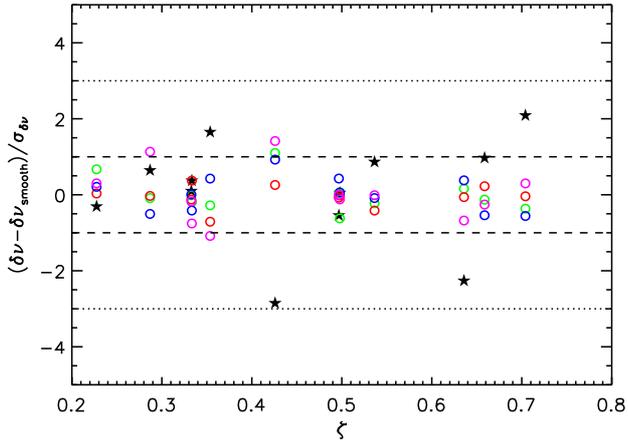}
\end{center}
\caption{Differences between the theoretical splittings that are obtained with a smooth rotation profile and the ones that are produced by discontinuous rotation profiles with depths of discontinuity of $r\ind{c}/R_\star=0.04$ (filled stars), 0.2 (green circles), 0.4 (red circles), 0.6 (blue circles), and 0.8 (purple circles) for star D (KIC8751420). The differences were normalized by the 1-$\sigma$ errors of the observations to emphasize the contribution of each mode to the $\chi^2$.
\label{fig_diff_split_875}}
\end{figure}

To understand this result, a closer inspection of the theoretical splittings (without noise) is instructive. Fig. \ref{fig_diff_split_875} shows the difference between the theoretical splittings of the input smooth profile and those of optimal discontinuous profiles with various values of $r\ind{c}$ for star D. We observe that for discontinuities between 0.2 and 1, the theoretical splittings do not vary much and are rather close to the splittings of the smooth profile. On the other hand, for discontinuities that are deeper (e.g. around $r\ind{c}=0.04\,R_\star$), the splittings of several modes significantly differ from the case of a smooth profile (for example the modes around 454 and 582 $\mu$Hz for star D, see Fig. \ref{fig_diff_split_875}), which explains why we can distinguish between these two types of profile. 


For each iteration, we also performed RLS inversions with a smoothness condition and estimated the agreement with the theoretical splittings by computing $\chi^2\ind{smooth}$. The distributions of $\chi^2\ind{smooth}-\chi^2\ind{disc}$ are shown in Fig. \ref{fig_histo_delta_chi2} (blue histograms). For all the stars, the values of $\chi^2\ind{smooth}$ are similar to those of $\chi^2\ind{disc}$. 
This was foreseeable since we saw from Fig. \ref{fig_rmin} that smooth rotation profiles produce splittings that resemble those of discontinuous profiles with discontinuities at certain depths. However, for stars D and F, if the input profile is smooth, $\chi^2\ind{smooth}$ is never larger than $\chi^2\ind{disc}$ by more than 0.8 and 0.9, respectively, whereas the observed values of $\Delta\chi^2$ are as large as 3.6 and 3.1 for these stars, which is much more consistent with the distribution found for discontinuous input profiles. For all the other stars, the observed $\Delta\chi^2$ are equally consistent with the two types of profiles.


To determine to what extent our results depend on the shape of the smooth rotation profile that is used as an input, we repeated the same simulations considering other smooth profiles (e.g. the optimal rotation profiles obtained by using the OLA method in Sect. \ref{sect_OLA}, or an optimal linear profile). We obtained results that are quantitatively very similar to the ones described above.



\end{appendix}

\end{document}